\documentclass[acmsmall,dvipsnames,x11names,svgnames,cleveref]{acmart}
\pdfoutput=1
\acmJournal{PACMPL}
\acmVolume{1}
\acmNumber{OOPSLA} 
\acmArticle{1}
\acmYear{2022}
\acmMonth{11}
\acmDOI{} 
\startPage{1}

\setcopyright{none}

\usepackage[utf8x]{inputenc}
\usepackage{graphicx}
\usepackage{booktabs}   
\usepackage{subcaption} 
\usepackage{comment}
\usepackage{adjustbox}
\usepackage{amsmath,amsfonts}
\usepackage{csquotes}
\usepackage{cleveref}
\usepackage{marvosym}
\usepackage{graphicx}
\usepackage{nameref}
\usepackage{stmaryrd}
\usepackage{xcolor}
\usepackage{xfrac}
\usepackage{xspace}
\usepackage{cancel}
\usepackage{tikz-cd}
\usepackage{thmtools}
\usepackage{thm-restate}
\usepackage{thm-autoref}
\usepackage{wrapfig}

\usepackage{listings,multicol}
\usepackage{tensor}
\usepackage{tikz}
\usetikzlibrary{patterns,decorations.pathreplacing,arrows,arrows.meta,decorations.pathmorphing,positioning,fit,trees,shapes,shadows,automata,calc}
\usepackage{pgfplots}

\usepackage{caption}
\usepackage{proof}
\usepackage{mathtools}

\usepackage[textwidth=19mm]{todonotes}

\newcommand{\Sup}{\reflectbox{\textnormal{\textsf{\fontfamily{phv}\selectfont S}}}\hspace{.2ex}}
\newcommand{\Inf}{\raisebox{.6\depth}{\rotatebox{-30}{\textnormal{\textsf{\fontfamily{phv}\selectfont \reflectbox{J}}}}\hspace{-.1ex}}}
\newcommand{\InfCaption}{\raisebox{1pt}{\rotatebox{-30}{\textnormal{\textsf{\fontfamily{phv}\selectfont \reflectbox{J}}}}\hspace{-.1ex}}}

\newcommand{\SupV}[1]{\Sup #1\colon}
\newcommand{\InfV}[1]{\Inf #1\colon}
\newcommand{\InfVCaption}[1]{\InfCaption #1\colon}

\newcommand{\todoin}[1]{\todo[inline]{#1}}

\newcommand{\qedtriangle}{\hfill\raisebox{-.15ex}{\rotatebox{90}{$\triangle$}}}

\newcommand{\incorrectness}[3]{\left[ \,{#1}\vphantom{#3}\, \right] \mathrel{#2} \left[ \, {#3}\vphantom{#1} \, \right]}
\newcommand{\hoare}[3]{\left\langle \,{#1}\vphantom{#3}\, \right\rangle \mathrel{#2} \left\langle \, {#3}\vphantom{#1} \, \right\rangle}

\newcommand{\sfsymbol}[1]{\textsf{\upshape {#1}}}

\newcommand{\ttsymbol}[1]{\texttt{\upshape {#1}}}

\newcommand{\wpsymbol}{\sfsymbol{wp}}
\newcommand{\boldwpsymbol}{\textbf{\sfsymbol{wp}}}
\renewcommand{\wp}[2]{\wpsymbol\,\llbracket#1\rrbracket\left(#2\right)}
\newcommand{\boldwp}[2]{\boldwpsymbol\,\boldsymbol{\llbracket#1\rrbracket\left(#2\right)}}

\newcommand{\wpC}[1]{\wpsymbol\llbracket#1\rrbracket}

\newcommand{\wpsymbolD}{\sfsymbol{wp}}
\newcommand{\wpD}[2]{\wpsymbolD\,\llbracket#1\rrbracket\left(#2\right)}
\newcommand{\wlpsymbolD}{\sfsymbol{wlp}}
\newcommand{\wlpD}[2]{\wlpsymbolD\,\llbracket#1\rrbracket\left(#2\right)}

\newcommand{\spsymbolD}{\sfsymbol{sp}}
\newcommand{\stpsymbolD}{\sfsymbol{stp}}
\newcommand{\spD}[2]{\spsymbolD\,\llbracket#1\rrbracket\left(#2\right)}
\newcommand{\stpD}[2]{\stpsymbolD\,\llbracket#1\rrbracket\left(#2\right)}

\newcommand{\spsymbol}{\sfsymbol{sp}}
\newcommand{\boldspsymbol}{\textbf{\sfsymbol{sp}}}
\renewcommand{\sp}[2]{\spsymbol\,\llbracket#1\rrbracket\left(#2\right)}
\newcommand{\boldsp}[2]{\boldspsymbol\,\boldsymbol{\llbracket#1\rrbracket\left(#2\right)}}
\newcommand{\spC}[1]{\spsymbol\llbracket#1\rrbracket}

\newcommand{\boldslp}[2]{\boldslpsymbol\,\boldsymbol{\llbracket#1\rrbracket\left(#2\right)}}
\newcommand{\slpsymbol}{\sfsymbol{slp}}
\newcommand{\boldslpsymbol}{\textbf{\sfsymbol{slp}}}
\newcommand{\slp}[2]{\slpsymbol\llbracket#1\rrbracket\left(#2\right)}
\newcommand{\slpC}[1]{\slpsymbol\llbracket#1\rrbracket}

\newcommand{\awpsymbol}{\sfsymbol{awp}}

\newcommand{\boldwlp}[2]{\boldwlpsymbol\,\boldsymbol{\llbracket#1\rrbracket\left(#2\right)}}
\newcommand{\wlpsymbol}{\sfsymbol{wlp}}
\newcommand{\boldwlpsymbol}{\textbf{\sfsymbol{wlp}}}
\newcommand{\wlp}[2]{\wlpsymbol\llbracket#1\rrbracket\left(#2\right)}
\newcommand{\wlpC}[1]{\wlpsymbol\llbracket#1\rrbracket}

\newcommand{\awlpsymbol}{\sfsymbol{awlp}}

\newcommand{\conditionalPair}[2]{{\let\oldarraystretch\arraystretch}\renewcommand{\arraystretch}{1}~\holter{~\raisebox{.5ex}{${#1}$}~}{~\raisebox{.125ex}{${#2}$}~}~\renewcommand{\arraystretch}{\oldarraystretch}}

\newcommand{\annotate}[1]{\boldsymbol{\annocolor{\!\fatslash\!\!\!\fatslash~~\vphantom{G'} {#1}}}}

\newcommand{\eqannotate}[1]{\annocolor{\!\!\hspace{.55ex}{}^{\annocolor{{=}}}{\!\!\!{\fatslash}\!\!{\fatslash}~~\hspace{.5ex}\vphantom{G'} {\boldsymbol{#1}}}}}

\newcommand{\guard}{\ensuremath{\varphi}} 
\newcommand{\iguard}{\ensuremath{\iverson{\guard}}} 
\newcommand{\inguard}{\ensuremath{\iverson{\neg\guard}}} 
\newcommand{\ee}{\ensuremath{e}} 

\newcommand{\SKIP}{\ttsymbol{skip}}

\newcommand{\DIVERGE}{\ensuremath{\textnormal{\texttt{diverge}}}}

\newcommand{\AssignSymbol}{\coloneqq}
\newcommand{\ASSIGN}[2]{\ensuremath{#1 \AssignSymbol #2}}

\newcommand{\AVAILLOC}[1]{\PosNats}
\usepackage{actuarialangle}

\newcommand{\COMPOSE}[2]{\ensuremath{{#1}{\,\fatsemi}~ {#2}}}

\newcommand{\NDCHOICE}[2]{\ensuremath{\left\{\, {#1} \,\right\}\mathrel{\Box}\left\{\, {#2} \,\right\}}}

\newcommand{\IFSYMBOL}{\ensuremath{\textnormal{\texttt{if}}}}
\newcommand{\IF}[1]{\ensuremath{\IFSYMBOL\,\left(\, {#1} \,\right)\,\{}}
\newcommand{\ELSESYMBOL}{\ensuremath{\textnormal{\texttt{else}}}}
\newcommand{\ELSE}{\ensuremath{\}\,\ELSESYMBOL\,\{}}
\newcommand{\ITE}[3]{\ensuremath{\IFSYMBOL\,\left(\, {#1} \,\right)\,\left\{\, {#2} \,\right\}\,\ELSESYMBOL\,\left\{\, {#3} \,\right\}}}

\newcommand{\WHILESYMBOL}{\ensuremath{\textnormal{\texttt{while}}}}
\newcommand{\WHILE}[1]{\ensuremath{\WHILESYMBOL \left(\, {#1} \,\right)\left\{\right.}}

\newcommand{\WHILEDO}[2]{\ensuremath{\WHILESYMBOL \left(\, {#1} \,\right)\left\{\, {#2} \,\right\}}}

\newcommand{\ngcl}{\textnormal{\sfsymbol{nGCL}}\xspace}   
\newcommand{\Vars}{\ensuremath{\mathsf{Vars}}\xspace}   

\newcommand{\R}{\ensuremath{\mathbb{R}}}

\newcommand{\PosNats}{\ensuremath{\mathbb{N}_{>0}}\xspace}
\newcommand{\Ints}{\ensuremath{\mathbb{Z}}\xspace}

\newcommand{\Reals}{\mathbb{R}}
\newcommand{\PosReals}{\mathbb{R}_{\geq 0}}

\newcommand{\PosRealsInf}{\mathbb{R}_{\geq 0}^\infty}
\newcommand{\minfty}{{-}\infty}
\newcommand{\pinfty}{{+}\infty}
\newcommand{\Rinf}{\ensuremath{\R^{{\pm}\infty}}}

\newcommand{\A}{\mathbb{A}}
\newcommand{\B}{\mathbb{B}}

\newcommand{\iverson}[1]{\left[ {#1} \right]}

\newcommand{\indicator}[1]{\left[ {#1} \right]}

\newcommand{\subst}[2]{\left[ {#1} \middle/ {#2}\right]}
\newcommand{\statesubst}[2]{\left[ {#1} \mapsto {#2}\right]}

\newcommand{\eval}[1]{\ensuremath{\llbracket {#1} \rrbracket}}
\newcommand{\sem}[2]{\ensuremath{\llbracket {#1} \rrbracket}(#2)}

\newcommand{\States}{\Sigma}

\newcommand{\true}{\mathsf{true}}
\newcommand{\false}{\mathsf{false}}
\newcommand{\mydot}{\text{{\Large\textbf{.}}~}}

\newcommand{\qiff}{\quad\textnormal{iff}\quad}
\newcommand{\qqiff}{\qquad\textnormal{iff}\qquad}

\newcommand{\qand}{\quad\textnormal{and}\quad}
\newcommand{\qqand}{\qquad\textnormal{and}\qquad}

\newcommand{\qqimplies}{\qquad\textnormal{implies}\qquad}

\newcommand{\morespace}[1]{~{}#1{}~}

\newcommand{\ppreceq}{~{}\preceq{}~}
\newcommand{\qpreceq}{\quad{}\preceq{}\quad}
\newcommand{\ssucceq}{~{}\succeq{}~}
\newcommand{\qsucceq}{\quad{}\succeq{}\quad}

\newcommand{\eeq}{~{}={}~}
\newcommand{\ccup}{~{}\cup{}~}

\newcommand{\pplus}{~{}+{}~}

\newcommand{\vvee}{~{}\vee{}~}

\newcommand{\ccurlyvee}{~{}\curlyvee{}~}
\newcommand{\ccurlywedge}{~{}\curlywedge{}~}

\newcommand{\qmid}{\quad{}|{}\quad}

\newcommand{\qeq}{\quad{}={}\quad}
\newcommand{\qqeq}{\qquad{}={}\qquad}
\newcommand{\lleq}{~{}\leq{}~}

\newcommand{\iimplies}{~{}\implies{}~}
\newcommand{\mmodels}{~{}\models{}~}
\newcommand{\nnotmodels}{~{}\not\models{}~}

\newcommand{\setcomp}[2]{\left\{\, {#1} ~\middle|~ {#2} \,\right\}}

\definecolor{webgreen}{rgb}{0,.5,0}
\newcommand{\gray}[1]{\textcolor{gray}{#1}}
\newcommand{\lightgray}[1]{\textcolor{lightgray}{#1}}
\newcommand{\black}[1]{\textcolor{black}{#1}}

\newcommand{\blue}[1]{\textcolor{DodgerBlue3}{#1}}
\newcommand{\orange}[1]{\textcolor{orange}{#1}}

\newcommand{\annocolor}[1]{\textcolor{webgreen}{#1}}

\newcommand{\cemph}[1]{\blue{#1}}

\newcommand{\underdot}[1]{%
	\tikz[baseline=(todotted.base)]{
		\node[inner sep=1pt,outer sep=0pt] (todotted) {#1};
		\draw[densely dotted] (todotted.south west) -- (todotted.south east);
	}%
}%
\newcommand{\cloze}[1]{\underdot{\phantom{#1}}}

\newcounter{computationarrowsone}
\newcounter{computationarrowstwo}

\newcounter{sarrow}

\newcommand{\lfp}{\ensuremath{\textnormal{\sfsymbol{lfp}}~}}
\newcommand{\gfp}{\ensuremath{\textnormal{\sfsymbol{gfp}}~}}

\newcommand{\powerset}[1]{\ensuremath{\mathcal{P}(#1)}}

\newcommand{\Conf}{\ensuremath{\textnormal{\texttt{Conf}}}}

\newcommand{\correctpassword}{\texttt{"oopsla2022"}}

\AtBeginDocument{\theoremstyle{acmdefinition}\newtheorem{remark}[theorem]{Remark}}

\begin{document}

\title{Quantitative Strongest Post}         
\subtitle{A Calculus for Reasoning about the Flow of Quantitative Information}                     


\author{Linpeng Zhang}
\authornote{Both authors contributed equally to this research.}
\email{linpeng.zhang.20@ucl.ac.uk}
\orcid{0000-0002-1485-327X}
\affiliation{%
	\institution{University College London}
	\city{London}
	\country{United Kingdom}
}

\author{Benjamin Lucien Kaminski}
\authornotemark[1]
\email{b.kaminski@ucl.ac.uk}
\orcid{0000-0001-5185-2324}
\affiliation{%
	\institution{Saarland University, Saarland Informatics Campus}
	\city{Saarbrücken}
	\country{Germany}
}
\affiliation{%
	\institution{University College London}
	\city{London}
	\country{United Kingdom}
}
\begin{abstract}
We present a novel \textbf{strongest-postcondition-style calculus for quantitative reasoning} about non-deterministic programs with loops.
Whereas existing quantitative weakest pre allows reasoning about the value of a quantity \emph{after} a program terminates on a given \emph{initial state}, 
quantitative strongest post allows reasoning about the value that a quantity had \emph{before} the program was executed and reached a given \emph{final state}.
We show how strongest post enables reasoning about the \emph{flow of quantitative information through programs}.

Similarly to weakest \emph{liberal} preconditions, we also develop a \emph{quantitative strongest liberal post}. 
As a byproduct, we obtain the entirely unexplored notion of \textbf{\emph{strongest liberal postconditions}} and show how these foreshadow a potential new program logic --- \textbf{\emph{partial incorrectness logic}} --- which would be a more liberal version of O'Hearn's recent incorrectness logic.
%
%
%
%
\end{abstract}



\maketitle              
%






\section{Introduction}

%

\subsubsection*{Partial Correctness} 

Already in one of the earliest works on program verification, \citet{turinglargeroutine} separates reasoning about partial correctness and termination.
Partial correctness means that the program is correct, \emph{if} it terminates.
Nontermination is in that sense deemed \enquote{correct} behavior.
\emph{Hoare triples}~\cite{Hoa69} capture partial correctness formally: 
Given program $C$ and predicates $G, F$, we say that $\hoare{G}{C}{F}$ is \emph{valid for partial correctness}, if from every state $\sigma$ satisfying precondition $G$, $C$~\emph{either} terminates in some state satisfying postcondition $F$, \emph{or} $C$ does \emph{not} terminate on $\sigma$.

A different approach to partial correctness are the \emph{weakest liberal preconditions} of \citet{DBLP:journals/cacm/Dijkstra75}:
Given program $C$ and postcondition $F$, the weakest liberal precondition is the weakest (largest) predicate $\wlp{C}{F}$, such that starting from any state $\sigma$ satisfying the precondition~$\wlp{C}{F}$,\linebreak $C$ \emph{either} terminates in some state satisfying the postcondition $F$, \emph{or} $C$ does not terminate on~$\sigma$. 
$\wlp{C}{\cloze{F}}$ is a called a backward-moving \emph{predicate transformer semantics}, because it transforms a postcondition (a~predicate) $F$ into a precondition (another~predicate) $\wlp{C}{F}$.

A different predicate transformer semantics are the forward-moving \emph{strongest postconditions} of \citet{Dijkstra1990}: 
they transform a precondition $G$ into the strongest (smallest) predicate~$\sp{C}{G}$, such that $\sp{C}{G}$ contains all states that can be reached by executing $C$ on some state satisfying the precondition $G$.
Hoare triples, weakest liberal preconditions, and strongest postconditions are strongly related by the following well-known fact:%
\begin{align*}
	\hoare{G}{C}{F} \text{ is valid for part.~corr.} \qqiff G \implies \wlp{C}{F} \qqiff \sp{C}{G} \implies F~.	
\end{align*}%
Having a choice between $\wlpsymbol$ and $\spsymbol$ is beneficial because sometimes the partial correctness proof can be easier in the, say, forward direction than in the backward direction.

\subsubsection*{Quantitative Verification}

\emph{Backward-moving} predicate transformers have been generalized to \emph{real-valued-function transformers}, first by \citet{DBLP:journals/jcss/Kozen85}, in order to reason about probabilistic programs, e.g.~ about the \emph{probability} that some postcondition will be satisfied after program termination.
For the forward direction, \citet{claire90} presented a counterexample to the existence of probabilistic strongest postconditions.
While we also cannot handle probabilistic programs, we will in this paper develop a \emph{quantitative strongest post transformer for reasoning about nondeterministic programs}. 

Intuitively, quantitative predicate-transformer-style calculi lift reasoning%
\begin{align*}
	\textnormal{from} \qquad \textit{predicates } F\colon \textsf{States} \to \{\true,\, \false\} \qquad\textnormal{to}\qquad \textit{quantities } f\colon \textsf{States} \to \Rinf~,
\end{align*}%
i.e.\ functions $f$ that associate a \emph{real number} (or $\pinfty$ or $\minfty$) to each state.
Given a \emph{post}quantity $f$ associating a number to \emph{final} states, our backward-moving weakest liberal pre transformer $\wlp{C}{f}\colon \textsf{States} \to \Rinf$ associates numbers to \emph{initial} states, so that $\wlp{C}{f}(\sigma)$ \mbox{\emph{anticipates}} what value $f$ will have after $C$ terminates on~$\sigma$ (and $\wlpsymbol$ \emph{anticipates}~$\pinfty$ if $C$ does not terminate~on~$\sigma$).

\begin{wrapfigure}[5]{r}{0.18\textwidth}
  \begin{minipage}{.99\linewidth}
    \abovedisplayskip=0pt%
    \belowdisplayskip=0pt%
    \vspace*{-1em}
\begin{align*}
	&\annotate{2  x + 2}\\
	&\ASSIGN{x}{x + 1}\\
	&\annotate{2 x}
\end{align*}%
\normalsize%
  \end{minipage}%
\end{wrapfigure}%
For example, what is the anticipated value of $2  x$ after executing the assignment $\ASSIGN{x}{x + 1}$?
Our quantitative weakest liberal pre calculus will push the \enquote{assertion} $2  x$ backward through the program, obtaining the annotations on the right (read from bottom to top).
Indeed, given an \emph{initial} value $x_\sigma = 5$ for the program variable~$x$, the \emph{final} value of the expression $2  x$ will be $2  x_\sigma + 2 = 2 \cdot 5 + 2 = 12$.

While counterintuitive --- since $\wlpsymbol$ moves backwards ---, $\wlpsymbol$ acts like a weather \emph{forecast}: 
Given the current state $\sigma$ of the global atmosphere, a~function $f$ mapping atmosphere state to the temperature in Auckland, and an (algorithmic) description $C$ of how the atmosphere evolves within 24 hours, $\wlp{C}{f}(\sigma)$ anticipates \emph{now} what the temperature in \mbox{Auckland will be \emph{tomorrow}}.

In this paper, we develop a \emph{quantitative strongest post transformer $\spsymbol$} with as strong a connection (more precisely: a \emph{Galois connection}) to quantitative $\wlpsymbol$ as in the qualitative case, namely%
\begin{align*}
	g \ppreceq \wlp{C}{f} \qqiff \sp{C}{g} \ppreceq f~.
\end{align*}%
Dually to $\wlpsymbol$, our forward-moving strongest post transformer acts like a weather \emph{backcast}: 
Given the current global atmosphere state~$\tau$, $\sp{C}{f}(\tau)$ \emph{retrocipates} now what the temperature in Auckland was \emph{yesterday}.
Speaking in terms of programs and quantities, given a \emph{pre}quantity $f$ associating a number to \emph{initial} states, \mbox{$\sp{C}{f}\colon \textsf{States} \to \Rinf$}~associates numbers to \emph{final} states, such that $\sp{C}{f}(\tau)$ \emph{retrocipates} what value $f$ had in an initial state before $C$ terminated in $\tau$
(and $\spsymbol$ \emph{retrocipates}~$\minfty$ if $\tau$ is not reachable by executing $C$ on some initial state).

\begin{wrapfigure}[5]{r}{0.18\textwidth}
  \begin{minipage}{.99\linewidth}
    \abovedisplayskip=0pt%
    \belowdisplayskip=0pt%
    \vspace*{-1em}
\begin{align*}
	&\annotate{2  x}\\
	&\ASSIGN{x}{x + 1}\\
	&\annotate{2 x - 2}
\end{align*}%
\normalsize%
  \end{minipage}%
\end{wrapfigure}%
For example, what is the retrocipated value of $2  x$ before the assignment $\ASSIGN{x}{x + 1}$?
Our quantitative strongest post calculus will push the \enquote{assertion} $2  x$ forward through the program, obtaining the annotations on the right (read from top to bottom).
Indeed, given a \emph{final} value $x_\tau = 5$ for the program variable~$x$, the \emph{initial} value of the expression $2  x$ must have been $2  x_\tau - 2 = 2 \cdot 5 - 2 = 8$.

Notably, our quantitative strongest post transformer provides some notion of \emph{flow of quantitative information through the program}:
If we start the above program with initial value $x_\sigma = 4$ for $x$, then we have initially $2 x_\sigma = 2 \cdot 4 = 8$.
After the execution of the program, the final value of $x$ is $x_\tau = 5$.
The expression $2 x - 2$ evaluated in $x_\tau$ is again $2 \cdot x_\tau = 2 \cdot 5 - 2 = 8$.
In that sense, our quantitative $\spsymbol$ takes a quantity --- for instance: a secret value --- and propagates through the program an expression which \emph{preserves} the value of the initial quantity. 
Given some final state, we can hence read off what the quantity was initially and so reason about quantitative flow and leakage of information.

\subsection*{Contributions and Organization}

\emph{Not} being our main contribution, we present in Sec.~\ref{se:weakest-pre} quantitative $\wpsymbol$ and $\wlpsymbol$.
Differently from~\cite{McIverM05,quantitative_sl,benni_diss}, our quantitative transformers act on \emph{signed} unbounded quantities in $\Rinf$, whereas traditional probabilistic $\wlpsymbol$ act on $[0,\, 1]$ and \mbox{$\wpsymbol$ on $\PosRealsInf$}.

In \Cref{se:sp}, we present our \emph{main contribution}: a novel quantitative strongest post transformer $\spsymbol$ as described above.
Moreover, we provide a quantitative strongest \emph{liberal} post transformer $\slpsymbol$, which gives a different value than $\spsymbol$ to \emph{unreachable} states (whereas $\wlpsymbol$ gives a different value than $\wpsymbol$ to \emph{nonterminating} states). 
We study essential properties of all our transformers in \Cref{se:healthiness} and show how they embed reasoning about predicates \`{a} la~\citet{Dijkstra1990}.

In \Cref{se:relationship}, we show that $\slpsymbol$ has as tight a (Galois) connection to $\wpsymbol$ as $\spsymbol$ to $\wlpsymbol$, namely
\begin{align*}
	\wp{C}{f} \ppreceq g \qqiff  f \ppreceq \slp{C}{g}~.
\end{align*}%
When restricting to predicates, our $\slpsymbol$ transformer yields the novel notion of \emph{strongest liberal postconditions}, which is entirely unexplored in the literature.
While it is known that strongest postconditions are tightly connected with the recent \emph{incorrectness logic} of \citet{OHearn19}, we show how $\slpsymbol$ foreshadows a new program logic --- \emph{partial incorrectness logic}.
We also hint at two further new program logics: one of \emph{necessary liberal preconditions} and one of \emph{necessary liberal postconditions}.

In \Cref{se:loops}, we present proof rules for loops for all four quantitative transformers.
In \Cref{se:examples} we demonstrate efficacy of $\spsymbol$ and $\slpsymbol$ for reasoning about the flow of quantitative information.

\section{Nondeterministic Programs}
The syntax of the \emph{nondeterministic guarded command language} ($\ngcl$) à la Dijkstra is given by%
%
\begin{align*}
	C ~{}~{}\,{}\Coloneqq{}~{}~{}\, \ASSIGN{x}{\ee}	\qmid \COMPOSE{C}{C} \qmid \NDCHOICE{C}{C} \qmid \ITE{\guard}{C}{C} \qmid \WHILEDO{\guard}{C}~.
\end{align*}%
%
where $x\in\Vars$ is a variable, $\ee$ is an arithmetic expression and $\guard$ is a predicate. 
%
%
A program state~$\sigma$ is a function that assigns an integer to each program variable. 
The set of program states is given by %
	$\States = \setcomp{\sigma}{\sigma:\Vars\to\Ints}$.
%
Given a program state $\sigma$, we denote by $\sigma(\xi)$ the evaluation of an arithmetic or Boolean expression~$\xi$ in~$\sigma$, i.e. the value that is obtained by evaluating $\xi$ after replacing any occurrence of any variable~$x$ in~$\xi$ by the value~$\sigma(x)$. 
Moreover, we denote by $\sigma\subst{x}{v}$ a new state that is obtained from $\sigma$ by setting the valuation of the variable~$x\in\Vars$ to~$v\in\Ints$. 
Formally: 
$\sigma\subst{x}{v}(y) = v$, if $y = x$; and $\sigma\subst{x}{v}(y) = \sigma(y)$, otherwise.
%

We assign meaning to our nondeterministic \ngcl-statements in terms of a denotational \emph{collecting} semantics (as is standard in program analysis, see~\cite{CousotC77,hecht,RY20}), i.e.\ we have as input a \emph{set of initial states} and as output the \emph{set of reachable states}.%
\begin{definition}[Collecting Semantics for $\ngcl$ Programs]%
	Let $\Conf = \powerset{\States}$ be the set of \emph{program configurations}, i.e.\ a single configuration is a \emph{set of program states}; and let $\eval{\guard}S = \{\sigma\mid\sigma\in S \land \sigma \mmodels \guard\}$ be a filtering of a program configuration to only those states where the predicate $\guard$ holds. 

The collecting semantics $\eval{C}\colon \Conf \to \Conf$ of an $\ngcl$ program $C$ is defined inductively by%
\allowdisplaybreaks%
\begin{align*}
    \eval{\ASSIGN{x}{\ee}} S  \eeq & \{\sigma\subst{x}{\sigma(\ee)}\mid \sigma \in S\}\tag{assignment}\\
    \eval{\COMPOSE{C_1}{C_2}} S \eeq & (\eval{C_2}\circ \eval{C_1}) S \tag{sequential composition} \\
    \eval{\ITE{\guard}{C_1}{C_2}} S  \eeq & (\eval{C_1}\circ \eval{\guard}) S \cup (\eval{C_2}\circ \eval{\neg\guard}) S \tag{conditional choice}\\
    	\eval{\WHILEDO{\guard}{C}} S \eeq & \eval{\neg \guard}\bigl(\lfp X\mydot S \ccup \bigl(\eval{C} \circ \eval{\guard} \bigr) X \bigr) \tag{loop}\\
    \eval{\NDCHOICE{C_1}{C_2}} S  \eeq &\eval{C_1} S \cup \eval{C_2}S~. \tag{nondeterministic choice}    
\end{align*}%
\allowdisplaybreaks[0]%
By slight abuse of notation, we write $\eval{C}(\sigma)$ for $\eval{C}\{\sigma\}$. 
For more details, see Appendix~\ref{se:while-semantics}.
\qedtriangle
\end{definition}%

\section{Weakest Pre}
\label{se:weakest-pre}

We develop novel weakest (liberal) pre calculi \'{a} la \citet{DBLP:journals/cacm/Dijkstra75} for \emph{quantitative reasoning} about nondeterministic programs.
While we repeat that the weakest pre calculi are not our main contribution (that being the quantitative strongest post calculi), we believe that weakest pre calculi are easier to understand and provide the necessary intuition for moving from the Boolean to the quantitative realm.
We first shortly recap Dijkstra's classical weakest preconditions before we lift them to a quantitative setting.
Thereafter, we lift weakest \emph{liberal} preconditions to quantities.

\subsection{Classical Weakest Preconditions}
\label{sec:sub-dijkstra-wp}

Dijkstra's weakest precondition calculus employs \emph{predicate transformers} of type%
\begin{align*}
	\wpC{C}\colon\quad \B \morespace{\to} \B~, \qquad \textnormal{where}\quad \B \eeq \{0,\, 1\}^\States~,
\end{align*}%
which associate to each nondeterministic program $C$ a mapping from predicates (sets of program states) to predicates.
Somewhat less common, we consider here an \emph{angelic} setting, where the nondeterminism is resolved to our advantage.\footnote{Considering an angelic setting allows us not only to show that our transformers enjoy several properties, but also to provide tight connections between quantitative weakest preconditions and quantitative strongest postconditions.%
}
Specifically, the \emph{angelic} weakest precondition transformer $\wpC{C}$ maps a \emph{post}{\-}condition~$\psi$ over final states to a \emph{pre}condition $\wp{C}{\psi}$ over initial states, such that executing the program $C$ on an initial state satisfying $\wp{C}{\psi}$ guarantees that $C$~\emph{\underline{can}}\footnote{Recall that $C$ is a \emph{nondeterministic} program. For the (standard) \emph{demonic} setting as well as for deterministic programs, we can \emph{replace \enquote{can} by \enquote{will}}.}~terminate in a final state satisfying~$\psi$.
More symbolically, recalling that $\eval{C}(\sigma)$ is the set of all final states reachable after termination of $C$ on $\sigma$,%
\begin{align*}
	\sigma \mmodels \wp{C}{\psi} \qqiff \exists\, \tau \in \eval{C}(\sigma)\colon \quad \tau \mmodels \psi~.
\end{align*}%
While the above is a \emph{set perspective} on $\wpsymbol$, an equivalent perspective on $\wpsymbol$ is a \emph{map perspective}, see \Cref{fig:wp-map}: 
\begin{figure}[t]
\begin{subfigure}[t]{.45\textwidth}
\begin{center}
	\begin{adjustbox}{max width=.8\textwidth}
		\begin{tikzpicture}[node distance=4mm, decoration={snake,pre=lineto,pre length=.5mm,post=lineto,post length=1mm, amplitude=.2mm}]
					\draw[use as bounding box,white] (-4.55,-0.75) grid (3.25, 4.25);
					\node (sigma) at (0, 4) {\Large$\boldsymbol{~\sigma}$};
					\node[gray,inner sep=0pt, outer sep=0pt] (firstplus) at (0,3) {\LARGE$\Box$};
					
					\node[gray,inner sep=0pt, outer sep=0pt] (branch) at (-0.225, 1.4) {\LARGE$\Box$};
					
					\node (tau1) at (-2, 0) {$\bullet$};
					\node (tau2) at (-.5, .25) {$\bullet$};
					\node (tau3) at (1, .125) {$\bullet$};
					\node[inner sep=0pt] (taudots) at (2.5, 0) {$\ddots$};
					
					\node[below of=tau1] {$\psi(\tau_1)$};
					\node[below of=tau2] {$\psi(\tau_2)$};
					\node[below of=tau3] {$\psi(\tau_3)$};

					\node (Exp) at (-3.1, -0.25) {\huge $\boldsymbol{\bigvee}$\Large~$\boldsymbol{\Bigl[}$};
					\node at (3.1, -0.25) {\Large $\boldsymbol{\Bigr]}$};

					\node[gray] (C) at (0.5, 1.5) {$C$};
					
					\draw[gray,decorate,thick] (sigma) -- (firstplus);
					\draw[gray,decorate,->,thick] (firstplus) -- (tau1);
					
					\draw[gray,decorate,thick] (firstplus) -- (branch);
					
					\draw[gray,decorate,->,thick] (branch) -- (tau2);
					\draw[gray,decorate,->,thick] (branch) -- (tau3);
					\draw[gray,decorate,thick] (firstplus) -- (taudots);
					
					\draw[decorate,lightgray] (tau1) -- (tau2) -- (tau3) -- (taudots);
					
					\draw[gray] (sigma) edge[|->,bend right=45,above left,thick] node {\textcolor{black}{\Large$\wp{C}{\psi}$}} (Exp);
				\end{tikzpicture}
	\end{adjustbox}
\end{center}
\caption{\footnotesize\textbf{Weakest preconditions:}
Given initial state $\sigma$, $\wp{C}{\psi}$ determines all final states $\tau_i$ reachable from executing $C$ on $\sigma$, evaluates $\psi$ in those states, and returns the disjunction~($\vee$) over all these truth values.}
\label{fig:wp-map}
\end{subfigure}
\hfill
\begin{subfigure}[t]{.45\textwidth}
\begin{center}
	\begin{adjustbox}{max width=.8\textwidth}
		\begin{tikzpicture}[node distance=4mm, decoration={snake,pre=lineto,pre length=.5mm,post=lineto,post length=1mm, amplitude=.2mm}]
					\draw[use as bounding box,white] (-4.55,-0.75) grid (3.25, 4.25);
					\node (sigma) at (0, 4) {\Large$\boldsymbol{~\sigma}$};
					\node[gray,inner sep=0pt, outer sep=0pt] (firstplus) at (0,3) {\LARGE$\Box$};
					
					\node[gray,inner sep=0pt, outer sep=0pt] (branch) at (-0.225, 1.4) {\LARGE$\Box$};							
					\node (tau1) at (-2, 0) {$\bullet$};
					\node (tau2) at (-.5, .25) {$\bullet$};
					\node (tau3) at (1, .125) {$\bullet$};
					\node[inner sep=0pt] (taudots) at (2.5, 0) {$\ddots$};
					
					\node[below of=tau1] {$f(\tau_1)$};
					\node[below of=tau2] {$f(\tau_2)$};
					\node[below of=tau3] {$f(\tau_3)$};

					\node (Exp) at (-3.1, -0.25) {\huge $\boldsymbol{\bigcurlyvee}$\Large~$\boldsymbol{\Bigl[}$};
					\node at (3.1, -0.25) {\Large $\boldsymbol{\Bigr]}$};

					\node[gray] (C) at (0.5, 1.5) {$C$};
					
					\draw[gray,decorate,thick] (sigma) -- (firstplus);
					\draw[gray,decorate,->,thick] (firstplus) -- (tau1);
					
					\draw[gray,decorate,thick] (firstplus) -- (branch);
					
					\draw[gray,decorate,->,thick] (branch) -- (tau2);
					\draw[gray,decorate,->,thick] (branch) -- (tau3);
					\draw[gray,decorate,thick] (firstplus) -- (taudots);
					
					\draw[decorate,lightgray] (tau1) -- (tau2) -- (tau3) -- (taudots);
					
					\draw[gray] (sigma) edge[|->,bend right=45,above left,thick] node {\textcolor{black}{\Large$\wp{C}{f}$}} (Exp);
				\end{tikzpicture}
	\end{adjustbox}
\end{center}%
\caption{\footnotesize\textbf{Quantitative weakest pre:}
Given initial state~$\sigma$, $\wp{C}{f}$ determines all final states $\tau_i$ reachable from executing $C$ on $\sigma$, evaluates $f$ in those states, and returns the supremum~($\curlyvee$) over all these quantities.}
\label{fig:wp-general}
\end{subfigure}	
\caption{(Angelic) weakest preconditions and quantitative weakest pres.}
\label{fig:wp-views}
\end{figure}%
The postcondi{\-}tion~\mbox{$\psi\colon \States \to \{0,\, 1\}$} maps program states to truth values.
The predicate $\wp{C}{\psi}$ is then a map that takes as input an initial state $\sigma$, determines for each reachable final state $\tau \in \eval{C}(\sigma)$ the (truth) value $\psi(\tau)$, takes a disjunction over all these truth values, and finally returns the truth value of that disjunction.
More symbolically,%
\begin{align*}
	\wp{C}{\psi}(\sigma) \qqeq\quad \bigvee_{\mathclap{\tau \in \eval{C}(\sigma)}} \quad \psi(\tau)~.
\end{align*}%
%
%
It is this map perspective which we will now gradually lift to a quantitative setting.
For that, we first need to leave the realm of Boolean valued  predicates and move to \emph{real-valued} functions.

\subsection{Quantities}


For our development here, we are interested in \emph{signed} quantities.
Such quantities form --- just like first-order logic for weakest preconditions --- the \emph{assertion \enquote{language} of our quantitative calculi}.%
%
%
\begin{definition}[Quantities]
\label{def:quantities}
	The set of all \emph{quantities} is defined by%
	\begin{align*}
		\A \eeq \setcomp{f}{f\colon \States \to \Rinf}
	\end{align*}%
	i.e.\ the set of all functions $f\colon \States \to \Rinf$ associating an \emph{extended real} (i.e.\ either a proper real number, or $\minfty$, or $\pinfty$) to each program state. 
	The point-wise order%
	\begin{align*}
		f \ppreceq g \qqiff \forall\, \sigma \in \States\colon \quad f(\sigma) \lleq g(\sigma)
	\end{align*}%
	renders $\langle \A,\, {\preceq}\rangle$ a complete lattice with join $\curlywedge$ and meet $\curlyvee$, given point-wise by%
	\begin{align*}
		f \curlywedge g \eeq \lambda \sigma\mydot \min \bigl\{f(\sigma),\, g(\sigma) \bigr\} \qqand f \curlyvee g \eeq \lambda \sigma\mydot \max \bigl\{f(\sigma),\, g(\sigma)\bigr\}~.
	\end{align*}%
	Joins and meets over arbitrary subsets exist. 
	When we write $a \curlyvee b \curlywedge c$, we assume that $\curlywedge$ binds stronger than $\curlyvee$, so we read that as $a \curlyvee (b \curlywedge c)$. \qedtriangle
\end{definition}%
\begin{remark}[Signed Quantities]
\label{rem:sign}
\citet{DBLP:journals/jcss/Kozen85} also considers \emph{signed} functions for reasoning about probabilistic programs.
However, Kozen's induction rule for while loops only applies to non-negative functions, see \cite[page 168]{DBLP:journals/jcss/Kozen85}. 
\citet{DBLP:conf/lics/KaminskiK17} have rules for probabilistic loops and signed functions, but their machinery is quite involved and their rule for loops is more involved than simple induction.
Our development in this paper is --- on the \mbox{plus-side ---} comparatively simple, but --- as a trade-off --- we cannot handle probabilistic programs. \qedtriangle
\end{remark}

\subsection{Quantitative Weakest Pre}
\label{sec:sub-wp}

We now define a calculus \'{a} la Dijkstra for formal reasoning about the value of a quantity $f \in \A$ after execution of a nondeterministic program.
For that, we generalize the \emph{map perspective} of weakest preconditions to quantities. 
Instead of a post\emph{condition}, we now have a post{\-}\emph{quantity}~$f\colon \States \to \Rinf$ mapping (final) program states to extended reals.
$\wp{C}{f}\colon \States \to \Rinf$ is then a function that takes as input an initial state $\sigma$, determines all final states $\tau$ reachable from executing $C$ on $\sigma$, evaluates the postquantity $f(\tau)$ in each final state $\tau$, and finally returns the supremum over all these so-determined quantities, see \Cref{fig:wp-general}.
If the program is completely \emph{deterministic} and if $C$ terminates on input $\sigma$, then $\wp{C}{f}(\sigma)$ \emph{\underline{antici}p\underline{ates}} the single possible value that $f$ will have, evaluated in the final state that is reached after executing $C$ on $\sigma$.

One of the main advantages of Dijkstra's calculus is that the weakest preconditions can be defined by induction on the program structure, thus allowing for \emph{compositional reasoning}.
Indeed, the same applies to our quantitative setting.%
\begin{definition}[Quantitative Weakest Pre]
\label{def:wp}
	The \emph{weakest pre transformer}%
	\begin{align*}
		\wpsymbol\colon\quad \ngcl \to (\A \to \A)
	\end{align*}%
	is defined inductively according to the rules in \textnormal{\Cref{table:wp}} (middle column).
	We call the function%
	\begin{align*}
	\Phi_f(X) \eeq  \iverson{\neg \guard} \curlywedge f \ccurlyvee \iverson{\guard} \curlywedge \wp{C}{X}~,
	\end{align*}
	whose \emph{least} fixed point defines the weakest pre $\wp{\WHILEDO{\guard}{C}}{f}$, the \emph{$\wpsymbol$--characteristic function} (of $\WHILEDO{\guard}{C}$ with respect to $f$).\qedtriangle%
%
\end{definition}%
\noindent{}%
\begin{table*}[t]
	\begin{adjustbox}{max width=\textwidth}
	\renewcommand{\arraystretch}{1.5}
	\begin{tabular}{@{\hspace{.5em}}l@{\hspace{2em}}l@{\hspace{2em}}l@{\hspace{.5em}}}
		\hline\hline
		$\boldsymbol{C}$			& $\boldwp{C}{f}$ & $\boldwlp{C}{f}$\\
		\hline
		$\DIVERGE$				& $\minfty$ & $\pinfty$	\\
		$\ASSIGN{x}{\ee}$			& $f\subst{x}{\ee}$ & \lightgray{$f\subst{x}{\ee}$} \\
		$\COMPOSE{C_1}{C_2}$		& $\wp{C_1}{\vphantom{\big(}\wp{C_2}{f}}$ & \lightgray{$\wlp{C_1}{\vphantom{\big(}\wlp{C_2}{f}}$}\\
		$\NDCHOICE{C_1}{C_2}$		& $\wp{C_1}{f} \ccurlyvee \wp{C_2}{f}$ & $\wlp{C_1}{f} \ccurlywedge \wlp{C_2}{f}$\\
		$\ITE{\guard}{C_1}{C_2}$		& $\iverson{\guard} \curlywedge \wp{C_1}{f} \ccurlyvee \iverson{\neg \guard} \curlywedge \wp{C_2}{f}$ & \lightgray{$\iverson{\guard} \curlywedge \wlp{C_1}{f} \ccurlyvee \iverson{\neg \guard} \curlywedge \wlp{C_2}{f}$} \\
		$\WHILEDO{\guard}{C'}$		& $\lfp  X\mydot \quad \iverson{\neg \guard} \curlywedge f \ccurlyvee \iverson{\guard} \curlywedge \wp{C'}{X}$ & $\gfp  X\mydot \quad \iverson{\neg \guard} \curlywedge f \ccurlyvee \iverson{\guard} \curlywedge \wlp{C'}{X}$\\[.25em]
		\hline\hline
	\end{tabular}%
	\end{adjustbox}%
	\vspace{.5em}
	\caption{Rules for $\wpsymbol$ and $\wlpsymbol$. 
			$\lfp g\mydot \Phi(g)$ and $\gfp g\mydot \Phi(g)$ denote the least and greatest fixed point of $\Phi$.}%
	\label{table:wp}
	\label{table:wlp}
	\vspace{-1em}
\end{table*}%
\noindent{}%
Let us show for some of the rules how the quantitative weakest pre semantics can be developed and understood analogously to Dijkstra's classical weakest preconditions.

%

\subsubsection*{\textbf{Assignment.}}

The weakest pre\emph{condition} of an assignment is given by%
\begin{align*}
	\wp{\ASSIGN{x}{e}}{\psi} \eeq \psi\subst{x}{e}~,
\end{align*}%
where $\psi\subst{x}{e}$ is the replacement of every occurrence of variable $x$ in the postcondition $\psi$ by the expression $e$.
For \emph{quantitative} weakest pre, we can do something completely analogous, except that we do not have a syntax like first-order logic for the postquantities at hand.\footnote{For probabilistic programs, an expressive and relatively complete (with respect to taking weakest preexpectations) \emph{syntax} for expressing functions (expectations) of type $\States \to \PosRealsInf$ has been presented in~\cite{DBLP:journals/pacmpl/BatzKKM21}.}
Still, we can define semantically what it means to \enquote{syntactically replace} every \enquote{occurrence} of $x$ in $f$ by $e$ --- and with it the quantitative weakest pre of an assignment --- as follows:
\begin{align*}
	\wp{\ASSIGN{x}{e}}{f} \eeq f\subst{x}{e} \morespace{\coloneqq} \lambda\, \sigma\mydot f\Bigl( \sigma\statesubst{x}{\sigma(e)} \Bigr)~.
\end{align*}%
So what is the value of $f$ in the final state reached after executing the assignment~$\ASSIGN{x}{e}$ on initial state $\sigma$? 
It is precisely $f$, but evaluated at the final state $\sigma\statesubst{x}{\sigma(e)}$ --- the state obtained from $\sigma$ by updating variable $x$ to value $\sigma(e)$.

\subsubsection*{\textbf{Nondeterministic Choice.}}

When \enquote{executing} the nondeterministic choice $\NDCHOICE{C_1}{C_2}$ on some initial state $\sigma$, \emph{either} $C_1$ \emph{or}~$C_2$ will be executed, chosen nondeterministically.
Hence, the execution will reach either a final state in which executing~$C_1$~on~$\sigma$ terminates or a final state in which executing~$C_2$~on~$\sigma$ terminates (or no final state if both computations diverge).

Denotationally, the \emph{angelic} weakest pre\emph{condition} of $\NDCHOICE{C_1}{C_2}$ \mbox{is given by}%
\begin{align*}
	\wp{\NDCHOICE{C_1}{C_2}}{\psi} \eeq \wp{C_1}{\psi} \morespace{\vee} \wp{C_2}{\psi}~.
\end{align*}%
Indeed, whenever an initial state $\sigma$ satisfies the precondition $\wp{C_1}{\psi} \vee \wp{C_2}{\psi}$, then \mbox{--- either by} executing $C_1$ or by executing~\mbox{$C_2$ ---} it is possible that the computation will terminate in some final state satisfying the postcondition $\psi$.

Quantitatively, what is the anticipated value of $f$ after termination of either~$C_1$ or~$C_2$? 
Since $C_1$ and $C_2$ could both terminate but very well yield different values for $f$, we need to accommodate for two different numbers.
In the maximizing spirit of \emph{angelic} $\wpsymbol$, we also maximize and select as \emph{quantitative }weakest pre of $\NDCHOICE{C_1}{C_2}$ the \emph{largest} possible final value of~$f$ via the meet%
\begin{align*}
	\wp{\NDCHOICE{C_1}{C_2}}{f} \eeq \wp{C_1}{f} \ccurlyvee \wp{C_2}{f}~.
\end{align*}%
%

\subsubsection*{\textbf{Diverge.}}

$\DIVERGE$ is a shorthand for $\WHILEDO{\true}{\SKIP}$ --- the certainly diverging loop.
Denotationally, the weakest pre\emph{condition} of $\DIVERGE$ is given by%
\begin{align*}
	\wp{\DIVERGE}{\psi} \eeq \false~.
\end{align*}%
As there is \emph{no} initial state that satisfies $\false$, this simply tells us that there is no initial on which $\DIVERGE$ could possibly terminate in any final state satisfying $\psi$.

Note that the predicate $\false$ is the \emph{least element} in the Boolean lattice.
When lifting this to a quantitative setting, we also assign the least element.
Hence, 
\begin{align*}
	\wp{\DIVERGE}{f} \eeq \minfty~.
\end{align*}%
Another explanation goes by considering again the \emph{angelic}, i.e.\ maximizing, aspect of quantitative weakest pre:
What is the maximal value that we can anticipate for $f$ \emph{after} $\DIVERGE$ has terminated?
Since $\DIVERGE$ does not terminate at all (but we are still forced to assign some \enquote{number} to this situation), the largest value that we can possibly anticipate is the absolute minimum: $\minfty$.%
\begin{remark}[Quantitative Weakest Pre and Nontermination]
\label{rem:nontermination-wp}
In some sense, $\minfty$ is \emph{the value of nontermination} in quantitative $\wpsymbol$.
Note that it is more tedious to detect nontermination by standard weakest preconditions:
Consider e.g.\ the program $\DIVERGE$ and postcondition \enquote{$x \textnormal{ is odd}$}.
Then
\begin{align*}
	\wp{\DIVERGE}{x \textnormal{ is odd}} &\eeq \false~.
%
\intertext{On the other hand, for the \emph{terminating} program $\ASSIGN{x}{2\cdot x}$, we also have}
%
	\wp{\ASSIGN{x}{2\cdot x}}{x \textnormal{ is odd}} \eeq 2\cdot x \textnormal{ is odd} &\eeq \false~.
\end{align*}%
Thus, $\wp{C}{\psi}(\sigma) = \false$ is not a sufficient criterion for detecting nontermination of $C$ on~$\sigma$. 
$\false$ merely tells us that the program \emph{either} does not terminate \emph{or} it fails to establish the postcondition. To distinguish the two cases, one needs to check, \emph{additionally}, whether $\sigma$ terminates, i.e., whether $\wp{C}{\true}(\sigma)$ holds.

In \emph{our} quantitative $\wpsymbol$ calculus, given any non-infinite postquantity $f$ our wp transformer distinguishes whether the program terminates or not \emph{in one go}. Indeed, if $\minfty \ppreceq f \ppreceq \pinfty$ and $\wp{C}{f}(\sigma) = 0$, then definitely $C$ terminates on $\sigma$ and $f$ assumes value 0 after termination of $C$ on $\sigma$. For instance, for postquantity $x$ we have%
\begin{align*}
	\wp{\DIVERGE}{x} \eeq \minfty
	\qqand
	\wp{\ASSIGN{x}{2\cdot x}}{x} \eeq 2\cdot x~,
\end{align*}%
and can thus read off that the program $\DIVERGE$ indeed does not terminate, whereas, since $x > \minfty$, we can see that $\ASSIGN{x}{2\cdot x}$ does always terminate.
\qedtriangle
\end{remark}%

\subsubsection*{\textbf{Conditional Choice.}}

When executing $\ITE{\varphi}{C_1}{C_2}$ on some initial state $\sigma$, the branch $C_1$ is executed $\sigma$ satisfies the predicate $\varphi$ and otherwise~$C_2$ is executed.

Denotationally, the weakest precondition of $\ITE{\varphi}{C_1}{C_2}$ is given by%
\begin{align*}
	\wp{\ITE{\varphi}{C_1}{C_2}}{\psi} \eeq \varphi\wedge\wp{C_1}{\varphi} \morespace{\vee} \neg\varphi\wedge\wp{C_2}{\varphi}~,
\end{align*}%
where --- as usual --- $\wedge$ binds stronger than $\vee$.
Indeed, whenever an initial state~$\sigma$ satisfies the above precondition then \emph{either} $\sigma \models \varphi$ and then --- since then $\sigma$ must also satisfy $\wp{C_1}{\psi}$ --- executing~$C_1$ can terminate in a final state satisfying $\varphi$, \emph{or} $\sigma \not\models \varphi$ and then --- since then $\sigma$ must also satisfy $\wp{C_2}{\psi}$ --- executing~$C_2$ can terminate in a final state satisfying $\varphi$.

In order to mimic the above in a quantitative setting, we make use of so called \emph{Iverson brackets}~\cite{Knuth1992}.
Usually, these turn a predicate $\varphi$ into an indicator function $\iverson{\varphi}_\textsf{\tiny std}\colon \States \to \{0,\, 1\}$, which map a state $\sigma$ to $1$ or $0$, depending on whether $\sigma \models \varphi$ or not.
In our extended real setting, however, we need to slightly adapt the Iverson brackets as follows:%
\begin{definition}[Extended Iverson Brackets]
\label{def:extended-iverson}
	For a predicate $\varphi$, we define the extended Iverson bracket $\iverson{\varphi}\colon \States \to \{\minfty,\, \pinfty\}$ by%
	\abovedisplayskip=-.5\baselineskip%
	\begin{align*}
		\iverson{\guard}(\sigma) \eeq \begin{cases}
			\pinfty & \text{ if } \sigma \mmodels \guard\\[.25em]
			\minfty & \text{ otherwise.}
		\end{cases}\tag*{\raisebox{-0.75\baselineskip}{\qedtriangle}}
	\end{align*}%
	\normalsize%
\end{definition}%
\noindent{}%
Intuitively, this choice is motivated by the fact that $\minfty,\pinfty$ are respectively the bottom and top element of the lattice, and equipped with $\curlyvee,\curlywedge$, they behave exactly as the boolean values $\true,\false$ with $\vee,\wedge$.
Using these Iverson brackets, we define the \emph{quantitative} weakest pre of conditional \mbox{choice by}%
\begin{align*}
	\wp{\ITE{\varphi}{C_1}{C_2}}{f} \eeq \iverson{\varphi} \curlywedge \wp{C_1}{f} \ccurlyvee \iverson{\neg\varphi} \curlywedge \wp{C_2}{f}~.
\end{align*}%
(Recall that $\curlywedge$ binds stronger than $\curlyvee$.)
If the current program state $\sigma$ satisfies~$\varphi$, then $\iverson{\varphi}$ evaluates to $\pinfty$ --- the \emph{greatest} element of $\A$. 
Taking a minimum ($\curlywedge$) with $\wp{C_1}{f}$ will thus yield exactly $\wp{C_1}{f}$.
$\iverson{\neg\varphi}$, on the other hand, then evaluates to $\minfty$ --- the \emph{smallest} element of $\A$.
Taking a minimum with any other lattice element will again yield $\minfty$.
Finally, we then take a maximum~($\curlyvee$) between $\wp{C_1}{f}$ and $\minfty$, yielding $\wp{C_1}{f}$.
This is precisely the quantity that we would expect to anticipate for $f$, if $\sigma \models \varphi$, because then $C_1$ is executed and $\wp{C_1}{f}$ anticipates the value of $f$ after execution of $C$.
The situation for~$\sigma \not\models \varphi$ is completely dual, yielding $\wp{C_2}{f}$.
Indeed, depending on whether an initial state satisfies $\varphi$ or not, the quantitative weakest pre anticipates \emph{either}~$\wp{C_1}{f}$ \emph{or} $\wp{C_2}{f}$.%
\begin{remark}
\label{rem:iversons}
We note that our $\wpsymbol$ rule for conditional choice is different from e.g.~\cite{DBLP:journals/jcss/Kozen85,DBLP:series/mcs/McIverM05,benni_diss}, who use standard instead of extended Iverson brackets, multiplication instead of minimum, and summation instead of maximum, i.e.%
\begin{align*}
	\wp{\ITE{\varphi}{C_1}{C_2}}{f} = \iverson{\guard}_\textsf{\tiny std}\cdot \wp{C_1}{f} + \iverson{\neg\guard}_\textsf{\tiny std}\cdot \wp{C_2}{f}~,
\end{align*}%
This rule, however, would fail in our context of signed quantities because of issues with ${+}\infty \cdot {-}\infty$.
\qedtriangle
\end{remark}%

\subsubsection*{\textbf{Sequential Composition.}}

What is the anticipated value of $f$ after executing $\COMPOSE{C_1}{C_2}$, i.e.\ the value of $f$ after first executing $C_1$ and then $C_2$?
To answer this, we first anticipate the value of $f$ after execution of $C_2$ which gives~$\wp{C_2}{f}$.
Then, we anticipate the value of the intermediate quantity $\wp{C_2}{f}$ after execution of $C_1$, yielding %
%
	$\wp{\COMPOSE{C_1}{C_2}}{f} = \wp{C_1}{\wp{C_2}{f}}$.
%

\subsubsection*{\textbf{Looping.}}

The quantitative weakest pre of a loop $\WHILEDO{\varphi}{C}$ is defined as a least fixed point of the \emph{$\wpsymbol$--characteristic function} $\Phi_f\colon \A\to\A$.
This function is chosen in a way so that iterating $\Phi_f$ on the least element of the lattice $\minfty$ essentially yields an ascending chain of loop unrollings%
\begin{align*}
	\Phi_f(\minfty) &\eeq \wp{\mathtt{if} ({\varphi}) \{\DIVERGE\}}{f}\\
	\Phi_f^2(\minfty) &\eeq \wp{\mathtt{if} ({\varphi}) \{\COMPOSE{C}{\mathtt{if} ({\varphi}) \{\DIVERGE\}}\}}{f}\\
	\Phi_f^3(\minfty) &\eeq \wp{\mathtt{if} ({\varphi}) \{\COMPOSE{C}{\mathtt{if} ({\varphi}) \{\COMPOSE{C}{\mathtt{if} ({\varphi}) \{\DIVERGE\}\}}}\}}{f}
\end{align*}%
and so on, whose supremum is the least fixed point of $\Phi_f$.

%
%
\begin{restatable}[Soundness of \textnormal{$\wpsymbol$}]{theorem}{wpsoundness}%
	\label{thm:wp-soundness}%
		For all programs $C$ and initial states $\sigma$,
		\begin{align*}
			\wp{C}{f}(\sigma)\qeq
				\bigcurlyvee_{\tau\in\eval{C}(\sigma)}f(\tau)~. 
		\end{align*}
\end{restatable}%
\noindent{}%
Intuitively, for a given postquantity $f$ and initial state $\sigma$, $\wp{C}{f}(\sigma)$ is the \emph{supremum} over all the values that $f$ can assume measured in the final states reached after successful termination of the program $C$ on initial state $\sigma$.
In case of no terminating state, i.e.\ $\eval{C}(\sigma) = \emptyset$, that supremum becomes~$\minfty$ --- the absolute minimal value.
In particular, if~\mbox{$\forall\, \tau\colon f(\tau) > \minfty$}, then \mbox{$\wp{C}{f}(\sigma) = \minfty$} unambiguously indicates \emph{nontermination} of $C$ on input $\sigma$.

\subsection{Weakest Liberal Pre}
\label{sec:sub-wlp}

Besides weakest preconditions, Dijkstra also defines weakest \emph{liberal} preconditions.
The \emph{weakest liberal precondition transformer} is again of type%
\begin{align*}
	\wlpC{C}\colon\quad \B \morespace{\to} \B~,
\end{align*}%
associating to each nondeterministic program $C$ a mapping from predicates to predicates.
For reasons of duality, we now consider a \emph{demonic} setting, where the nondeterminism is resolved to our \emph{dis}{\-}advantage.
The difference from \emph{non}liberal weakest preconditions, however, is that \emph{nonterminating} behavior is deemed \emph{good behavior} (i.e.\ as if the program terminated in a state satisfying the post{\-}condition).
Specifically, the \emph{demonic} weakest liberal precondition transformer $\wlpC{C}$ maps a \emph{post}{\-}condition~$\psi$ over final states to a \emph{pre}condition $\wlp{C}{\psi}$ over initial states, such that executing~$C$ on an initial state satisfying $\wlp{C}{\psi}$ guarantees that $C$ will either not terminate, or terminate in a final state satisfying~$\psi$.
More symbolically, recalling that $\eval{C}(\sigma)$ is the set of all final states reachable after termination of $C$ on $\sigma$,%
\begin{align*}
	\sigma \mmodels \wlp{C}{\psi} \qqiff \forall\, \tau \in \eval{C}(\sigma)\colon \quad \tau \mmodels \psi~,
\end{align*}%
where the right-hand-side of the implication is vacuously true if $\eval{C}(\sigma) = \emptyset$, i.e.\ if $C$ does not terminate on $\sigma$.
From the \emph{map perspective}, $\wlp{C}{\psi}$ is a function that takes as input an initial state $\sigma$, determines for each reachable final state~\mbox{$\tau \in \eval{C}(\sigma)$} the (truth) value $\psi(\tau)$, and returns a \emph{conjunction} over all these truth values.
More symbolically,%
\begin{align*}
	\wlp{C}{\psi}(\sigma) \qqeq\quad \bigwedge_{\mathclap{\tau \in \eval{C}(\sigma)}} \quad \psi(\tau)~,
\end{align*}%
where the conjunction over an empty set is --- as is standard --- given by $\true$.

Just like a conjunction in some sense minimizes truth values, our quantitative weakest liberal pre should also minimize, while at the same time assigning a maximal value to nontermination.
This is captured by the following transformer:
\begin{definition}[Quantitative Weakest Liberal Pre]
\label{def:wlp}
	The \emph{quantitative weakest liberal pre transformer}%
	\begin{align*}
		\wlpsymbol\colon\quad \ngcl \to (\A \to \A)
	\end{align*}%
	is defined inductively according to the rules in \textnormal{\Cref{table:wlp}} (right column).
	We call the function%
	\begin{align*}
		\Phi_f(X) \eeq \iverson{\neg \guard} \curlywedge f \ccurlyvee \iverson{\guard} \curlywedge \wlp{C}{X}~,
	\end{align*}
	whose \emph{greatest} fixed point defines the weakest liberal pre $\wlp{\WHILEDO{\guard}{C}}{f}$, the \emph{$\wlpsymbol$--characteristic function} (of $\WHILEDO{\guard}{C}$ with respect to $f$).\qedtriangle
\end{definition}%
\noindent{}%
The rules for assignments, sequential composition, and conditional choice are the same as for $\wpsymbol$.
This is unsurprisingly so, since those rules pertain neither to nontermination nor to nondeterminism.
Let us thus go over the rules for the language constructs, where the rules for $\wlpsymbol$ and $\wpsymbol$ differ.

\subsubsection*{\textbf{Diverge.}}

Since $\DIVERGE$ is certainly nonterminating and liberal pre\emph{conditions} deem this good behavior, the weakest liberal precondition of $\DIVERGE$ \mbox{is given by}%
\begin{align*}
	\wlp{\DIVERGE}{\psi} \eeq \true~.
\end{align*}%
Note that $\true$ is the \emph{greatest element} in the Boolean lattice.
When moving to quantities, we also assign to nonterminating behavior the greatest element, i.e.%
\belowdisplayskip=-.5\baselineskip%
\begin{align*}
	\wlp{\DIVERGE}{f} \eeq \pinfty~.
\end{align*}%
\normalsize%
\begin{remark}[Quantitative Weakest Liberal Pre and Nontermination]%
\label{rem:nontermination-wlp}
Analogously to $\minfty$ being the \emph{the value of nontermination} in $\wpsymbol$ (see \Cref{rem:nontermination-wp}), $\pinfty$ is \emph{the value of nontermination} in $\wlpsymbol$.\qedtriangle
\end{remark}%

\subsubsection*{\textbf{Nondeterministic Choice.}}

Since weakest \emph{liberal} pre is \emph{demonic}, we ask in $\wlpsymbol$ for the \emph{minimal} anticipated value of $f$ after termination of $C_1$ or $C_2$. Hence the rule is dually given by the meet
\begin{align*}
	\wlp{\NDCHOICE{C_1}{C_2}}{f} \eeq \wlp{C_1}{f} \ccurlywedge \wlp{C_2}{f}~.
\end{align*}%
Notice that if either $C_1$ or $C_2$ yield $\pinfty$ because of nontermination, the $\wlpsymbol$ above will select as value the respective other branch if that one terminates.

\subsubsection*{\textbf{Looping.}}

The weakest liberal pre of a loop $\WHILEDO{\varphi}{C}$ is defined as a greatest fixed point of the \emph{$\wlpsymbol$--characteristic function} $\Phi_f\colon \A\to\A$.
This function is chosen in a way so that iterating $\Phi_f$ on the greatest element of the lattice $\pinfty$ essentially yields a descending chain of loop unrollings%
\begin{align*}
	\Phi_f(\pinfty) &\eeq \wlp{\mathtt{if} ({\varphi}) \{\DIVERGE\}}{f}\\
	\Phi_f^2(\pinfty) &\eeq \wlp{\mathtt{if} ({\varphi}) \{\COMPOSE{C}{\mathtt{if} ({\varphi}) \{\DIVERGE\}}\}}{f}\\
	\Phi_f^3(\pinfty) &\eeq \wlp{\mathtt{if} ({\varphi}) \{\COMPOSE{C}{\mathtt{if} ({\varphi}) \{\COMPOSE{C}{\mathtt{if} ({\varphi}) \{\DIVERGE\}\}}}\}}{f}
\end{align*}
and so on, whose infimum is the greatest fixed point of $\Phi_f$.

%
%
\begin{restatable}[Soundness of \textnormal{$\wlpsymbol$}]{theorem}{wlpsoundness}%
	\label{thm:wlp-soundness}%
		For all programs $C$ and states $\sigma\in\States$,
		\begin{align*}
	\wlp{C}{f}(\sigma)\qeq
	\bigcurlywedge_{\tau\in\eval{C}(\sigma)}f(\tau)~. 
\end{align*}
\end{restatable}%
\noindent{}%
Intuitively, for a given postquantity $f$ and initial state $\sigma$, the quantitative weakest liberal pre $\wlp{C}{f}(\sigma)$ is the \emph{infimum} over all values that $f$ can assume measured in the final states after termination of the program $C$ on initial state~$\sigma$.
In case of no terminating state, i.e.\ $\eval{C}(\sigma) = \emptyset$, that infimum automatically becomes $\pinfty$ --- the absolute maximal value.
In particular, if~\mbox{$\forall\, \tau\colon f(\tau) < \pinfty$}, then $\wlp{C}{f}(\sigma) = \pinfty$ unambiguously indicates \emph{nontermination} of $C$ \mbox{on input $\sigma$}.%

\section{Strongest Post}
\label{se:sp}


We now present our main contribution: A lifting of the strongest postcondition calculus of \citet{Dijkstra1990} to quantities and a completely novel (quantitative) strongest \emph{liberal} post calculus. 
To the best of our knowledge, a strongest liberal post(condition) has never been proposed before, not even in the \emph{qualitative} setting.\footnote{Although some authors do use the term \enquote{strongest liberal postcondition}, 
see~\Cref{se:relatedwork:slp} for a comparison.}
We again start by recapping the classical calculus.

\subsection{Classical Strongest Postconditions}
Dijkstra and Scholten's strongest postcondition calculus employs \emph{predicate transformers} of type%
\begin{align*}
	\spC{C}\colon\quad \B \morespace{\to} \B~, \qquad \textnormal{where}\quad \B \eeq \States \to \{0,\, 1\}~,
\end{align*}%
which associate to each nondeterministic program $C$ a mapping from predicates (sets of program states) to predicates. Strongest post transformers, analogously to the collecting semantics, characterize the set states that \emph{can} be reached, so that an \emph{angelic} setting is chosen to resolve nondeterminism to our advantage.
Concretely, the \emph{angelic} strongest postcondition transformer $\spC{C}$ maps a \emph{pre}{\-}condition~$\psi$ over initial states to a \emph{post}condition $\sp{C}{\psi}$ over final states, such that every state in the postcondition \emph{is reachable} from some initial state satisfying $\psi$.
This corresponds exactly with the definition of the collecting semantics $\eval{C}(\sigma)$:
In fact, 
\begin{align*}
	\tau \mmodels \sp{C}{\psi} \qqiff 
	\exists\,\sigma \textnormal{ with } \tau \in \eval{C}(\sigma)\colon \quad \sigma \mmodels \psi~.
\end{align*}%
As we did for weakest pre, let us provide a \emph{map perspective} on strongest postconditions, see \Cref{fig:sp-map}.
\begin{figure}[t]
	\begin{subfigure}[t]{.45\textwidth}
		\begin{center}
			\begin{adjustbox}{max width=.8\textwidth}
				\begin{tikzpicture}[node distance=4mm, decoration={snake,pre=lineto,pre length=.5mm,post=lineto,post length=1mm, amplitude=.2mm}]
							\draw[use as bounding box,white] (-4.55,-0.4) grid (3.25, 4.25);
							\node (tau) at (0, 0) {\Large$\boldsymbol{~\tau}$};
							\node[gray,inner sep=0pt, outer sep=0pt] (firstplus) at (-1,2) {\LARGE$\Box$};
							
							\node[black,inner sep=0pt, outer sep=0pt] (branch) at (0.3, 2.1) {$\bullet$};

							\node (sigma1) at (-2, 4) {$\bullet$};
							\node (sigma2) at (-.5, 3.75) {$\bullet$};
							\node (sigma3) at (1, 3.875) {$\bullet$};
							\node[inner sep=0pt] (sigmadots) at (2.5, 4) {$\ddots$};
							\node[inner sep=0pt] (taudots) at (-1.5, 0.8) {$\ddots$};

							\node[above of=sigma1] {$\psi(\sigma_1)$};
							\node[above of=sigma2] {$\psi(\sigma_2)$};
							\node[above of=sigma3] {$\psi(\sigma_3)$};

							\node (Exp) at (-3.1, 4.25) {\huge $\boldsymbol{\bigvee}$\Large~$\boldsymbol{\Bigl[}$};
							\node at (3.1, 4.25) {\Large $\boldsymbol{\Bigr]}$};

							\node[gray] (C) at (0.5, 1.5) {$C$};
							
							\draw[gray,decorate,->,thick] (firstplus)-- (tau);
							\draw[gray,decorate,->,thick] (sigmadots)-- (tau);
							\draw[gray,decorate,thick] (firstplus) -- (sigma1);
							
							\draw[gray,decorate,->,thick] (firstplus) -- (taudots);
							
							\draw[gray,decorate,thick] (branch) -- (sigma2);
							\draw[gray,decorate,thick] (branch) -- (sigma3);
							\draw[gray,decorate,->,thick] (branch) -- (tau);
							
							\draw[decorate,lightgray] (sigma1) -- (sigma2) -- (sigma3) -- (sigmadots);
							
							\draw[gray] (tau) edge[|->,bend left=45,below left,thick] node {\textcolor{black}{\Large$\sp{C}{\psi}$}} (Exp);
						\end{tikzpicture}
			\end{adjustbox}
		\end{center}%
	\caption{\footnotesize\textbf{Strongest postconditions:}
	Given final state $\tau$, $\sp{C}{\psi}$ determines all initial states $\sigma_i$ that can reach~$\tau$ by executing $C$, evaluates $\psi$ in those states, and returns the disjunction over all these truth values.}
	\label{fig:sp-map}
	\end{subfigure}
	\hfill
	\begin{subfigure}[t]{.45\textwidth}
		\begin{center}
			\begin{adjustbox}{max width=.8\textwidth}
				\begin{tikzpicture}[node distance=4mm, decoration={snake,pre=lineto,pre length=.5mm,post=lineto,post length=1mm, amplitude=.2mm}]
							\draw[use as bounding box,white] (-4.55,-0.4) grid (3.25, 4.25);
							\node (tau) at (0, 0) {\Large$\boldsymbol{~\tau}$};
							\node[gray,inner sep=0pt, outer sep=0pt] (firstplus) at (-1,2) {\LARGE$\Box$};
							
							\node[black,inner sep=0pt, outer sep=0pt] (branch) at (0.3, 2.1) {$\bullet$};

							\node (sigma1) at (-2, 4) {$\bullet$};
							\node (sigma2) at (-.5, 3.75) {$\bullet$};
							\node (sigma3) at (1, 3.875) {$\bullet$};
							\node[inner sep=0pt] (sigmadots) at (2.5, 4) {$\ddots$};
							\node[inner sep=0pt] (taudots) at (-1.5, 0.8) {$\ddots$};

							\node[above of=sigma1] {$f(\sigma_1)$};
							\node[above of=sigma2] {$f(\sigma_2)$};
							\node[above of=sigma3] {$f(\sigma_3)$};

							\node (Exp) at (-3.1, 4.25) {\huge $\boldsymbol{\bigcurlyvee}$\Large~$\boldsymbol{\Bigl[}$};
							\node at (3.1, 4.25) {\Large $\boldsymbol{\Bigr]}$};

							\node[gray] (C) at (0.5, 1.5) {$C$};
							
							\draw[gray,decorate,->,thick] (firstplus)-- (tau);
							\draw[gray,decorate,->,thick] (sigmadots)-- (tau);
							\draw[gray,decorate,thick] (firstplus) -- (sigma1);
							
							\draw[gray,decorate,->,thick] (firstplus) -- (taudots);
							
							\draw[gray,decorate,thick] (branch) -- (sigma2);
							\draw[gray,decorate,thick] (branch) -- (sigma3);
							\draw[gray,decorate,->,thick] (branch) -- (tau);
							
							\draw[decorate,lightgray] (sigma1) -- (sigma2) -- (sigma3) -- (sigmadots);
							
							\draw[gray] (tau) edge[|->,bend left=45,below left,thick] node {\textcolor{black}{\Large$\sp{C}{f}$}} (Exp);
						\end{tikzpicture}
			\end{adjustbox}
		\end{center}%
	\caption{\footnotesize\textbf{Quantitative strongest post:}
	Given final state~$\tau$, $\sp{C}{f}$ determines all initial states $\sigma_i$ that can reach~$\tau$ by executing $C$, evaluates $f$ in those states, and returns the supremum~($\curlyvee$) over all these quantities.}
	\label{fig:sp-general}
	\end{subfigure}
	\caption{Angelic strongest postconditions and quantitative strongest posts.}
	\label{fig:sp-views}
	\end{figure}%
From this perspective, the precondition $\psi\colon \States \to \{0,\, 1\}$ maps program states to truth values.
The predicate $\sp{C}{\psi}$ is then a map that takes as input a \emph{final} state $\tau$, determines for all initial states~$\sigma$ that can reach $\tau$ the (truth) value~$\psi(\sigma)$, and returns the disjunction ($\vee$) over all these truth values:%
%
\begin{align*}
	\sp{C}{\psi}(\tau) \qqeq\quad \bigvee_{\mathclap{\sigma \textnormal{ with } \tau \in \eval{C}(\sigma)}} \quad \psi(\sigma)~.
\end{align*}%
In other words: \emph{Given a final state $\tau$, $\sp{C}{\psi}(\tau)$ \underline{retro}dicts whether before executing $C$ the predicate~$\psi$ could have been true}.
In the following, we define quantitative strongest post and strongest liberal post calculi which \emph{\underline{retroc}ip\underline{ate}} values of signed quantities \emph{before} the execution of a nondeterministic program (whereas $\wpsymbol$ and $\wlpsymbol$ \emph{anticipate} values after the execution).


%
%
\subsection{Quantitative Strongest Post}

Let us generalize the \emph{map perspective} of strongest postconditions to quantities. 
Instead of a pre\emph{condition}, we now have a pre\emph{quantity} $f\colon \States \to \Rinf$.
$\sp{C}{f}\colon \States \to \Rinf$ is then a function that takes as input a \emph{final} state $\tau$, determines all initial states $\sigma$ that can reach $\tau$ by executing~$C$, evaluates the prequantity $f(\sigma)$ in each of those initial states $\sigma$, and finally returns the supremum over all these so-determined quantities, see \Cref{fig:sp-general}.
As a transformer, we obtain the following:%
%
%
%
\begin{definition}[Quantitative Strongest Post]%
	\label{def:sp}%
		The \emph{strongest post transformer}%
		\begin{align*}
			\spsymbol\colon\quad \ngcl \to (\A \to \A)
		\end{align*}%
		is defined inductively according to the rules in \textnormal{\Cref{table:sp}} (middle column).
		We call the function%
		\begin{align*}
			\Psi_f(X) \eeq  f \ccurlyvee \sp{C}{\iguard\curlywedge X}~,
		\end{align*}
		whose \emph{least} fixed point is used to define $\sp{\WHILEDO{\guard}{C}}{f}$, the $\spsymbol$--charac{\-}teristic function of $\WHILEDO{\guard}{C}$ with respect to $f$.\qedtriangle
	\end{definition}%
	\noindent{}%
\begin{table*}[t]
	\begin{adjustbox}{max width=\textwidth}
	\renewcommand{\arraystretch}{1.5}
	\begin{tabular}{@{\hspace{.5em}}l@{\hspace{2em}}l@{\hspace{2em}}l@{\hspace{.5em}}}
		\hline
		$\boldsymbol{C}$			& $\boldsp{C}{f}$ & $\boldslp{C}{f}$\\
		\hline
		$\DIVERGE$				& $\minfty$ 														& $\pinfty$	\\
		$\ASSIGN{x}{\ee}$			& $\SupV{\alpha}~ \iverson{x = e\subst{x}{\alpha}} \ccurlywedge f\subst{x}{\alpha}$ 	& $\InfV{\alpha}~ \iverson{x \neq e\subst{x}{\alpha}} \ccurlyvee f\subst{x}{\alpha}$\\
		$\COMPOSE{C_1}{C_2}$		& $\sp{C_2}{\vphantom{\big(}\sp{C_1}{f}}$ & \lightgray{$\slp{C_2}{\vphantom{\big(}\slp{C_1}{f}}$}\\
		$\NDCHOICE{C_1}{C_2}$		& $\sp{C_1}{f} \ccurlyvee \sp{C_2}{f}$ & $\slp{C_1}{f} \ccurlywedge \slp{C_2}{f}$\\
		$\ITE{\guard}{C_1}{C_2}$		& $\sp{C_1}{\iverson{\guard} \curlywedge f} \ccurlyvee \sp{C_2}{\iverson{\neg \guard} \curlywedge f}$ & $\slp{C_1}{\inguard \curlyvee f} \ccurlywedge \slp{C_2}{\iguard\curlyvee f}$\\
 		$\WHILEDO{\guard}{C'}$		& $\iverson{\neg\guard} \curlywedge \bigl( \lfp Y\mydot f \curlyvee \sp{C'}{\iverson{\guard} \curlywedge Y}\bigr)$ & $\iguard \curlyvee \bigl( \gfp Y\mydot f \ccurlywedge \slp{C'}{\inguard \curlyvee Y}\bigr)$\\
		%
		%
		%
		%
		%
		\hline\hline
	\end{tabular}%
	\end{adjustbox}%
	\vspace{.5em}
	\caption{Rules for $\spsymbol$ and $\slpsymbol$. $\lfp g\mydot \Psi(g)$ and $\gfp g\mydot \Psi(g)$ denote the least and greatest fixed point of $\Phi$. $\InfVCaption{\alpha} f(\alpha)$ and $\SupV{\alpha} f(\alpha)$ denote the infimum and supremum of $f(\alpha)$ ranging over all values of $\alpha$.}
	\label{table:sp}
	\label{table:slp}
	\vspace{-1em}
\end{table*}%
\noindent{}%
Again, let us go over some of the rules for quantitative $\spsymbol$ and show how they can be developed and understood analogously to strongest postconditions.


\subsubsection*{\textbf{Assignment.}}
Dijkstra and Scholten's strongest postcondition of an assignment is given by%
\begin{align*}
	\sp{\ASSIGN{x}{e}}{\psi} \qqeq \exists\, \alpha\colon \quad \gray{\underbrace{\black{x=e\subst{x}{\alpha}}}_{\textnormal{(1)}}} \morespace{\wedge} \gray{\underbrace{\black{\psi\subst{x}{\alpha}}}_{\textnormal{(2)}}}~.
\end{align*}%
Intuitively, the quantified $\alpha$ represents an \emph{initial} value that $x$ could have had \emph{before} executing the assignment.
(If at all possible), the $\alpha$ is chosen in a way so that
\begin{enumerate}
	\item $x$ has in the final state the value of expression $e$ but evaluated using $x$'s initial value $\alpha$, and
	\item the precondition $\psi$ was true in the initial state where $x$ had value $\alpha$.
\end{enumerate}
For quantities, we note that, regarding (1), there could have been multiple valid initial values $\alpha$ for~$x$; for instance, before the execution of $\ASSIGN{x}{10}$, \emph{any} initial value $\alpha$ is valid. 
Our intuition is that, in order to preserve backward compatibility, we substitute the \emph{existential} quantifier with a \emph{supremum} (denoted by the $\Sup$ \enquote{quantifier}, cf.~\cite{DBLP:journals/pacmpl/BatzKKM21}), thus obtaining the supremum of $f\subst{x}{\alpha}$ ranging over all valid initial values $\alpha$ of $x$:%
%
\begin{align*}
	\sp{\ASSIGN{x}{e}}{f} \qqeq \SupV{\alpha}\quad \iverson{x = e\subst{x}{\alpha}} \ccurlywedge f\subst{x}{\alpha}~.
\end{align*}%
Let us consider a few examples.
First, consider%
\begin{align*}
	\sp{\ASSIGN{x}{x + 1}}{x}  \qeq \SupV{\alpha}~ \iverson{x = \alpha + 1} \ccurlywedge \alpha \qeq \SupV{\alpha}~ \iverson{\alpha = x - 1} \ccurlywedge \alpha \qeq x - 1~.
\end{align*}%
For a final state $\tau(x) = 10$, this gives us $\tau(x) - 1 = 10 - 1 = 9$ which is indeed \emph{the} initial value that the prequantity $x$ must have had if the final state after executing $\ASSIGN{x}{x + 1}$ is $\tau(x) = 10$.

As another example, consider%
\begin{align*}
	\sp{\ASSIGN{x}{10}}{x}  \qeq \SupV{\alpha}~ \iverson{x = 10} \ccurlywedge \alpha \qeq \iverson{x = 10} \ccurlywedge \infty \qeq \iverson{x = 10}~.
\end{align*}%
For the final state $\tau(x) = 10$, this gives us $\iverson{10 = 10} = \iverson{\true} = {+}\infty$ which is indeed the \emph{least upper bound} (angelic!) on the initial value of $x$ if the final state after executing $\ASSIGN{x}{10}$ is $\tau$.
In other words: by evaluating $\iverson{x = 10}$ in $\tau$, we know that $\tau$ was reachable, but we have no information on what maximal value $x$ could have had initially, which is sensible because $\ASSIGN{x}{10}$ forgets any \mbox{initial value of $x$}.
For final state $\tau'(x) = 9$, on the other hand, we get $\iverson{9 = 10} = \iverson{\false} = {-}\infty$ which is the \emph{value of unreachability} in $\spsymbol$ (cf.~also the next paragraph on divergence).
Indeed, the final state after executing $\ASSIGN{x}{10}$ cannot ever be $\tau'$.

\subsubsection*{\textbf{Diverge.}}

The strongest postcondition of $\DIVERGE$ is given by%
\begin{align*}
	\sp{\DIVERGE}{\psi} \eeq \false~,
\end{align*}%
the \emph{least element} in the Boolean lattice.
Since there is \emph{no} state that satifies $\false$, this simply tells us that there is no final state reachable by executing $\DIVERGE$.

For quantities, we also assign the least element and hence get%
%
\begin{align*}
	\sp{\DIVERGE}{f} \eeq \minfty~.
\end{align*}%
\normalsize%
Another explanation goes by considering again the \emph{angelic}, i.e.\ maximizing, aspect of strongest post:
What is the maximal value that we can retrocipate for $f$ \emph{before} $\DIVERGE$ has terminated in some final state $\tau$?
Since $\DIVERGE$ does not terminate at all and hence no such $\tau$ could have been reached (but we are still forced to assign some \enquote{number} to this situation), the largest value that we can possibly retrocipate is the absolute minimum: $\minfty$.%
\begin{remark}[Quantitative Strongest Post and Unreachability]
\label{rem:sp-value-of-unreach}
	Dually to values of nontermination in $\textsf{w(l)p}$ (see Remarks~\ref{rem:nontermination-wp} and \ref{rem:nontermination-wlp}), $\minfty$ is in that sense \emph{the value of unreachability} in $\spsymbol$.\qedtriangle
\end{remark}%

\subsubsection*{\textbf{Nondeterministic Choice.}}

The \emph{angelic} strongest post\emph{condition} of $\NDCHOICE{C_1}{C_2}$ is given by%
\begin{align*}
	\sp{\NDCHOICE{C_1}{C_2}}{\psi} \eeq \sp{C_1}{\psi} \morespace{\vee} \sp{C_2}{\psi}~.
\end{align*}%
Indeed, the set of reachable states starting from initial states satisfying $\psi$ is the union of the reachable set after executing $C_1$ and the ones after executing $C_2$.

In a quantitative setting, where we want to retrocipate the value of a quantity~$f$ before executing either $C_1$ or $C_2$, we \emph{angelically} maximize between the two retrocipated quantities:%
%
\begin{align*}
	\sp{\NDCHOICE{C_1}{C_2}}{f} \eeq \sp{C_1}{f} \ccurlyvee \sp{C_2}{f}~.
\end{align*}%

\subsubsection*{\textbf{Conditional Choice.}}

The strongest post\emph{condition} of $\ITE{\varphi}{C_1}{C_2}$ is given by%
\begin{align*}
	\sp{\ITE{\varphi}{C_1}{C_2}}{\psi} \eeq \sp{C_1}{\varphi\wedge\psi} \morespace{\vee}\sp{C_2}{\neg\varphi\wedge\psi}~,
\end{align*}%
So to determine the set of reachable states starting from precondition $\psi$, we split the precondition into two disjoint ones --- $\varphi\wedge\psi$ assumes that the guard is true and we execute $C_1$, whereas $\neg\varphi\wedge\psi$ assumes the guard to be false and we execute~$C_2$. 
Thereafter, we union the \mbox{so-obtained reachable sets}.

Similarly for our quantitative strongest post calculi, we make use of the extended Iverson brackets and thus, the denotational strongest post of the conditional choice is:
\begin{align*}
	\sp{\ITE{\varphi}{C_1}{C_2}}{f} \eeq  \sp{C_1}{\iverson{\varphi} \curlywedge f} \ccurlyvee \sp{C_2}{ \iverson{\neg\varphi} \curlywedge f}~.
\end{align*}%
Intuitively, $\sp{C_1}{\iverson{\varphi} \curlywedge f}$ is the supremum of $f$ measured in all initial states before the execution of $C_1$ satisfying $\varphi$; and analogously for $\sp{C_2}{\iverson{\neg\varphi} \curlywedge f}$.
By then taking $\curlyvee$, we finally obtain the maximum initial quantity that $f$ could have had before the execution of the conditional choice.

\subsubsection*{\textbf{Sequential Composition.}}

What is the retrocipated value of $f$ before executing $\COMPOSE{C_1}{C_2}$?
For this, we first retrocipate the value of $f$ before executing $C_1$ which gives $\sp{C_1}{f}$.
Then, we retrocipate the value $\sp{C_1}{f}$ before executing $C_2$, yielding 
%
	\mbox{$\sp{\COMPOSE{C_1}{C_2}}{f} = \sp{C_2}{\sp{C_1}{f}}$}.%
%

\subsubsection*{\textbf{Looping.}}
The strongest post of a loop $\WHILEDO{\varphi}{C}$ is characterized using the least fixed point of the so-called \emph{$\spsymbol$--characteristic function} $\Psi_f\colon \A\to\A$. 
As for weakest pre, the function is chosen so that by Kleene's fixpoint theorem, the least fixed point corresponds to iterating on the least element of the lattice $\minfty$, which yields an ascending chain of loop unrollings
\begin{align*}
	\inguard \curlywedge\Psi_f(\minfty) &\eeq \sp{\mathtt{if} ({\varphi}) \{\DIVERGE\}}{f}\\
	\inguard \curlywedge\Psi_f^2(\minfty) &\eeq \sp{\mathtt{if} ({\varphi}) \{\COMPOSE{C}{\mathtt{if} ({\varphi}) \{\DIVERGE\}}\}}{f}\\
	\inguard \curlywedge\Psi_f^3(\minfty) &\eeq \sp{\mathtt{if} ({\varphi}) \{\COMPOSE{C}{\mathtt{if} ({\varphi}) \{\COMPOSE{C}{\mathtt{if} ({\varphi}) \{\DIVERGE\}\}}}\}}{f}
\end{align*}
and so on, where the guard is needed to filter only those states that exit the loop; we finally obtain as strongest post%
\belowdisplayskip=-.5\baselineskip%
\begin{align*}
	\sp{\WHILE{\guard}{C}}{f} \qeq \inguard \ccurlywedge \lfp \Psi_f~.
\end{align*}%

\normalsize%
%
%
%
\begin{restatable}[Soundness of $\spsymbol$]{theorem}{spsoundness}%
	\label{thm:sp-soundness}%
	For all programs $C$ and final states $\tau$,
	\begin{align*}
		\sp{C}{f}(\tau) \qeq\quad \bigcurlyvee_{\mathclap{\sigma  \textnormal{ with } \tau\in \eval{C}(\sigma)}}\quad
		f(\sigma)~. 
	\end{align*}
\end{restatable}
\noindent{}%
Intuitively, for a given prequantity $f$ and final state $\tau$, $\spC{f}(\tau)$ is the \emph{supremum} over all the values that $f$ can assume in those initial states $\sigma$ from which executing $C$ terminates in $\tau$.
In case that the final state $\tau$ is \emph{unreachable}, i.e.\ $\forall \sigma\colon \tau\notin \eval{C}(\sigma)$, that supremum automatically becomes $\minfty$ --- the absolute minimal value.
In particular, if~\mbox{$\forall\, \sigma\colon f(\sigma) > \minfty$}, then $\sp{C}{f}(\tau) = \minfty$ unambiguously indicates \emph{unreachability} of $\tau$ by executing $C$ on any input $\sigma$.

	\subsection{Quantitative Strongest Liberal Post}
	Although Dijkstra does not define strongest \emph{liberal} postconditions, we believe that a reasonable choice for a quantitative strongest liberal post transformer is to take the \emph{infimum} over all prequantities. 
	Restricting to predicates, we thereby also obtain a novel \emph{strongest liberal postcondition transformer} of type %
		$\slpC{C}\colon \B \to \B$
	%
	associating to each nondeterministic program $C$ a mapping from predicates to predicates.
	Since $\slpsymbol$ is associated with the infimum, we will consider a \emph{demonic} setting, where the nondeterminism is resolved to our \emph{dis}advantage.
	Whereas weakest liberal pre, in contrast to the \emph{non}-liberal transformers, deems \emph{non-termination} good behavior, strongest liberal post deems \emph{unreachability} good behavior.
	
	Specifically, the \emph{demonic} strongest liberal postcondition transformer $\slpC{C}$ maps a \emph{pre}{\-}condition~$\psi$ over initial states to a \emph{post}condition $\slp{C}{\psi}$ over final states, such that for a given final state $\tau$ satisfying $\slp{C}{\psi}$, all initial states that can reach $\tau$ satisfy the precondition $\psi$.
	More symbolically, recalling that $\eval{C}(\sigma)$ is the set of all final states reachable after termination of $C$ on $\sigma$,%
	\begin{align*}
		\tau \mmodels \slp{C}{\psi} \qqiff \forall\, \sigma \textnormal{ with } \tau \in \eval{C}(\sigma)\colon \quad \sigma \mmodels \psi~,
	\end{align*}%
	where the right-hand-side of the implication is vacuously true if $\tau$ is unreachable.
	From a \emph{map perspective} on $\slpsymbol$, the predicate $\slp{C}{\psi}$ is a function that takes as input a final state $\tau$, determines for each initial state $\sigma$ that can reach $\tau$, i.e., $\tau \in \eval{C}(\sigma)$, the (truth) value $\psi(\sigma)$, takes a \emph{conjunction} over all these truth values, and finally returns the truth value of that conjunction.
	More symbolically,%
	\begin{align*}
		\slp{C}{\psi}(\tau) \qqeq\quad \bigwedge_{\mathclap{\sigma \textnormal{ with } \tau \in \eval{C}(\sigma)}} \quad \psi(\sigma)~,
	\end{align*}%
	where the conjunction over an empty set is defined --- as is standard --- as $\true$.
	For quantities, we essentially replace $\wedge$ by $\curlywedge$ and define the following quantitative strongest liberal post transformer:%
	\begin{definition}[Quant.~Strongest Liberal Post]%
		\label{def:slp}%
		The \emph{quantitative strongest liberal post transformer}%
		\begin{align*}
			\slpsymbol\colon\quad \ngcl \to (\A \to \A)
		\end{align*}%
		is defined inductively according to the rules in \textnormal{\Cref{table:slp}} (right column).
		We call the function%
		\begin{align*}
			\Psi_f(X) \eeq f \ccurlywedge \sp{C}{\inguard\curlyvee X}~,
				\end{align*}
		whose \emph{greatest} fixed point is used to define $\slp{\WHILEDO{\guard}{C}}{f}$, the $\slpsymbol$--characteristic function of $\WHILEDO{\guard}{C}$ with respect to $f$. \qedtriangle	
	\end{definition}%
	\noindent{}%
Let us thus go over the language constructs where the rules for $\slpsymbol$ and $\spsymbol$ differ and explain both strongest liberal postconditions and quantitative strongest liberal post.

\subsubsection*{\textbf{Assignment.}}
The strongest liberal post\emph{condition} of an assignment is given by%
\begin{align*}
	\slp{\ASSIGN{x}{e}}{\psi} \qqeq \forall\, \alpha\colon \quad \gray{\underbrace{\black{x\neq e\subst{x}{\alpha}}}_{\textnormal{(1)}}} \morespace{\vee} \gray{\underbrace{\black{\psi\subst{x}{\alpha}}}_{\textnormal{(2)}}}~.
\end{align*}%
Intuitively, the quantified $\alpha$ represents \emph{candidates for initial values} of $x$ \emph{before} executing the assignment.
For each such candidate $\alpha$, it must be true that
\begin{enumerate}
	\item $\alpha$ is in fact \emph{not} a valid initial value for $x$, i.e.\ $x$ does \emph{not} have in the final state the value of expression $e$ evaluated using the candidate value $\alpha$ for $x$, or
	\item $\alpha$ \emph{is} valid and the precondition $\psi$ was true in the initial state where $x$ had value $\alpha$.
\end{enumerate}%
Intuitively, (1) captures that strongest liberal postconditions deem unreachability good behavior, because if some state is not reachable by executing $\ASSIGN{x}{e}$, then $x \neq e\subst{x}{\alpha}$ is true for all $\alpha$ and hence the strongest liberal post evaluates to $\true$.

For quantities, dually to the strongest \emph{non-liberal} post, we now substitute the \emph{universal} quantifier with an \emph{infimum} (denoted by the $\Inf$ \enquote{quantifier}~\cite{DBLP:journals/pacmpl/BatzKKM21}) and the $\vee$ with a $\curlyvee$, thus obtaining
\begin{align*}
	\slp{\ASSIGN{x}{e}}{f} \qqeq \InfV{\alpha}\quad \iverson{x \neq e\subst{x}{\alpha}} \ccurlyvee f\subst{x}{\alpha}
\end{align*}%
Let us again consider a few examples.
First, one can convince oneself that %
\begin{align*}
	\slp{\ASSIGN{x}{x + 1}}{x} \eeq x - 1 \eeq \sp{\ASSIGN{x}{x + 1}}{x}~.
\end{align*}%
$\slpsymbol = \spsymbol$ is not surprising in this case, because \emph{every} state $\tau(x) = \beta$ is reachable by executing $\ASSIGN{x}{x+1}$, namely by starting from initial state $\sigma(x) = \beta - 1$.
As another example, consider%
\begin{align*}
	\slp{\ASSIGN{x}{10}}{x}  \qeq \InfV{\alpha}~ \iverson{x \neq 10} \ccurlyvee \alpha \qeq \iverson{x \neq 10} \ccurlyvee {}\infty \qeq \iverson{x \neq 10}~.
\end{align*}%
For the final state $\tau(x) = 10$, this gives us $\iverson{10 \neq 10} = \iverson{\false} = {-}\infty$ which is indeed the \emph{greatest lower bound} (demonic!) on the initial value of $x$ if the final state after executing $\ASSIGN{x}{10}$ is $\tau$.
In other words: by evaluating $\iverson{x \neq 10}$ in $\tau$, we know that $\tau$ was reachable, but we have no information on what minimal value $x$ could have had initially, which is sensible because $\ASSIGN{x}{10}$ forgets any \mbox{initial value of $x$}.
For final state $\tau'(x) = 9$, on the other hand, we get $\iverson{9 \neq 10} = \iverson{\true} = {+}\infty$ which is the \emph{value of unreachability} in $\slpsymbol$ (cf.~also the next paragraph on divergence).
Indeed, the final state after executing $\ASSIGN{x}{10}$ cannot ever be $\tau'$.

\subsubsection*{\textbf{Diverge.}}

Since $\DIVERGE$ is certainly nonterminating, i.e.\ it reaches no final state, and since liberal post deems nonreachability good behavior, the quantitative strongest liberal post assigns the greatest element, i.e.\ %
	$\slp{\DIVERGE}{f} = \pinfty$.
%
%
\begin{remark}[Quantitative Strongest Liberal Post and Unreachability]
	Analogously to $\minfty$ being the value of unreachability in $\spsymbol$ (cf.~\Cref{rem:sp-value-of-unreach}), $\pinfty$ is \emph{the value of unreachability} in $\slpsymbol$.\qedtriangle
\end{remark}%

\subsubsection*{\textbf{Nondeterministic Choice.}}

The \emph{demonic} strongest liberal postcondition of $\NDCHOICE{C_1}{C_2}$ is%
\begin{align*}
	\slp{\NDCHOICE{C_1}{C_2}}{\psi} \eeq \slp{C_1}{\psi} \morespace{\wedge} \slp{C_2}{\psi}~.
\end{align*}%
Indeed, $\slp{C_i}{\psi}$ contains all final states $\tau$ such that all initial states $\sigma$ that can reach $\tau$ by executing~$C_i$ satisfy $\psi$. By intersecting $\sp{C_1}{\psi}$ and $\sp{C_2}{\psi}$ we ensure the \emph{stronger} requirement that all initial states $\sigma$ that can reach $\tau$ by executing $C_1$ \emph{or} $C_2$ satisfy $\psi$.


In a quantitative setting, where we want to retrocipate the value of a quantity $f$ before executing $C_1$ or $C_2$, we \emph{demonically} minimize the possible initial value and hence take as strongest post%
\begin{align*}
	\slp{\NDCHOICE{C_1}{C_2}}{f} \eeq \slp{C_1}{f} \ccurlywedge \slp{C_2}{f}~.
\end{align*}%

\subsubsection*{\textbf{Conditional Choice}}
The \emph{demonic} strongest liberal postcondition of $\NDCHOICE{C_1}{C_2}$ is given by%
\begin{align*}
	\slp{\ITE{\guard}{C_1}{C_2}}{\psi} \eeq \slp{C_1}{\neg\guard\vee\psi} \morespace{\wedge} \slp{C_2}{\guard\vee\psi}~,
\end{align*}%
Indeed, since the disjunction can be seen as an implication, $\slp{C_1}{\neg\guard\vee\psi}$ contains all final states $\tau$ such that, all initial states that satisfy $\guard$ (sic!) and that can reach $\tau$ by executing $C_1$ do also satisfy $\psi$. Similarly, $\slp{C_2}{\guard\vee\psi}$ contains all final states $\tau$ such that, all initial states that satisfy $\neg\guard$  (sic!) and that can reach $\tau$ by executing $C_2$ do also satisfy $\psi$. 
By intersecting the postconditions $\slp{C_1}{\neg\guard\vee\psi}$ and $\slp{C_2}{\guard\vee\psi}$, we obtain exactly all those final states $\tau$ such that, all initial states that, \emph{either} satisfy $\guard$ and can reach $\tau$ by executing $C_1$, or satisfy $\neg\guard$ and can reach $\tau$ by executing $C_2$ do also satisfy the precondition~$\psi$.

Similarly for our quantitative strongest post calculi, we make use of the \emph{extended Iverson brackets} and thus, the quantitative strongest liberal post of the conditional choice is%
\begin{align*}
	\slp{\ITE{\varphi}{C_1}{C_2}}{f} \eeq  \slp{C_1}{\iverson{\neg\varphi} \curlyvee f} \ccurlywedge \slp{C_2}{ \iverson{\varphi} \curlyvee f}~.
\end{align*}%
Intuitively, $\slp{C_1}{\iverson{\neg\varphi} \curlyvee f} $ characterizes the infimum of $f$ measured in all initial states before the execution of $C_1$ satisfying $\varphi$; and analogously for $\slp{C_2}{\iverson{\varphi} \curlyvee f}$.
By taking $\curlywedge$, we obtain exactly the minimum initial quantity that $f$ could have had before executing the conditional choice.

\subsubsection*{\textbf{Looping}}
For a loop $\WHILEDO{\guard}{C}$, $\slpsymbol$ is characterized using the greatest fixed point of the so-called \emph{$\slpsymbol$--characteristic function} $\Psi_f\colon \A\to\A$. 
As for weakest liberal pre, the function is chosen so that by Kleene's fixpoint theorem, the greatest fixed point corresponds to iterating on the top element of the lattice $\pinfty$, which yields a descending chain of loop unrollings
\begin{align*}
	\iguard \curlyvee \Psi_f(\pinfty) &\eeq \slp{\mathtt{if} ({\varphi}) \{\DIVERGE\}}{f}\\
	\iguard \curlyvee\Psi_f^2(\pinfty) &\eeq \slp{\mathtt{if} ({\varphi}) \{\COMPOSE{C}{\mathtt{if} ({\varphi}) \{\DIVERGE\}}\}}{f}\\
	\iguard \curlyvee\Psi_f^3(\pinfty) &\eeq \slp{\mathtt{if} ({\varphi}) \{\COMPOSE{C}{\mathtt{if} ({\varphi}) \{\COMPOSE{C}{\mathtt{if} ({\varphi}) \{\DIVERGE\}\}}}\}}{f}
\end{align*}
and so on. Since our strongest liberal postcondition considers unreachability as “good behavior”, we join the Kleene's iterates with all the final states where the guard still hold and obtain as strongest liberal post:%
\belowdisplayskip=-.5\baselineskip%
\begin{align*}
	\slp{\WHILEDO{\guard}{C}}{f} \qeq \iguard \ccurlyvee \gfp \Psi_f~.
\end{align*}%
\normalsize%
%
%
%
%
\begin{restatable}[Soundness of \textnormal{$\slpsymbol$}]{theorem}{slpsoundness}%
	\label{thm:slp-soundness}%
	For all programs $C$ and states $\tau\in\States$,
	\begin{align*}
	\slp{C}{f}(\tau) \qeq \quad \bigcurlywedge_{\sigma \textnormal{ with } \tau\in \eval{C}{\sigma}} \quad f(\sigma)
\end{align*}
\end{restatable}
\noindent{}%
Intuitively, for a given prequantity $f$ and final state $\tau$, the $\slp{C}{f}(\tau)$ is the \emph{infimum} over all values that $f$ can assume measured in the initial states $\sigma$, so that executing $C$ on $\sigma$ terminates in $\tau$.
In case that the final state $\tau$ is \emph{unreachable}, i.e.\ $\forall \sigma\colon \tau\notin \eval{C}(\sigma)$, that infimum becomes $\pinfty$ --- the absolute maximum value.
In particular, if~\mbox{$\forall\, \sigma\colon f(\sigma) < \pinfty$}, then $\sp{C}{f}(\tau) = \pinfty$ unambiguously indicates \emph{unreachability} of $\tau$ by executing $C$ on any input $\sigma$.

\section{Healthiness Properties of Quantitative Transformers}
\label{se:healthiness}

Our quantitative transformers enjoy of several so-called \emph{healthiness properties}, some of which are analogous to Dijkstra's, Kozen's, or McIver \& Morgan's calculi.
We furthermore present several dualities between our transformers and how to embed classical into quantitative reasoning.

\subsection{Healthiness Properties}
\label{se:wpwlp-properties}

\begin{restatable}[Healthiness Properties of Quantitative Transformers]{theorem}{wpwlpspslphealthiness}%
	\label{thm:wpwlpspslphealthiness}%
	For all programs $C$, the non-liberal transformers $\wpC{C}$ and $\spC{C}$ satisfy the following properties:
	\begin{enumerate}
		\item
		\label{thm:wpwlpspslphealthiness:conjunctive} 
			Quantitative universal conjunctiveness:\quad \hspace{.4em}For any set of quantities \mbox{$S \subseteq \A$},%
			\begin{align*}
				\wp{C}{\scalebox{1.25}{$\curlyvee$} S} \eeq \scalebox{1.25}{$\curlyvee$}~ \wp{C}{S} \qand \sp{C}{\scalebox{1.25}{$\curlyvee$} S} &\eeq \scalebox{1.25}{$\curlyvee$}~ \sp{C}{S}~.
			\end{align*}%

		\item \label{thm:wpwlpspslphealthiness:strictness}
		Strictness:\quad \hspace{1.6em}$\wp{C}{\minfty} \eeq \minfty$ \qand $\sp{C}{\minfty} \eeq \minfty$
	\end{enumerate}
	The liberal transformers $\wlpC{C}$ and $\slpC{C}$ satisfy the following properties:
	\begin{enumerate}
		\setcounter{enumi}{2}
		\item \label{thm:wpwlpspslphealthiness:disjunctive}
		Quantitative universal disjunctiveness:\quad \hspace{.4em}For any set of quantities \mbox{$S \subseteq \A$},
		\begin{align*}
			\wlp{C}{\scalebox{1.25}{$\curlywedge$} S} \eeq \scalebox{1.25}{$\curlywedge$}~ \wlp{C}{S} \qand \slp{C}{\scalebox{1.25}{$\curlywedge$} S} &\eeq \scalebox{1.25}{$\curlywedge$}~ \slp{C}{S}~.
		\end{align*}
		\item \label{thm:wpwlpspslphealthiness:costrictness} 
		Costrictness:\quad \hspace{1.15em}$\wlp{C}{\pinfty} \eeq \pinfty$ \qand $\slp{C}{\pinfty} \eeq \pinfty$  
	\end{enumerate}
	All quantitive transformers are monotonic, i.e.%
	\label{thm:wpwlpspslphealthiness:mono}
	\begin{align*}
		f \ppreceq g \qqimplies \sfsymbol{ttt}\,\llbracket C \rrbracket\left(f\right) \ppreceq \sfsymbol{ttt}\,\llbracket C \rrbracket\left(g\right)~, \quad \textnormal{for } \sfsymbol{ttt} \in \{\wpsymbol,\, \wlpsymbol,\, \spsymbol,\, \slpsymbol\}~.
	\end{align*}
\end{restatable}%
\noindent{}%
Quantitative universal conjunctiveness of $\wpsymbol$/$\spsymbol$ as well as disjunctiveness of $\wlpsymbol$ are quantitative analogues to Dijkstra and Scholten's original calculi, whereas disjunctiveness of $\slpsymbol$ is novel (since $\slpsymbol$ is novel) and fits well into this picture of duality. Note that quantitative universal conjunctiveness (disjunctiveness) implies $\omega$-(co)continuity, which in turn ensures that Kleene's fixed point theorem guarantees the existence of least (greatest) fixed points for defining weakest/strongest (liberal) pre/post of loops. 
Monotonicity (implied by continuity) also ensures existence of fixed points but fixed point iteration may stabilize only at ordinals higher than $\omega$ for non-(co)continuous functions.

Strictness of $\wpsymbol$, i.e.\ $\wp{C}{\orange{\minfty}} = \blue{\minfty}$, says that the anticipated value of~$\orange{\minfty}$ after executing~$C$ is $\blue{\minfty}$ if the program terminates, and otherwise yields $\wpsymbol$'s value of nontermination: $\blue{\minfty}$.
Strictness of $\spsymbol$, i.e.\ $\sp{C}{\blue{\minfty}} = \orange{\minfty}$, says that $\orange{\minfty}$ retrocipates the value of~$\blue{\minfty}$ if the final state is reachable, and otherwise yields $\spsymbol$'s value of unreachability: $\orange{\minfty}$.
Explanations for costrictness are analogous.

The predicate interpretation of (co)strictness is also preserved: 
Since $\minfty=\indicator{\false}$ and $\pinfty = \indicator{\true}$ and hence $\wp{C}{\indicator{\false}}=\indicator{\false}$ and $\wlp{C}{\indicator{\true}}=\indicator{\true}$, strictness of quantitative $\wpC{C}$ means that $C$ cannot terminate in some $\tau \in \emptyset$;
strictness of $\spC{C}$ that no $\tau$ is reachable by executing $C$ on any $\sigma \in \emptyset$; 
costrictness of $\wlpC{C}$ that on all states $C$~either terminates or not; 
and costrictness of $\slpC{C}$ (novelly) that all states are either reachable by executing $C$ or unreachable.
%
%
%

Sub- and superlinearity have been studied by Kozen, McIver \& Morgan, and Kaminski for probabilistic $\sfsymbol{w(l)p}$ transformers.
Our transformers similarly also obey linearity.%
\begin{restatable}[Linearity]{theorem}{wpwlpspslplinear}%
	\label{thm:wpwlpspslplinear}%
	For all programs $C$, $\wpC{C}$ and $\spC{C}$ are sublinear, and $\wlpC{C}$ and $\slpC{C}$ are superlinear, i.e.\ for all \mbox{$f, g \in \A$} and \emph{non-negative} constants $r\in\PosReals$,%
	\belowdisplayskip=0pt%
	\begin{align*}
		\wp{C}{r\cdot f +  g} &\ppreceq r\cdot\wp{C}{f} + \wp{C}{g}~,\\
		\sp{C}{r\cdot f +  g} &\ppreceq r\cdot\sp{C}{f} + \sp{C}{g}~,\\[.5em]
		r\cdot\wlp{C}{f} + \wlp{C}{g} &\ppreceq \wlp{C}{r\cdot f +  g}~, \quad \textnormal{and}\\
		r\cdot\slp{C}{f} + \slp{C}{g} &\ppreceq \slp{C}{r\cdot f +  g} ~.
	\end{align*}%
	\normalsize%
\end{restatable}%
\noindent{}%
%
%

\subsection{Relationship between Qualitative and Quantitative Transformers}
\label{se:sp-predicates}
%
%
%

Our calculi subsume both the classical ones of \citet{Dijkstra1990} and our definition of strongest liberal postcondition for predicates by means of our extended Iverson brackets:%
\begin{restatable}[Embedding Classical into Quantitative Transformers]{theorem}{wpwlpspslpembeddingpredicates}%
	\label{thm:wpwlpspslpembeddingpredicates}%
	For all \underline{deterministic} programs $C$ and predicates $\psi$, we have%
	\begin{align*}
		\wp{C}{\indicator{\psi}} 
		\eeq \indicator{\wpD{C}{\psi}} \qqand
		\wlp{C}{\indicator{\psi}}  \eeq\indicator{\wlpD{C}{\psi}}~, 
		\intertext{and for \underline{all} programs $C$ and predicates $\psi$, we have}
		\sp{C}{\indicator{\psi}} 
		\eeq \indicator{\spD{C}{\psi}} \qqand
		\slp{C}{\indicator{\psi}}  \eeq\indicator{\slp{C}{\psi}}~.
	\end{align*}
\end{restatable}%
\noindent{}%
%
%
%
From a predicate perspective, $\sp{C}{\psi}$ contains final states $\tau$ that are reachable from at least one initial state satisfying $\psi$, whereas $\slp{C}{\psi}$ requires that every initial state that may end in $\tau$ satisfies $\psi$. Hence, we have a fundamentally dual meaning of the word \emph{liberal}:
\begin{itemize}
	\item $\wlpsymbol$, differently from $\wpsymbol$, provides \emph{preconditions} containing all \emph{diverging} initial states, but contains no state that can terminate outside the postcondition.
	\item $\slpsymbol$, differently from $\spsymbol$, provides \emph{postconditions} containing all \emph{unreachable} final states, but contains no state that can be reached from outside the precondition.
\end{itemize}%
Let us also consider two other examples: 
$\sp{C}{\indicator{\true}}$ is the indicator function of the reachable states. 
If $\sp{C}{\indicator{\true}}=\indicator{\false}$ (i.e.~$\sp{C}{\pinfty} = \minfty$), no state is reachable and hence $C$ diverges on every input.
Similarly, $\slp{C}{\indicator{\false}}$ is the indicator function of all states that are either reachable from an initial state satisfying $\false$ (of which there are none) or which are unreachable. 
Thus, if $\slp{C}{\indicator{\false}} = \indicator{\true}$ (i.e.~$\slp{C}{\minfty} = \pinfty$) then all states are unreachable, meaning $C$ diverges on every input.
Put shortly, 
\begin{align*}
	\sp{C}{\pinfty} \eeq \minfty \qiff \slp{C}{\minfty} \eeq \pinfty~.
\end{align*}%
Finally, we note that the \emph{quantitative weakest pre calculi} of \citet[Section 2.3]{benni_diss}, restricted to \emph{deterministic non-probabilistic programs} are even simply subsumed by the fact that we consider a larger lattice, namely quantities of type $f\colon \States \to \Rinf$ instead of $f\colon \States \to \PosRealsInf$.

\subsection{Relationship between Liberal and Non-liberal Transformers}
%
%
\begin{restatable}[Liberal--Non-liberal Duality]{theorem}{wpwlpspslpduality}
	\label{thm:wpwlpspslpduality}%
	For any program $C$ and quantity $f$, we have
	\begin{align*}
		\wp{C}{f}\eeq-\wlp{C}{-f}~ 
		\qqand 
		\sp{C}{f}\eeq-\slp{C}{-f}~.
	\end{align*}%
\end{restatable}%
\noindent{}%
The duality for weakest pre is very similar to $\wp{C}{\psi} = \neg\wlp{C}{\neg\psi}$ in Dijkstra's classical calculus and $\wp{C}{f} = 1 - \wlp{C}{1 - f}$ for 1-bounded functions $f$ in Kozen's and McIver \& Morgans development for probabilistic programs.

When considering only \emph{deterministic} programs $C$ (i.e.\ \emph{syntactically} without nondeterministic choices), then executing $C$ on initial state~$\sigma$ will either terminate in a \emph{single} final state (i.e.\ $\sem{C}{\sigma}=\{\tau\}$, for some $\tau$), or diverge (i.e.\ $\sem{C}{\sigma}=\emptyset$), meaning that $\sem{C}{\cloze{$\sigma$}}$ becomes a proper (partial) function.
Hence, in case of termination, supremum and infimum of the final values of $f$ coincide:%
\begin{corollary}
	\label{cor:wp-wlp-deterministic}
	If a deterministic program $C$ terminates on an input $\sigma$, 
	then for all quantities $f$,%
	\begin{align*}
		\wp{C}{f}(\sigma) \eeq\wlp{C}{f}(\sigma)~,
	\end{align*}%
	and otherwise \qquad
	%
		$\wp{C}{f}(\sigma) \eeq \minfty$ \qand $\wlp{C}{f}(\sigma)  \eeq \minfty$~.
	%
\end{corollary}%
%
\noindent{}%
As a direct consequence of Corollary~\ref{cor:wp-wlp-deterministic},
for postquantities everywhere smaller than $\pinfty$ (which is not restrictive since values of program variables are finite), we can precisely detect whether a given initial state has terminated or not.
\citet[Remark 2.12]{benni_diss}, in contrast, cannot easily distinguish whether a certain initial state does not terminate, or whether the anticipated value is $0$.

Note that dual results for $\spsymbol$ and $\slpsymbol$ do \emph{not} hold since even for deterministic programs the fiber of the concrete semantics is not a function: multiple initial states can terminate in a single \mbox{final state $\tau$}.
%

%
%
	%
	%
%
%
%

\section{Correctness and Incorrectness Reasoning}
\label{se:relationship}

\subsection{Galois Connections between Weakest Pre and Strongest Post}
\label{se:galois-connections}
The classical strongest postcondition is the left adjoint to the weakest liberal precondition \cite[Section 12]{Dijkstra1990}, i.e.\ the transformers $\wlpsymbol$ and $\spsymbol$ form the Galois connection%
\begin{align*}
	G \iimplies \wlp{C}{F} \qqiff \sp{C}{G} \iimplies F~, \tag{$\dagger$}
\end{align*}%
which intuitively is true because $G \implies \wlpD{C}{F}$ means that starting from $G$ the program $C$ will either diverge or terminate in a state satisfying $F$, and $\spD{C}{G}\implies F$ means that starting from $G$ any state reachable by executing $C$ satisfies $F$.

The above Galois connection is preserved in our quantitative setting; in fact, by substituting the partial order $\implies$ on predicates with the partial order $\preceq$ on $\A$ we obtain:%
\begin{restatable}[Galois Connection between $\wlpsymbol$ and $\spsymbol$]{theorem}{wlpsprelationship}%
	\label{thm:wlp-sp-relationship}%
	For all $C \in \ngcl$ and $g,f \in \A$:
	\begin{align*}
	g\ppreceq \wlp{C}{f}  \qqiff		\sp{C}{g}\ppreceq f~.
\end{align*}%
\end{restatable}%
\noindent{}%
%
%
%
As $\wlpsymbol$ is for partial correctness, \Cref{thm:wlp-sp-relationship} shows that $\spsymbol$ is also suitable for partial correctness.
One may now wonder whether there exists a strongest post transformer that is tightly related to $\wpsymbol$, and hence, to total correctness. 
Unfortunately,~\citet[Section 12]{Dijkstra1990} show that there cannot exist a predicate transformer $\stpsymbolD$ --- a \emph{\enquote{strongest total postcondition}} --- such that%
\begin{align*}
	G \iimplies \wpD{C}{F} \qqiff \stpD{C}{G} \iimplies F~.
\end{align*}%
Categorically, that negative result is a consequence of the fact that we are requiring $\wpsymbol$ to be a \emph{right adjoint functor}, and a necessary condition for that is to preserve all infima, but this is not true since $\wpsymbol$ is not costrict. 
Despite this negative result, since $\wpsymbol$ preserves all suprema (cf.~\Cref{thm:wpwlpspslphealthiness}~(\ref{thm:wpwlpspslphealthiness:conjunctive})), we argue that $\wpsymbol$ is instead a \emph{left adjoint functor} and show that its right adjoint is exactly $\slpsymbol$:%
%
%
\begin{restatable}[Galois Connection between \textnormal{$\wpsymbol$} and \textnormal{$\slpsymbol$}]{theorem}{wpslprelationship}%
	\label{thm:wp-slp-relationship}%
	For all $C \in \ngcl$ and $g,f \in \A$:
	\begin{align*}
		\wp{C}{f} \ppreceq g \qqiff f\ppreceq \slp{C}{g} 
	\end{align*}%
\end{restatable}%
\noindent{}%
Let us provide an intuition on this connection, for simplicity only with \enquote{predicates} $\indicator{F}$ and $\indicator{G}$:
$\iverson{F}\preceq \slp{C}{\iverson{G}}$ means that every final state satisfying $F$ is either reached only by states satisfying $G$ or unreachable. 
This is equivalent to saying that all initial states terminating in $F$ must satisfy $G$, which is precisely expressed by $\wp{C}{\iverson{F}} \preceq \iverson{G}$.

\subsection{Resolving Nondeterministic Choice: Angelic vs.~Demonic}

Our choices of how to resolve nondeterminism are motivated by establishing dualities between weakest pre and strongest post presented in Section~\ref{se:galois-connections}. 
The only thing we take for granted is that the standard definition of $\spsymbol$ is angelic, thus characterizing the \enquote{set of reachable states}.
Indeed, if $\spsymbol$ is angelic, then we are (provably) also forced to make $\wpsymbol$ angelic, and both $\wlpsymbol$ and $\slpsymbol$ demonic -- otherwise, duality would break.
We can also come up with an intuition for these choices:
Both, angelic $\wpsymbol$ and demonic $\wlpsymbol$ transformers try to \emph{avoid nontermination}, if at all possible, whereas angelic $\spsymbol$ and demonic $\slpsymbol$ try to \emph{avoid unreachability}.

By dualizing all resolutions of nondeterminism one would obtain the following intuition: 
Demonic $\wpsymbol$ and angelic $\wlpsymbol$ transformers try to \emph{drive the execution towards nontermination} (more standard for both $\wpsymbol$ and $\wlpsymbol$), whereas demonic $\spsymbol$ and angelic $\slpsymbol$ try to \emph{establish unreachability} (less standard for $\spsymbol$, whereas $\slpsymbol$ is novel anyway).
We leave it as future work to study whether this dual situation would also preserve the Galois connections of Section~\ref{se:galois-connections}.

\subsection{Strongest Post and Incorrectness Logic}

A Hoare triple $\hoare{G}{C}{F}$ is \emph{valid for partial correctness} iff $G \implies \wlp{C}{F}$ or (equivalently, see~($\dagger$) in \Cref{se:galois-connections}) $\sp{C}{G} \implies F$ holds.
Somewhat recently, a different kind of triples have been proposed, first by
\citet{VK11} under the name \emph{reverse Hoare logic} for studying reachability specifications.
A few years, \citet{OHearn19} rediscovered those triples under the name \emph{incorrectness logic} and used them for explicit error handling.
\citet{Bruni2021ALF} provide a logic parametrized by an abstract interpretation that, through a notion of local completeness, can prove both correctness and incorrectness.

In this section we 
show, first, the relationship between our strongest post transformer and incorrectness triples~\cite{VK11,OHearn19}; then, more importantly, we argue that such triples deal with \emph{total incorrectness} and hint at novel \emph{partial incorrectness} triples.

\subsubsection*{\textbf{(Total) Incorrectness}}
In the sense of~\citet{VK11}, an \emph{incorrectness triple}%
\begin{align*}
	\incorrectness{G}{C}{F} \textnormal{ is valid} 
	\qqiff
	\forall\, \tau \models F \quad\exists\,\sigma \textnormal{ with } \tau\in\sem{C}{\sigma} \colon\quad \sigma\mmodels G~.
\end{align*}%
In other words, the set of states $F$ is an underapproximation of the set of states reachable by executing $C$ on some state in $G$, i.e., $F\subseteq \sp{C}{G}$~\cite[Definition 1]{OHearn19}.
The term \emph{incorrectness logic} originates from the fact that if $\incorrectness{G}{C}{F}$ is valid and $F$ contains an error state, then this \emph{error state is guaranteed to be reachable from $G$}. 
Since our quantitative strongest post transformer subsumes the classical one, we can (re)define incorrectness triples by substituting predicates with extended Iverson brackets and obtain the following equivalent definition:%
%
%
%
%
%
%
\begin{definition}[Incorrectness Triples]
	\label{def:total-incorrectness}
	For predicates $G,F$ and program $C$, the \emph{incorrectness triple}
	\begin{align*}
		\incorrectness{G}{C}{F} \textnormal{ is \emph{valid for (total) incorrectness}} \qqiff \indicator{F}\ppreceq \sp{C}{\indicator{G}}~. \tag*{\qedtriangle}
	\end{align*}
\end{definition}

\subsubsection*{\textbf{Partial Incorrectness}}
We argue that the aforementioned triples deal with \emph{total incorrectness} by providing novel triples for \emph{partial incorrectness}. 
Recall that a Hoare triple $\hoare{G}{C}{F}$ is valid for total correctness if $G\implies \wp{C}{F}$.
By replacing $\wpsymbol$ with $\wlpsymbol$, we can define partial correctness triples: 
$\hoare{G}{C}{F}$ is valid for partial correctness if $G\implies \wlp{C}{F}$. 
By mimicking the above, we define \emph{partial incorrectness} by replacing $\spsymbol$ with $\slpsymbol$ in Definition~\ref{def:total-incorrectness}:%
\begin{definition}[Partial Incorrectness]
	\label{def:partial-incorrectness}
	For predicates $G,F$ and program $C$, the \emph{incorrectness triple}
	\begin{align*}
		\incorrectness{G}{C}{F} \textnormal{ is \emph{valid for partial incorrectness}} \qqiff \indicator{F}\ppreceq \slp{C}{\indicator{G}}~. \tag*{\qedtriangle}
	\end{align*}
\end{definition}%
\noindent{}%
By definition of $\slpsymbol$,%
\begin{align*}
	\incorrectness{G}{C}{F} \textnormal{ is val.~for part.~incorr.} 
	\qqiff
	\forall\, \tau \models F \quad \forall\,\sigma \textnormal{ with } \tau\in\sem{C}{\sigma} \colon\quad \sigma\mmodels G~.
\end{align*}%
%
In other words, only if the state $\tau$ is \emph{reachable}, then the triple guarantees that $\tau$ is reached only from initial states $\sigma$ that satisfy $G$. 
Note that this is dual to the relationship between total and partial correctness: 
with partial incorrectness, to have \emph{full} information on \emph{initial states} we require an additional \emph{proof of reachability} on \emph{final states} (whereas with partial correctness, to obtain full information on \emph{final states} we require an additional \emph{proof of termination} on \emph{initial states}).

We also note that, due to the Galois between $\wpsymbol$ and $\slpsymbol$ (Theorem~\ref{thm:wp-slp-relationship}) we have
\begin{align*}
		\incorrectness{G}{C}{F} \text{ is valid for \emph{partial incorrectness}} \qqiff \wp{C}{\indicator{F}}\ppreceq \indicator{G}~.
\end{align*}
	This implies that $G$ is an overapproximation of the set of states that end up in $F$, and corresponds to the notion of \emph{necessary preconditions} studied by~\citet{Cousot13}. In particular, if an initial state $\sigma \not\models G$, then $\sigma$ is guaranteed to not terminate in $F$ ($\sigma$ could also diverge).

\subsubsection*{\textbf{Other Triples}}

\begin{wrapfigure}[10]{r}{0.55\textwidth}
  \begin{minipage}{.99\linewidth}%
  \vspace*{-1.45em}%
  \begin{center}%
    \footnotesize%
    \renewcommand{\arraystretch}{1.2}
    \begin{tabular}{rcl@{\qquad}l}
    	\multicolumn{3}{c}{\hspace{-1.1em}\textbf{implication}}& \textbf{defines}\\ \midrule
	$G$ 				& $\implies$ 			& $\wp{C}{F}$ 			& total correctness \\
	\blue{$G$} 		& \blue{$\implies$} 		& \blue{$\wlp{C}{F}$} 		& \blue{partial correctness} \\
	\orange{$\wp{C}{F}$} 	& \orange{$\implies$} 	& \orange{$G$} 			& \orange{partial incorrectness} \\
	$\wlp{C}{F}$ 		& $\implies$ 			& $G$ 				& ??? \\
	$F$ 				& $\implies$ 			& $\sp{C}{G}$ 			& (total) incorrectness \\
	\orange{$F$} 		& \orange{$\implies$} 	& \orange{$\slp{C}{G}$} 	& \orange{partial incorrectness} \\
	\blue{$\sp{C}{G}$} 	& \blue{$\implies$} 		& \blue{$F$} 			& \blue{partial correctness} \\
	$\slp{C}{G}$ 		& $\implies$ 			& $F$ 				& \raisebox{.5ex}{\textquestiondown\textquestiondown\textquestiondown}
\end{tabular}
\end{center}
  \end{minipage}%
\end{wrapfigure}%
We note that the naming conventions \emph{correctness} and \emph{incorrectness} may not necessarily always be appropriate. 
First of all, we argue that incorrectness triples~\cite{VK11,OHearn19} can be used to prove \emph{good behavior}: 
for instance, a triple $\incorrectness{G}{C}{F}$ where $F$ contains good states, ensures that every (\emph{good}) state in $F$ is reachable from precondition $G$. 
Rather than \emph{correctness} versus \emph{incorrectness}, we believe that the fundamental difference between the triples is that correctness triples provide information on the behavior of \emph{initial} states satisfying \emph{preconditions}, whereas incorrectness triples guarantee reachability properties on \emph{final} states satisfying \emph{postconditions}. 

Secondly, note that our transformers can define two additional triples other than total (partial) correctness (incorrectness), for which the current naming conventions are insufficient.
So far, we have the picture depicted in the table above.
The two blue and the two orange lines define the same notion due to the Galois connections between $\wlpsymbol$/$\spsymbol$ and $\wpsymbol$/$\slpsymbol$.
For ??? and \raisebox{.5ex}{\textquestiondown\textquestiondown\textquestiondown}, however, there are no appropriate names (let alone program logics) yet.
We can say, however, that ??? gives rise to a notion of \emph{necessary liberal preconditions}, in the sense that (1) $G$ contains all initial states $\sigma$ that \emph{diverge}, and (2) whenever $\sigma \not\models G$, then $\sigma$ is guaranteed to \emph{terminate} in \mbox{a state $\tau\not\models F$}. 
\raisebox{.5ex}{\textquestiondown\textquestiondown\textquestiondown}, on the other hand, provides \emph{necessary liberal postconditions}, meaning that (1) $F$ contains all \emph{unreachable} states, and every final state $\tau \not\models F$ is guaranteed to be \emph{reachable} from some initial state $\sigma\not\models G$. 

Following the terminology from above, which is inspired from the naming \emph{necessary preconditions} of~\citet{Cousot13}, we can state that%
\begin{itemize}
	\item
		total correctness triples provide \emph{sufficient preconditions};
	\item
		total incorrectness triples provide \emph{sufficient postconditions};
	\item
		partial correctness triples provide \emph{sufficient liberal preconditions} (or \emph{necessary postconditions});
	\item
		partial incorrectness triples provide \emph{sufficient liberal postconditions} (or \emph{necessary preconditions}).
\end{itemize}%
We also note that even the terminology for the predicate transformers, \emph{strongest} post- and \emph{weakest} precondition, might be imprecise. Indeed, as pointed by~\citet{OHearn19}, such terminology is tied with the classical aim of Hoare logic to find either the smallest (strongest) set of necessary (overapproximating) postconditions or the largest (weakest) set of sufficient (underapproximating) preconditions. 
The strongest postcondition can be seen also as the \emph{weakest sufficient postcondition}, whereas the weakest precondition is the \emph{strongest necessary precondition}. 
Switching to our liberal predicate transformers, our strongest liberal post computes the \emph{strongest necessary liberal postcondition} or, equivalently, the \emph{weakest sufficient liberal postcondition}. 
Finally, our weakest liberal pre computes the \emph{weakest sufficient liberal precondition} or the \emph{strongest necessary liberal precondition}.

\subsubsection*{\textbf{Duality}}
As a consequence of the liberal--non-liberal duality of Theorem~\ref{thm:wpwlpspslpduality}, we have
\begin{align*}
	G \iimplies\wp{C}{F} \qqiff  \wlp{C}{\neg{F}} \iimplies \neg{G}~.
\end{align*}
In other words, the triples connected to ??? are the contrapositive of total correctness triples. 
Similarly, {\textquestiondown\textquestiondown\textquestiondown} is the contrapositive of total incorrectness, whereas partial incorrectness is the contrapositive of partial correctness. 
This implies (interestingly) that only three kind of triples fundamentally cannot be stated in terms of other triples. 
Nevertheless, we would argue that it is still useful to work with, e.g.\ ??? triples, depending on the verification aim, especially in the context of \emph{explainable verification}: 
For example, if one is interested in inferring necessary preconditions, it would certainly appear easier and more natural to work and think directly with partial incorrectness, instead of complementing both the \emph{sufficient liberal preconditions} obtained via partial correctness and the original postcondition. The resulting proof and annotations, directly in terms of \emph{necessary preconditions}, will be much easier to understand for a working programmer.

\section{Loops rules}
\label{se:loops}

\begin{restatable}[Induction Rules for Loops]{theorem}{quantitativeinductionrules}
	\label{thm:quantitativeinductionrules}
	The following proof rules for loops are valid:
	\allowdisplaybreaks%
	\begin{align*}
		&\infer[\mathrm{while-}\wlpsymbol]{
			g \qpreceq \wlp{\WHILEDO{\guard}{C}}{f}
		}{
			g \qpreceq i \qpreceq \iverson{\neg \guard} \curlywedge f \ccurlyvee \iverson{\guard} \curlywedge \wlp{C}{i}
		}\\[1em]
		&\infer[\mathrm{while-}\spsymbol]{
			\sp{\WHILEDO{\guard}{C}}{g} \qpreceq f
		}{
			g \ccurlyvee \sp{C}{\iverson{\guard} \curlywedge i} \qpreceq i 
			\qqand 
			\inguard\curlywedge i \qpreceq f}\\[1em]
%
		&\infer[\mathrm{while-}\wpsymbol]{
			\wp{\WHILEDO{\guard}{C}}{f}\ppreceq g
		}{
			\iverson{\neg \guard} \curlywedge f \ccurlyvee \iverson{\guard} \curlywedge \wp{C}{i} \qpreceq i \qpreceq g}\\[1em]
		&\infer[\mathrm{while-}\slpsymbol]{
            f \qpreceq \slp{\WHILEDO{\guard}{C}}{g}
            }
		{i \qpreceq g \ccurlywedge \slp{C}{\inguard \curlyvee i} \qqand f \qpreceq \iguard\curlyvee i}
	\end{align*}%
	\allowdisplaybreaks[0]%
\end{restatable}%
\noindent{}%
The rule $\mathrm{while-}\spsymbol$ is novel.
The $\mathrm{while-}\wlpsymbol$ rule has already been investigated in~\cite[Section 5]{benni_diss} in a probabilistic setting, but in a more restricted lattice where quantities map to the unit interval.
Our definition of $\wlpsymbol$ is not probabilistic but for a more general lattice of unbounded signed quantities. 
Notice that $\mathrm{while-}\wlpsymbol$ and $\mathrm{while-}\spsymbol$ are tightly connected by a Galois connection (cf.~\Cref{thm:wlp-sp-relationship}), and by taking $g = \iverson{G}$ and $f = \iverson{F}$ for predicates $G, F$, we conclude for both rules the validity of the Hoare triple $\hoare{G}{\WHILEDO{\guard}{C}}{F}$ for partial correctness. Indeed, as standard in literature, the rule $\mathrm{while-}\wlpsymbol$ requires to find an invariant that satisfy two conditions:%
\begin{enumerate}
	\item $\iverson{G}\preceq \iverson{I}$, meaning that whenever precondition $G$ holds, then the invariant $I$ also holds.
	\item $\iverson{I} \preceq\inguard \curlywedge \iverson{F} \curlyvee \iverson{\guard} \curlywedge \wlp{C}{\iverson{I}}$, meaning that whenever $I$ holds, either the loop guard $\varphi$ does \emph{not} hold, but then postcondition $F$ holds; or $\guard$ \emph{does} hold, but then $I$ still holds after one iteration of the loop body (or the loop body itself diverges (think: nested loops)).
\end{enumerate}%
By induction, (2) ensures that, starting from $I$ and no matter how many loop iterations are executed, $I$ can only terminate in states that again satisfy $I$. 
Assuming termination, eventually $\neg\guard$ will hold and thus $I$ implies the postcondition $F$. 
(1) guarantees that the initial precondition $G$ implies $I$. 
Hence any state initially satisfying $G$ and on which the loop eventually terminates will do so in a final state satisfying postcondition $F$. 
The rule $\mathrm{while-}\spsymbol$ is analogous, but for forward reasoning.

The rule $\mathrm{while-}\wpsymbol$ has also been investigated by \citet{benni_diss} in a probabilistic setting but again in a more restricted lattice where quantities map to unsigned positive extended reals.
The rule $\mathrm{while-}\slpsymbol$ is completely novel (since $\slpsymbol$ is novel).
Again, by Galois connection and by taking as quantities the Iverson bracket of predicates $G, F$, we obtain for the last two rules as conclusion the validity of the triple $\incorrectness{G}{\WHILEDO{\guard}{C}}{F}$ for \emph{partial incorrectness} in the sense of Definition~\ref{def:partial-incorrectness}.
As for an intuition, recall that validity for partial incorrectness means here that $G$ is a \emph{necessary precondition} to end in a final state satisfying $F$ after termination of $\WHILEDO{\guard}{C}$. 
For proving this, the rule $\mathrm{while-}\wpsymbol$ requires to find an invariant $I$, such that:%
\begin{enumerate}
	\item $\iverson{I} \preceq \iverson{G}$, meaning that whenever invariant $I$ holds, then the precondition $G$ also holds.
	\item $\inguard \curlywedge \iverson{F} \preceq \iverson{I}$, meaning that if the loop has terminated in postcondition $F$, then $I$ holds;
	\item $\iverson{\guard} \curlywedge \wp{C}{\iverson{I}}\preceq \iverson{I}$, meaning that if the loop is in some state $\sigma$ in which the loop guard holds (i.e.\ the loop is about to be executed once more) and one loop iteration will terminate in some state where $I$ holds again, then $I$ holds for $\sigma$.
\end{enumerate}%
By induction, (2) and (3), which represent the first premise of $\mathrm{while-}\wpsymbol$, imply that $I$ is a necessary precondition for the loop to terminate in $F$. 
Indeed, starting from the base case (2), for the inductive step we assume that $I$ overapproximates those states terminating in $F$ after $n$ loop iterations. 
By (3), $I$ also contains $\iverson{\guard} \curlywedge \wp{C}{\iverson{I}}$, i.e., an overapproximation of those states terminating in $F$ after $n+1$ iterations. 
(1) guarantees that the precondition $G$ contains $I$ and hence $G$ is a necessary precondition for the loop to terminate in $F$. 
Again, the rule $\mathrm{while-}\slpsymbol$ is analogous, but forward.%
\begin{example}[Inductive Reasoning]
	\renewcommand{\mod}{\mathbin{{|}}}
	Consider the loop $\WHILEDO{x<10}{\ASSIGN{x}{x+4}}$. In order to show that $x \mod 4$ (read: $x$ is divisible by 4) is a \emph{necessary precondition} to terminate in postcondition $x=12$, it is sufficient to prove the partial incorrectness triple $\iverson{x=12}\preceq\slp{C}{\iverson{x \mod 4}}$. If we apply the inductive rule we obtain:
	\begin{align*}
		\infer[\mathrm{while-}\slpsymbol]{
            \iverson{x=12}\qpreceq \slp{\WHILEDO{x<10}{\ASSIGN{x}{x+4}}}{\iverson{x \mod 4}}
            }
		{i \ppreceq \iverson{x \mod 4}\curlywedge \slp{\ASSIGN{x}{x+4}}{\iverson{x\geq 10} \curlyvee i} \qqand \iverson{x=12} \ppreceq \iverson{x<10}\curlyvee i}
	\end{align*}%
	Now take as invariant $i=\iverson{x \mod 4}$. 
	As for the right premise, we can easily convince ourselves that $\iverson{x=12} \preceq \iverson{x<10}\curlyvee \iverson{x \mod 4}$ holds. 
	As for the left premise, we have%
	\begin{align*}
		\iverson{x \mod 4} \ccurlywedge \slp{\ASSIGN{x}{x+4}}{\iverson{x\geq 10} \curlyvee \iverson{x \mod 4}}
		\qeq &\iverson{x \mod 4} \ccurlywedge \bigl(\iverson{x - 4 \geq 10} \curlyvee \iverson{x - 4 \mod 4} \bigr)\\
		\qeq &\iverson{x \mod 4} \ccurlywedge \bigl(\iverson{x \geq 14} \curlyvee \iverson{x \mod 4} \bigr)\\
		\qeq &\iverson{x \mod 4} \qsucceq \iverson{x \mod 4} \qeq i~.
	\end{align*}
	Hence we can infer the conclusion of while--$\slpsymbol$ and we have proven that $\iverson{x \mod 4}$ is a necessary precondition for the loop to terminate in $\iverson{x = 12}$.\qedtriangle
\end{example}%
\noindent{}%
The forward transformers $\spsymbol$ and $\slpsymbol$ come with an additional induction rule: 
under certain premises, it allows to immediately conclude that the fixpoint of the characteristic function for a quantity $f$ is \emph{precisely} $f$ itself, i.e. the \emph{second} Kleene iterate.%
\begin{restatable}[]{proposition}{spslpconvergence}
	\label{prop:spslpconvergence}
	The following proof rules for loops are valid:%
	\begin{align*}
		\infer[]{
			\sp{\WHILEDO{\guard}{C}}{f} \qeq \iverson{\neg \guard} \curlywedge f
		}{
			\sp{C}{f}\qpreceq f
		}
		\qquad\quad
		\infer[]{
			\slp{\WHILEDO{\guard}{C}}{f} \qeq \iverson{\guard} \curlyvee f
		}{
			f\qpreceq \slp{C}{f}
		}
	\end{align*}
\end{restatable}%
\noindent{}%
An intuition of Proposition~\ref{prop:spslpconvergence} for $\spsymbol$ is the following: for a loop $\WHILEDO{\guard}{C}$, the premise $\sp{C}{f}\preceq f$ means that the value of $f$ \emph{retrocipated for one iteration} is lower than the original value of $f$. 
By induction, retrocipating $f$ for any number of iterations leads to a decreasing quantity. So what is the \emph{maximum initial value} that $f$ could have had? It is the initial quantity $f$, i.e.\ $\spsymbol$ \enquote{gets away} with not even entering the loop. 
The guard $\inguard$ in the conclusion is needed to ensure reachability. For $\slpsymbol$, retrocipating the execution of the loop increases the initial quantity $f$ - and hence the \emph{minimum initial value} of $f$ is again $f$ itself.%
\begin{example}
	Consider the loop $C = \WHILEDO{x<10}{\NDCHOICE{\ASSIGN{x}{x+1}}{\ASSIGN{x}{x+2}}}$ and the precondition $x\geq 0$. 
	To determine the set of states reachable from precondition $x\geq 0$, i.e.\ to determine $\sp{C}{\iverson{x\geq 0}}$, we first check the premise%
	\begin{align*}
		&\sp{\NDCHOICE{\ASSIGN{x}{x+1}}{\ASSIGN{x}{x+2}}}{\iverson{x\geq 0}} \\
		&\qeq  \iverson{x-1\geq 0}\curlyvee\iverson{x-2\geq 0}
		\qeq \iverson{x\geq 1}\curlyvee\iverson{x\geq 2} 
		\qeq \iverson{x \geq 1 }
		\qpreceq \iverson{x\geq0}
	\end{align*}%
	and thus conclude by \Cref{prop:spslpconvergence} that
	\begin{align*}
		\sp{C}{\iverson{x\geq 0}} \qpreceq \iverson{x\geq 10}\curlywedge \iverson{x\geq 0} \qeq \iverson{x\geq 10}
	\end{align*}
	This allows to include immediately that $x\geq 10$ is the \emph{strongest necessary postcondition} or, equivalently, the \emph{weakest sufficient postcondition}. 
	In particular, this result verifies that \emph{precisely} those final states with $x\geq 10$ are \emph{reachable} from initial states with $x\geq 0$. \qedtriangle
\end{example}

\section{Case Studies}%
\label{se:examples}

\begin{wrapfigure}[4]{r}{0.15\textwidth}
  \begin{minipage}{.99\linewidth}
  \vspace*{-2.1em}%
    \footnotesize%
    \abovedisplayskip=0pt%
    \belowdisplayskip=0pt%
\begin{align*}
	&\annotate{f}\\
	&C\\
	&\annotate{g}\\
	&\eqannotate{g'}
\end{align*}%
\normalsize%
  \end{minipage}%
\end{wrapfigure}%
In this section, we demonstrate the efficacy of quantitative strongest (liberal) post reasoning. 
We use the annotation style on the right
to express that $g = \sp{C}{f}$ (or that $g = \slp{C}{f}$, depending on the context) and furthermore that~$g' = g$. 
Full calculations of strongest posts are provided in Appendix~\ref{app:examples}.
\subsection{Quantitative Information Flow --- Loop Free}
\label{ex:1}

Consider the program $C_\mathit{flow} = \ITE{\mathit{hi}>7}{\ASSIGN{\mathit{lo}}{99}}{\ASSIGN{\mathit{lo}}{80}}$. 
	As usual in quantitative information flow, $hi$ is a secret and we want to ensure that, by observing the variable $\mathit{lo}$, one cannot infer information about $hi$.
	Below, we show $\spsymbol$ (left) and $\slpsymbol$ (right) annotations for prequantity $hi$, i.e.\ we indeed show how the initial value of $hi$ flows from the top to the bottom of the computation.%
\begin{center}
	\scriptsize
	\begin{minipage}{.48\textwidth}%
		\abovedisplayskip=0pt%
		\begin{align*}
			& \annotate{\mathit{hi}} \\
			& \IF{ \mathit{hi} > 7} \\
			& \qquad \annotate{\iverson{\mathit{hi} > 7} \ccurlywedge \mathit{hi}} \\
			& \qquad \ASSIGN{\mathit{lo}}{99} \\
			& \qquad \annotate{\SupV{\alpha}~ \iverson{\mathit{lo} = 99} \ccurlywedge \iverson{\mathit{hi} > 7} \ccurlywedge \mathit{hi}} \\
			& \qquad \eqannotate{\iverson{\mathit{lo} = 99} \ccurlywedge \iverson{\mathit{hi} > 7} \ccurlywedge \mathit{hi}} \\
			& \ELSE \\
			& \qquad \annotate{\iverson{\mathit{hi} \leq 7} \ccurlywedge \mathit{hi}} \\
			& \qquad \ASSIGN{\mathit{lo}}{80} \\
			& \qquad \annotate{\SupV{\alpha}~ \iverson{\mathit{lo} = 80} \ccurlywedge \iverson{\mathit{hi} \leq 7} \ccurlywedge \mathit{hi}} \\
			& \qquad \eqannotate{\iverson{\mathit{lo} = 80} \ccurlywedge \iverson{\mathit{hi} \leq 7} \ccurlywedge \mathit{hi}} \\
			& \} \\
			& \annotate{\bigl(\iverson{\mathit{lo} = 99} \curlywedge \iverson{\mathit{hi} > 7} \curlywedge \mathit{hi} \bigr)\curlyvee \bigl( \iverson{\mathit{lo} = 80} \curlywedge \iverson{\mathit{hi} \leq 7} \curlywedge \mathit{hi} \bigr)}
		\end{align*}%
		\normalsize%
	\end{minipage}%
	\quad
	\begin{minipage}{.48\textwidth}%
		\abovedisplayskip=0pt%
		\begin{align*}
			& \annotate{\mathit{hi}} \\
			& \IF{ \mathit{hi} > 7} \\
			& \qquad \annotate{\iverson{\mathit{hi} \leq 7} \ccurlyvee \mathit{hi}} \\
			& \qquad \ASSIGN{\mathit{lo}}{99} \\
			& \qquad \annotate{\InfV{\alpha}~ \iverson{\mathit{lo} \neq 99} \ccurlyvee \iverson{\mathit{hi} \leq 7} \ccurlyvee \mathit{hi}} \\
			& \qquad \eqannotate{\iverson{\mathit{lo} \neq 99} \ccurlyvee \iverson{\mathit{hi} > 7} \ccurlyvee \mathit{hi}} \\
			& \ELSE \\
			& \qquad \annotate{\iverson{\mathit{hi} > 7} \ccurlyvee \mathit{hi}} \\
			& \qquad \ASSIGN{\mathit{lo}}{80} \\
			& \qquad \annotate{\InfV{\alpha}~ \iverson{\mathit{lo} \neq 80} \ccurlyvee \iverson{\mathit{hi} > 7} \ccurlyvee \mathit{hi}} \\
			& \qquad \eqannotate{\iverson{\mathit{lo} \neq 80} \ccurlyvee \iverson{\mathit{hi} > 7} \ccurlyvee \mathit{hi}} \\
			& \} \\
			& \annotate{\bigl(\iverson{\mathit{lo} \neq 99} \curlyvee \iverson{\mathit{hi} \leq 7} \curlyvee \mathit{hi} \bigr)\curlywedge \bigl( \iverson{\mathit{lo} \neq 80} \curlyvee \iverson{\mathit{hi} > 7} \curlyvee \mathit{hi} \bigr)}
		\end{align*}%
		\normalsize%
	\end{minipage}%
\end{center}%
Let us first note that we can precisely infer the set of states that are reachable after executing $C_\mathit{flow}$ by recalling that for a prequantity $f$ strictly larger than $\minfty$, $\sp{C}{f}(\tau) = \minfty$ if and only if $\tau$ is unreachable.
When does the (left) expression $\bigl(\iverson{\mathit{lo} = 99} \curlywedge \iverson{hi > 7} \curlywedge hi \bigr) \ccurlyvee \bigl( \iverson{\mathit{lo} = 80} \curlywedge \iverson{hi \leq 7} \curlywedge hi \bigr)$ evaluate to something larger than $\minfty$?
This is precisely the case if either the final value of $\mathit{lo}$ is $99$ and $hi$ is larger than $7$, or if $\mathit{lo}$ is $80$ and $hi$ smaller or equal $7$.
The reachable states are thus given by%
\begin{align*}
	\{\tau\mid\sp{C}{\mathit{hi}}(\tau)\neq \minfty\}\eeq \{\tau\mid \tau(\mathit{lo})=99 \wedge \tau(\mathit{hi})> 7 \vvee \tau(\mathit{lo})=80 \wedge \tau(\mathit{hi})\leq 7\}~.
\end{align*}%
The same insight could have been achieved with $\slpsymbol$ by computing $\{\tau\colon\slp{C}{\mathit{hi}}(\tau)\neq \pinfty\}$.
%

Secondly, we can --- in a principled way --- construct from the $\spsymbol$ and $\slpsymbol$ annotations a function $\xi$ that, given the final value of only the observable variable $\mathit{lo}$ (which we denote $\mathit{lo}'$), returns the set containing an overapproximation of all possible \emph{initial} values of the quantity $\mathit{hi}$, namely:
\begin{align*}
	\xi(\mathit{lo}') \eeq & \setcomp{ \alpha }{ \tau \in \States,\quad \tau(\mathit{lo})=lo',\quad \slp{C_\mathit{flow}}{\mathit{hi}}(\tau)\lleq  \alpha \lleq\sp{C_\mathit{flow}}{\mathit{hi}}(\tau) }\\
	 \eeq &\begin{cases}
		\setcomp{\alpha }{ 7 < \alpha }, & \text{ if } lo'=99\\
		\setcomp{\alpha }{ \alpha \leq 7 }, & \text{ if } lo'=80\\
\emptyset,  & \text{ otherwise.}
	 \end{cases}
\end{align*}
Now, what can we infer about the secret initial value of $\mathit{hi}$ by observing only the final value $\mathit{lo}'$? If $lo'=99$, then $\mathit{hi}$ must be larger than $7$; if $\mathit{lo} = 90$, then $\mathit{hi}$ must be smaller or equal $7$, and otherwise this state was actually unreachable (and hence such a situation could have not been observed in the first place).
Hence, observing the final value of $\mathit{lo}$ leaks information about the secret $\mathit{hi}$.
In fact, by having used both $\spsymbol$ and $\slpsymbol$, the above gave us \emph{precisely} the entire information that is leaked about $\mathit{hi}$ from observing the final value of $\mathit{lo}$.
\subsection{Quantitative Information Flow for Loops}
\label{ex:2}
Consider the program $C_\mathit{while} = \COMPOSE{\ASSIGN{\mathit{hi}}{\mathit{hi}+5}}{\WHILEDO{\mathit{lo}<\mathit{hi}}{\ASSIGN{\mathit{lo}}{lo+1}}}$. Again, we show below the $\spsymbol$ (left) and $\slpsymbol$ (right) annotations for prequantity $\mathit{hi}$.

\begin{center}
	\scriptsize
	\begin{minipage}{.45\textwidth}%
		\begin{align*}
			& \annotate{\mathit{hi}}\\
			& \ASSIGN{\mathit{hi}}{\mathit{hi}+5}\\
			& \annotate{\mathit{hi}-5}\\ 
			& \WHILE{\mathit{lo}<\mathit{hi}}\\
			& \qquad\ASSIGN{\mathit{lo}}{lo+1} \quad \}\\
			& \annotate{\iverson{lo\geq \mathit{hi}}\curlywedge(\mathit{hi}-5)}
		\end{align*}%
	\end{minipage}%
	\qquad%
	\begin{minipage}{.45\textwidth}%
		\begin{align*}
			& \annotate{\mathit{hi}}\\
			& \ASSIGN{\mathit{hi}}{\mathit{hi}+5}\\
			& \annotate{\mathit{hi}-5}\\ 
			& \WHILE{\mathit{lo}<\mathit{hi}}\\
			& \qquad\ASSIGN{\mathit{lo}}{lo+1}\quad \}\\
			& \annotate{\iverson{\mathit{lo}< \mathit{hi}}\curlyvee (\mathit{hi}-5)}
		\end{align*}%
	\end{minipage}%
\end{center}%
%
For $\spsymbol$ and $\slpsymbol$ of the loop, the Kleene iteration stabilizes after 2 iterations, see Appendix~\ref{app:examples} for detailed computations.
There is no need for invariant, nor reasoning about limits, or anything alike.
Even more conveniently, we can alternatively apply \Cref{prop:spslpconvergence}: indeed, for instance for $\spsymbol$ we have $\sp{\ASSIGN{\mathit{lo}}{lo+1}}{\mathit{hi}-5}=\mathit{hi}-5\preceq \mathit{hi}-5$ and thus \Cref{prop:spslpconvergence} yields that $\spsymbol$ of the loop is \emph{precisely} $\iverson{lo\geq \mathit{hi}}\curlywedge(\mathit{hi}-5)$.

We construct (again) the function $\xi$ that, given the final value $\mathit{lo}'$ of the variable $\mathit{lo}$, returns an overapproximation of all possible \emph{initial} values of the quantity $\mathit{hi}$, 
and obtain $\xi(\mathit{lo}') \eeq \setcomp{\alpha}{\alpha\leq lo'-5}$.
Hence, by observing only the final value $\mathit{lo}'$ we infer that $\mathit{hi}$ must be at most $lo'-5$. In fact, any of such value $\alpha\leq lo'-5$ after being incremented by $5$ leads to a value that $\alpha'\leq lo'$, so without entering the loop, $C_\mathit{while}$ terminates with the correct final value $\mathit{lo}'$. Again, using both $\spsymbol$ and $\slpsymbol$, we obtain \emph{precisely} the entire information that is leaked about $\mathit{hi}$ from observing the final value of $\mathit{lo}$.

\subsubsection*{\textbf{Quantitative Information Flow for Loops using $\wpsymbol$.}}
\label{se:example-wp}
The set $\xi(\mathit{lo}')$ could have alternatively been determined with classical weakest preconditions: 
In fact, $\wp{C}{\iverson{\mathit{lo} = \mathit{lo}'}}$ is the set of all initial states that will end with a final state where $lo=lo'$, and by projecting only to the values of the variable $\mathit{hi}$ we obtain all initial values of $\mathit{hi}$. 
However, aside from a (perhaps subjective) elegance perspective, we point out that the computation of $\wp{C}{\iverson{\mathit{lo} = \mathit{lo}'}}$ is actually more involved:
the Kleene's iterates of the loop for $\wpsymbol$ stabilize only at $\omega$ -- not 2:%
\allowdisplaybreaks%
\begin{align*}
	\Phi(\false)&\eeq \iverson{lo\geq \mathit{hi}}\land\iverson{lo=lo'}\\
	\Phi^2(\false)&\eeq\iverson{lo\geq \mathit{hi}}\land\iverson{lo=lo'}\lor\iverson{lo'-1<\mathit{hi}\leq lo'}\land\iverson{lo=lo'-1}\\
	\Phi^3(\false)&\eeq\iverson{lo\geq \mathit{hi}}\land\iverson{lo=lo'}\lor\iverson{lo'-1<\mathit{hi}\leq lo'}\land\iverson{lo=lo'-1}\\
	&\quad\lor\iverson{lo'-1<\mathit{hi}\leq lo'}\land\iverson{lo=lo'-2}\\[-1em]
	\vdots\\[-.75em]
	\Phi^\omega(\false)&\eeq \iverson{\mathit{hi}\leq lo}\land\iverson{lo=lo'} \vee \Bigg(\bigvee_{n=1}^\omega \iverson{lo'-1<\mathit{hi}\leq lo'}\land\iverson{lo=lo'-n}\Bigg)
\end{align*}%
Reasoning about this requires some form of creativity or advanced technique: either reasoning about the limit, or finding an invariant plus a termination prove. 
Only after determining $\Phi^\omega(\false)$, one can perform the $\wpsymbol$ for the assignment, which again results in a huge formula. 
For $\spsymbol$ and $\slpsymbol$, the Kleene's iterates stabilize after 2 iterations (Appendix~\ref{app:examples}): no need for invariant nor reasoning about limits nor projections of huge formulas. 

\subsection{Automation}
Our calculi, in their full generality, cannot be fully automated, 
which is not surprising since our calculi can express both termination and reachability properties for a Turing-complete computational model -- both of which are well known to be undecidable~\cite{Turing1936,Rice53}. 
Nevertheless, we believe that our calculi are at least syntactically mechanizable. For this aim, we plan to investigate an expressive “assertion” language for quantities, such as the one proposed by~\citet{DBLP:journals/pacmpl/BatzKKM21} for quantitative reasoning about probabilistic programs. This would allow showing \emph{relative completeness} in the sense of~\citet{Cook1978}, i.e., decidability modulo checking whether $g\ppreceq f$ holds, where $g,f$ may contain suprema and infima. 
Similar problems (decidability modulo checking a logical implication) exist for classical predicate transformers and Hoare logic~\cite{Cook1978}.

We also point out that the main goal of our calculi is to provide a framework, on which future tools for (partially) automating quantitative $\wlpsymbol$/$\spsymbol$/$\slpsymbol$ proofs can ground. For example, it may well be possible to fully automate the transformers for some syntactic (e.g.\ linear) fragments of \ngcl.

\subsection{Partial Incorrectness Reasoning}
We now show an application of partial incorrectness triples and, hence, of our strongest liberal postconditions. 
Consider a program/system $C_\mathit{login}$ that takes as input a variable $\textit{password}$. 
If $\textit{password}$ contains the correct password, say $\correctpassword$, then $C_\mathit{login}$ terminates in a final state containing a boolean variable “\textit{access}” storing the value $\true$; otherwise, the program terminates with value $\textit{access}=\false$. Now, recall that%
\begin{align*}
	\slp{C_\mathit{login}}{\iverson{\textit{password}=\correctpassword}}
\end{align*}%
is a predicate characterizing those final states which are reached \emph{only} by initial states $\sigma$ with the correct password, i.e.~initial states with $\sigma(\textit{password})=\correctpassword$. If the partial incorrectness triple $\incorrectness{\textit{access}=\true}{C_\mathit{login}}{\textit{password}=\correctpassword}$, which translates to%
\begin{align*}
\iverson{\textit{access}=\true} \iimplies \slp{C}{\iverson{\textit{password}=\correctpassword}}~,
\end{align*}%
holds, then knowing the correct password is a \emph{necessary precondition} to access the system. 
In other words, validity of the partial incorrectness triple guarantees that no user without knowledge of the correct password can end up in a final state $\tau$ where $\tau(\textit{access})=\true$.

We also note that, by the Galois Connection of Theorem~\ref{thm:wp-slp-relationship}, one can check whether the partial incorrectness triple holds also by employing $\wpsymbol$:%
\begin{align*}
\wp{C}{\iverson{\textit{access}=\true}} \implies \iverson{\textit{password}=\correctpassword}
\end{align*}%
However, reasoning with $\slpsymbol$ may well (1) be more feasible in practice (as demonstrated in Section~\ref{se:example-wp}) as well as (2) more intuitive when reasoning about \emph{necessary preconditions}  to access a system.

\section{Related Work}
\label{se:relatedwork}
\subsubsection*{More General Predicate Transformers.}
~\citet{Aguirre2020WeakestPI} focus on an abstract theory of $\wpsymbol$ for \emph{loop-free} programs. In particular, our $\textsf{w(l)p}$, \emph{restricted to the fragment of loop-free programs}, can be derived by instantiating their Corollary 4.6 (for details, see Appendix~\ref{app:related-work}). 
\citet[Section 4.1]{Aguirre2020WeakestPI} also define an \emph{abstract} strongest postcondition as a left adjoint of their weakest precondition (without constructing it); we believe that, due to our Theorem~\ref{thm:wp-slp-relationship}, an abstract strongest liberal post can be defined dually as a right adjoint of their weakest precondition.
On the other hand, our definition of strongest post is explicitly given by induction on the program structure and not implicitly as an adjoint.
The difficulties with finding strongest posts for probabilistic programs demonstrate that an explicit definition of a strongest post is more than desirable.

\subsubsection*{Strongest Liberal Post}
\label{se:relatedwork:slp}
The term \emph{“strongest \underline{liberal} postcondition”} is sometimes used in the literature for the original \underline{non-liberal} strongest postcondition, see e.g.~\cite[Section 2.2]{Back88},~\cite[Section 0]{Jacobs85}, or~\cite[Definition 8]{Wulandari_2020}.
 In fact,~\cite[Section 2.2]{Back88} argues that the strongest postcondition is often denoted also as strongest liberal postcondition due to the relationship between weakest liberal pre. However, since $\wlpsymbol$ \enquote{allows} nontermination whereas $\wpsymbol$ does not, and analogously $\slpsymbol$ \enquote{allows} unreachability whereas $\spsymbol$ does not, we believe that our naming convention of $\slpsymbol$ and $\spsymbol$ is more appropriate and natural.

\subsubsection*{Information Flow Analysis}
Some previous work on information flow analysis use type systems \cite{Volpano97,Orbaek97}.
However, these are imprecise and may reject safe programs such as $\COMPOSE{\ASSIGN{\mathit{lo}}{\mathit{hi}}}{\ASSIGN{\mathit{lo}}{0}}$ due to a \emph{potential} flow from $\mathit{hi}$ to $\mathit{lo}$ \cite{Amtoft04}. 
A Hoare-like logic combined with abstract interpretation has been proposed by~\citet{Amtoft04}, but fails for simple programs such as~\cite[Section 9]{Amtoft04}, which instead can be easily detected with our $\sfsymbol{s(l)p}$ analysis. Other abstract interpretation-based techniques focus on the trace semantics~\cite{UrbanM-a,Cousot19a}.~\citet{UrbanCWZ19} verify dependency fairness of neural networks by applying a backward analysis to compute the set of input values that lead to a certain ouput value; this approach is similar to a $\wpsymbol$-based calculus with ghost variables, as shown in Example~\ref{se:example-wp}, and we speculate that $\spsymbol$-based approaches could also be applied and potentially lead to better performances (as shown in Example~\ref{ex:2}). In Security Concurrent Separation logic~\cite{Ernst19} the authors provide an extension of concurrent separation logic~\cite{CSL,Reynolds02} by adding sensitivity assertions which, roughly, assigns to a certain variable a certain degree of security; however, their proof system deals only with partial correctness and restricts to conditional statements and loops that cannot use sensitive variables, so that our examples from~\Cref{se:examples} cannot be covered by their logic. Differently from the aforementioned works, our framework provides \emph{quantitative details} about the amount of information flow, instead of a single boolean output, see~\cite{smith2009foundations} for an overview.


\section{Conclusion \& Future Work}
We have presented a novel \emph{quantitative strongest post calculus} that subsumes classical strongest postconditions.
Moreover, we developed a novel \emph{quantitative strongest \underline{liberal} post} calculus. 
Restricted to a Boolean setting, we obtain the -- to the best of our knowledge -- unexplored notion of \emph{strongest liberal postconditions} which ultimately lead to our definition of \emph{partial incorrectness}. 
The latter connection is justified by the fundamental Galois connection between $\slpsymbol$ and $\wpsymbol$, and the strong duality between total and partial correctness, but where we replace \emph{nontermination} with \emph{unreachability}. 
Finally, we notice that there are three additional Hoare-style triples that can be naturally defined using our transformers, and we identify a precise connection between \emph{partial incorrectness} and the so-called \emph{necessary preconditions}~\cite{Cousot13}.

As future work, we plan to investigate the newly observed Hoare triples and to provide novel proof systems for them. 
We also plan to extend our quantitative strongest calculi with heap manipulation, similarly to the work of~\cite{quantitative_sl} for weakest pre calculi; this could lead to connections with incorrectness separation logic~\cite{IncorrectnessSeparationLogic}.

Finally, we plan to deepen the applications of quantitative strongest post calculi to quantitative information flow, perhaps by establishing connections with abstract interpretation~\cite{CousotC77}. 
In fact, we believe that our \textsf{s(l)p} transformers can be viewed as sound approximations of the fiber of the concrete semantics. Examples~\ref{ex:1},~\ref{ex:2} go into this direction after-all, since the combination of our strongest and strongest liberal post calculi can be viewed as an \emph{interval abstraction}~\cite{CC76} of the possible initial values of a certain pre-quantity.


\bibliography{literature}

\pagebreak
\appendix
\section*{Appendix}

\section{Collecting Semantics of While-loops}
\label{se:while-semantics}
Let us explain the semantics of $\WHILEDO{\guard}{C}$. Let $S$ again be the set of input states.
First, we denote by $F_S$ the function%
\begin{align*}
	F_S(X) \eeq S \ccup \bigl(\eval{C} \circ \eval{\guard} \bigr) X~,
\end{align*}%
i.e.\ $F_S$ first applies the filtering with respect to the loop guard $\guard$ to its input $X$, then applies the semantics of the loop body $C$ to the filtered set, and finally unions that result with the given set of input states $S$.
Using $F_S$, the standard collecting semantics for while loops can be expressed as%
\begin{align*}
	\eval{\WHILEDO{\guard}{C}} S \eeq & \eval{\neg \guard}\bigl(\lfp X\mydot F_S(X) \bigr)~,
\end{align*}%
where the least fixed point above is understood with respect to the partial order of set inclusion, which renders the structure $\langle \Conf,\, {\subseteq}\rangle$ a complete lattice with least element $\emptyset$.
The least fixed point above filtered by $\neg \guard$ expresses exactly the set $\eval{\WHILEDO{\guard}{C}} S$ of final states reachable after termination of $\WHILEDO{\guard}{C}$ starting from any initial state in $S$. 
We remark that to determine the least fixed point of the continuous function~$F_S$, it is sufficient to apply Kleene's fixpoint theorem and, as a result, we have that the infinite ascending chain $\emptyset \subseteq F_S^1(\emptyset)\subseteq F_S^2(\emptyset)\subseteq\dots F_S^\omega(\emptyset)$, where $F_S^{i+1}(X) = F_S(F_S^i(X))$, converges in at most $\omega$ iterations. 

\begin{example}[Standard Collecting Semantics of While Loops]
	\label{ex:semantics-div}
	Assume there is only a single program variable $x$ and consider the configuration $S=\{\{x\mapsto 0\}, \{x\mapsto 8\}\}$.
	We now want to execute the loop $\WHILEDO{x>5}{\ASSIGN{x}{x+1}}$ on this configuration and collect the reachable states. 
	By our construction above, we have%
	\begin{align*}
		\eval{\WHILEDO{x>5}{\ASSIGN{x}{x+1}}} S &\eeq   \eval{x\leq 5}\bigl(\lfp X\mydot F_S(X) \bigr)~, \quad \textnormal{where}\\[.25em]
		F_S(X) \eeq S \ccup \bigl(\eval{C} \circ \eval{\guard} \bigr) X &\eeq \{\{x\mapsto 0\}, \{x\mapsto 8\}\} \ccup \setcomp{\sigma\subst{x}{x+1}}{\sigma \in X,~ \sigma(x) > 5}~,
	\end{align*}%
	and the Kleene iterates are:%
	\begin{align*}
		F(\emptyset) \eeq & \bigl\{\{x\mapsto 0\},\{x\mapsto 8\} \bigr\}\ccup \emptyset\\
		F^2(\emptyset) \eeq & \bigl\{\{x\mapsto 0\},\, \{x\mapsto 8\} \bigr\} \ccup \bigl\{\{x\mapsto 9\} \bigr\} \\
		F^2(\emptyset) \eeq & \bigl\{\{x\mapsto 0\},\, \{x\mapsto 8\} \bigr\} \ccup \bigl\{\{x\mapsto 9\},\, \{x\mapsto 10\} \bigr\} \\
		\vdots & \\
		F^\omega(\emptyset) \eeq & \bigl\{\{x\mapsto 0\}\bigr\} \ccup \setcomp{\vphantom{\big(} \{x \mapsto i\}}{i \geq 9}
	\end{align*}%
	After filtering $F^\omega(\emptyset)$ by the negation of the loop guard, we obtain the loop's collecting semantics%
	\begin{align*}
		\eval{\WHILEDO{x>5}{\ASSIGN{x}{x+1}}} S \eeq \eval{x\leq 5}\bigl(F^\omega(\emptyset) \bigr) \eeq \bigl\{\{x\mapsto 0\}\bigr\}~.	
		\tag*{\qedtriangle}
	\end{align*}
\end{example}%
\section{Proofs of Section~\ref{se:weakest-pre}}
\subsection{Proof of Soundness for \textnormal{$\wpsymbol$}, Thereom \ref{thm:wp-soundness}}
\label{proof:wp-soundness}

\wpsoundness*
\begin{proof}
	We prove Theorem~\ref{thm:wp-soundness} by induction on the structure of $C$.
	For the induction base, we have the atomic statements:%
	\paragraph{The effectless program $\SKIP$:} We have%
	\begin{align*}
		\wp{\SKIP}{f}(\sigma)
		\eeq & f (\sigma )\\
		\eeq &	\sup_{\tau \in \{\sigma\}}f (\tau)\\
		\eeq &	\sup_{\tau\in \eval{\SKIP} (\sigma)}f (\tau)~.
	\end{align*}%
	\paragraph{The assignment $\ASSIGN{x}{e}$:} We have%
	\begin{align*}
		\wp{\ASSIGN{x}{\ee}}{f}(\sigma)
		\eeq & f\subst{x}{\ee} (\sigma) \\
		\eeq & f(\sigma\subst{x}{\sigma(\ee)}) \\
		\eeq &	\sup_{\tau\in\{\sigma\subst{x}{\sigma(\ee)}\}}f (\tau)\\
		\eeq &	\sup_{\tau\in\eval{\ASSIGN{x}{\ee}} (\sigma)}f (\tau)~.
	\end{align*}%
	This concludes the proof for the atomic statements.%
	\paragraph{Induction Hypothesis:} 
	For arbitrary but fixed programs $C$, $C_1$, $C_2$, we proceed with the inductive step on the composite statements.%
	\paragraph{The sequential composition $\COMPOSE{C_1}{C_2}$:} We have%
	\begin{align*}
		\wp{\COMPOSE{C_1}{C_2}}{f}(\sigma) 
		\eeq & \wp{C_1}{\wp{C_2}{f}}(\sigma) \\
		\eeq & \sup_{\tau'\in\eval{C_1} (\sigma)} \wp{C_2}{f}(\tau') \tag{by I.H.~on $C_1$}\\
		\eeq & \sup_{\tau'\in\eval{C_1} (\sigma)\land \tau\in\eval{C_2} (\tau')} {f}(\tau) \tag{by I.H.~on $C_2$}\\
		\eeq &	\sup_{\tau\in\eval{C_2} (\eval{C_1} (\sigma))}f (\tau)\\
		\eeq &	\sup_{\tau\in\eval{\COMPOSE{C_1}{C_2}} (\sigma)}f (\tau)~.
	\end{align*}%
	\noindent{}%
	\paragraph{The conditional branching $\ITE{\guard}{C_1}{C_2}$:} We have%
	\begin{align*}
		&\wp{\ITE{\guard}{C_1}{C_2}}{f}(\sigma) \\
		& \eeq \bigl(\iverson{\guard}\curlywedge \wp{C_1}{f} \ccurlyvee \iverson{\neg \guard}\curlywedge \wp{C_2}{f}\big)(\sigma)\\
		& \eeq
		\begin{cases}
			\wp{C_1}{f}(\sigma) & \text{ if } \sigma\mmodels \guard\\
			\wp{C_2}{f}(\sigma) & \text{ otherwise }
		\end{cases}\\
		& \eeq
		\begin{cases}
			\sup_{\tau\in\eval{C_1} (\sigma)} f(\tau)  & \text{ if } \sigma\mmodels \guard\\
			\sup_{\tau\in\eval{C_2} (\sigma)} f(\tau)  & \text{ otherwise }
		\end{cases}\tag{by I.H.~on $C_1, C_2$}\\
		&\eeq	\sup_{\tau\in  (\eval{C_1} \circ \eval{\guard}) (\sigma) \cup ( \eval{C_2} \circ \eval{\neg\guard})(\sigma)}f (\tau)\\
		&\eeq	\sup_{\tau\in\eval{\ITE{\guard}{C_1}{C_2}} (\sigma)}f (\tau)~.
	\end{align*}%
	\noindent{}%
	\paragraph{The nondeterministic choice $\NDCHOICE{C_1}{C_2}$:} We have%
	\begin{align*}
		\wp{\NDCHOICE{C_1}{C_2}}{f}(\sigma)
		\eeq & \bigl(\wp{C_1}{f} \ccurlyvee \wp{C_2}{f}\big)(\sigma) \\
		\eeq &	\sup_{\tau\in  \eval{C_1} (\sigma)} f(\tau) \ccurlyvee \sup_{\tau\in  \eval{C_2} (\sigma)} f(\tau) \tag{by I.H.~on $C_1,C_2$}\\
		\eeq&	\sup_{\tau\in  \eval{C_1} (\sigma) \cup \eval{C_2} (\sigma)}f (\tau)\\
		\eeq&	\sup_{\tau\in\eval{\NDCHOICE{C_1}{C_2}} (\sigma)}f (\tau)~.
	\end{align*}%
	\noindent{}%
	
	\paragraph{The loop $\WHILEDO{\guard}{C}$:} 
	Let%
	\begin{align*}
		\Phi_{f}(X) &\eeq \iverson{\neg\guard}\curlywedge f \ccurlyvee \iverson{\guard}\curlywedge \wp{C}{X}~,
	\end{align*}%
	be the $\wpsymbol$-characteristic functions of the loop $\WHILEDO{\guard}{C}$ with respect to postanticipation $f$ and
	\begin{align*}
		F_S(X) &\eeq  S\cup (\eval{C}\circ\eval{\guard})X~,
	\end{align*}%
	be the collecting semantics characteristic functions of the loop $\WHILEDO{\guard}{C}$ with respect to any input $S\in\powerset{\Conf}$.
	We now prove by induction on $n$ that, for all $\sigma\in\States$%
	\begin{equation}
		\label{eq:wp-soundness-induction}
		\Phi_{f}^n(\minfty)(\sigma)  \eeq \sup_{\tau\in\eval{\neg\guard}  F_{\{\sigma\}}^n(\emptyset)}f (\tau)~.
	\end{equation}%
	For the induction base $n = 0$, consider the following:%
	\begin{align*}
		\Phi_{f}^0(\minfty)  (\sigma) 
		& \eeq  \minfty\\
		& \eeq \sup \emptyset \\
		& \eeq \sup_{\tau\in\emptyset} f(\tau)\\
		& \eeq \sup_{\tau\in\eval{\neg\guard}  F_{\{\sigma\}}^0(\emptyset)}f (\tau)~.
	\end{align*}%
	As induction hypothesis, we have for arbitrary but fixed $n$ and all $\sigma\in\States$,%
	\begin{align*}
		\Phi_{f}^n(\minfty)(\sigma)   \eeq \sup_{\tau\in\eval{\neg\guard}  F_{\{\sigma\}}^n(\emptyset)}f (\tau)~.
	\end{align*}%
	For the induction step $n \longrightarrow n + 1$, consider the following:%
	\begin{align*}
		&	\Phi_{f}^{n+1}(\minfty)(\sigma) \\
		& \eeq  \left(\iverson{\neg\guard}\curlywedge f\right)(\sigma) \ccurlyvee \left(\iverson{\guard}\curlywedge \wp{C}{\Phi_{f}^n(\minfty)}\right)(\sigma)  \\
		& \eeq  (\iverson{\neg\guard}\curlywedge f)(\sigma) \ccurlyvee
		\sup_{\tau\in  \eval{C} (\sigma)\land \sigma\mmodels\guard} \Phi_{f}^n(\minfty)(\tau) \tag{by I.H.\ on $C$} \\
		& \eeq 
		\begin{cases}
			\sup_{\tau\in  \eval{C} (\sigma)} \Phi_{f}^n(\minfty)(\tau) & \text{ if } \sigma\mmodels \guard\\
			f(\sigma) & \text{ otherwise }
		\end{cases}\\
		& \eeq 
		\begin{cases}
			\sup_{\tau\in  \eval{C} (\sigma)} ~
			\sup_{\tau'\in\eval{\neg\guard} ~ F_{\{\tau\}}^n(\emptyset)}f (\tau') & \text{ if } \sigma\mmodels \guard\\
			f(\sigma) & \text{ otherwise }
		\end{cases}
		\tag{by I.H.\ on $n$}  \\
		& \eeq 
		\begin{cases}
			\sup_{\tau'\in\eval{\neg\guard} ~ F_{\eval{C} (\sigma)}^n(\emptyset)}f (\tau') & \text{ if } \sigma\mmodels \guard\\
			f(\sigma) & \text{ otherwise }
		\end{cases} \\
		& \eeq 
		\begin{cases}
			\sup_{\tau'\in\eval{\neg\guard} ~ F_{(\eval{C} \circ \iguard )(\sigma)}^n(\emptyset)}f (\tau') & \text{ if } \sigma\mmodels \guard\\
			f(\sigma) & \text{ otherwise }
		\end{cases} \\
		& \eeq \sup_{
			\tau'\in\eval{\neg\guard}(\{\sigma\} \cup F_{(\eval{C}\circ\eval{\guard})(\sigma)}^n(\emptyset) ) } f(\tau')\\
		&
		\eeq \sup_{\tau\in\eval{\neg\guard}  F_{\{\sigma\}}^{n+1}(\emptyset)}f (\tau)~.
	\end{align*}
	
	This concludes the induction on $n$. Now we have:
	\begin{align*}
		\wp{\WHILEDO{\guard}{C}}{f}(\sigma) 
		\eeq & \bigl(\lfp  X\mydot \iverson{\neg \guard} \curlywedge f \ccurlyvee \iverson{\guard} \curlywedge \wp{C}{X}\bigr)(\sigma)\\
		\eeq & \sup_{n\in\mathbb{N}}\Phi_{f}^{n}(\minfty)(\sigma) \tag{By Kleene's fixpoint theorem}\\
		\eeq&	\sup_{n\in\mathbb{N}}~\sup_{\tau\in\eval{\neg\guard}  F_{\{\sigma\}}^n(\emptyset)}f (\tau)\tag{by Equation~\ref{eq:wp-soundness-induction}}\\
		\eeq&	\sup_{\tau\in	\cup_{n\in\mathbb{N}}(\eval{\neg\guard}  F_{\{\sigma\}}^n(\emptyset))}f (\tau)\\
		\eeq&	\sup_{\tau\in	\eval{\neg\guard} (\cup_{n\in\mathbb{N}}F_{\{\sigma\}}^n(\emptyset))}f (\tau) \tag{by continuity of $\eval{\neg\guard} $}\\
		\eeq&	\sup_{\tau\in\eval{\neg \guard} (\lfp X\mydot \{\sigma\} \cup (\eval{C} \circ \eval{\guard})X )}f (\tau)\tag{by Kleene's fixpoint theorem}\\		
		\eeq&	\sup_{\tau\in\eval{\WHILEDO{\guard}{C}} (\sigma)}f (\tau)~,
	\end{align*}
	and this concludes the proof.
\end{proof}

\subsection{Proof of Soundness for \textnormal{$\wlpsymbol$}, Thereom \ref{thm:wlp-soundness}}
\label{proof:wlp-soundness}

\wlpsoundness*
\begin{proof}
	We prove Theorem~\ref{thm:wlp-soundness} by induction on the structure of $C$.
	For the induction base, we have the atomic statements:%
	\paragraph{The effectless program $\SKIP$:} We have%
	\begin{align*}
		\wlp{\SKIP}{f}(\sigma)
		\eeq & f (\sigma )\\
		\eeq &	\inf_{\tau \in \{\sigma\}}f (\tau)\\
		\eeq &	\inf_{\tau\in \eval{\SKIP} (\sigma)}f (\tau)~.
	\end{align*}%
	\paragraph{The assignment $\ASSIGN{x}{e}$:} We have%
	\begin{align*}
		\wlp{\ASSIGN{x}{\ee}}{f}(\sigma)
		\eeq & f\subst{x}{\ee} (\sigma) \\
		\eeq & f(\sigma\subst{x}{\sigma(\ee)}) \\
		\eeq &	\inf_{\tau\in\{\sigma\subst{x}{\sigma(\ee)}\}}f (\tau)\\
		\eeq &	\inf_{\tau\in\eval{\ASSIGN{x}{\ee}} (\sigma)}f (\tau)~.
	\end{align*}%
	This concludes the proof for the atomic statements.%
	\paragraph{Induction Hypothesis:} 
	For arbitrary but fixed programs $C$, $C_1$, $C_2$, we proceed with the inductive step on the composite statements.%
	\paragraph{The sequential composition $\COMPOSE{C_1}{C_2}$:} We have%
	\begin{align*}
		\wlp{\COMPOSE{C_1}{C_2}}{f}(\sigma) 
		\eeq & \wlp{C_1}{\wlp{C_2}{f}}(\sigma) \\
		\eeq & \inf_{\tau'\in\eval{C_1} (\sigma)} \wlp{C_2}{f}(\tau') \tag{by I.H.~on $C_1$}\\
		\eeq & \inf_{\tau'\in\eval{C_1} (\sigma)\land \tau\in\eval{C_2} (\tau')} {f}(\tau) \tag{by I.H.~on $C_2$}\\
		\eeq &	\inf_{\tau\in\eval{C_2} (\eval{C_1} (\sigma))}f (\tau)\\
		\eeq &	\inf_{\tau\in\eval{\COMPOSE{C_1}{C_2}} (\sigma)}f (\tau)~.
	\end{align*}%
	\noindent{}%
	\paragraph{The conditional branching $\ITE{\guard}{C_1}{C_2}$:} We have%
	\begin{align*}
		&\wlp{\ITE{\guard}{C_1}{C_2}}{f}(\sigma) \\
		& \eeq \bigl(\iverson{\guard} \curlywedge \wlp{C_1}{f} \ccurlyvee \iverson{\neg \guard} \curlywedge \wlp{C_2}{f}\big)(\sigma)\\
		& \eeq
		\begin{cases}
			\wlp{C_1}{f}(\sigma) & \text{ if } \sigma\mmodels \guard\\
			\wlp{C_2}{f}(\sigma) & \text{ otherwise }
		\end{cases}\\
		& \eeq
		\begin{cases}
			\inf_{\tau\in\eval{C_1} (\sigma)} f(\tau)  & \text{ if } \sigma\mmodels \guard\\
			\inf_{\tau\in\eval{C_2} (\sigma)} f(\tau)  & \text{ otherwise }
		\end{cases}\tag{by I.H.~on $C_1, C_2$}\\
		&\eeq	\inf_{\tau\in  (\eval{C_1} \circ \eval{\guard}) (\sigma) \cup ( \eval{C_2} \circ \eval{\neg\guard})(\sigma)}f (\tau)\\
		&\eeq	\inf_{\tau\in\eval{\ITE{\guard}{C_1}{C_2}} (\sigma)}f (\tau)~.
	\end{align*}%
	\noindent{}%
	\paragraph{The nondeterministic choice $\NDCHOICE{C_1}{C_2}$:} We have%
	\begin{align*}
		\wlp{\NDCHOICE{C_1}{C_2}}{f}(\sigma)
		\eeq & \bigl(\wlp{C_1}{f} \ccurlywedge \wlp{C_2}{f}\big)(\sigma) \\
		\eeq &	\inf_{\tau\in  \eval{C_1} (\sigma)} f(\tau) \ccurlywedge \inf_{\tau\in  \eval{C_2} (\sigma)} f(\tau) \tag{by I.H.~on $C_1,C_2$}\\
		\eeq&	\inf_{\tau\in  \eval{C_1} (\sigma) \cup \eval{C_2} (\sigma)}f (\tau)\\
		\eeq&	\inf_{\tau\in\eval{\NDCHOICE{C_1}{C_2}} (\sigma)}f (\tau)~.
	\end{align*}%
	\noindent{}%
	
	\paragraph{The loop $\WHILEDO{\guard}{C}$:} 
	Let%
	\begin{align*}
		\Phi_{f}(X) &\eeq \iverson{\neg\guard}\curlywedge f \ccurlyvee \iverson{\guard}\curlywedge \wlp{C}{X}~,
	\end{align*}%
	be the $\wlpsymbol$-characteristic functions of the loop $\WHILEDO{\guard}{C}$ with respect to postanticipation $f$ and
	\begin{align*}
		F_S(X) &\eeq  S\cup (\eval{C}\circ\eval{\guard})X~,
	\end{align*}%
	be the collecting semantics characteristic functions of the loop $\WHILEDO{\guard}{C}$ with respect to any input $S\in\powerset{\Conf}$.
	We now prove by induction on $n$ that, for all $\sigma\in\States$%
	\begin{equation}
		\label{eq:wlp-soundness-induction}
		\Phi_{f}^n(\pinfty)(\sigma)  \eeq \inf_{\tau\in\eval{\neg\guard}  F_{\{\sigma\}}^n(\emptyset)}f (\tau)~.
	\end{equation}%
	For the induction base $n = 0$, consider the following:%
	\begin{align*}
		\Phi_{f}^0(\pinfty)  (\sigma) 
		& \eeq  \pinfty\\
		& \eeq \inf \emptyset \\
		& \eeq \inf_{\tau\in\emptyset} f(\tau)\\
		& \eeq \inf_{\tau\in\eval{\neg\guard}  F_{\{\sigma\}}^0(\emptyset)}f (\tau)~.
	\end{align*}%
	As induction hypothesis, we have for arbitrary but fixed $n$ and all $\sigma\in\States$,%
	\begin{align*}
		\Phi_{f}^n(\pinfty)(\sigma)   \eeq \inf_{\tau\in\eval{\neg\guard}  F_{\{\sigma\}}^n(\emptyset)}f (\tau)~.
	\end{align*}%
	For the induction step $n \longrightarrow n + 1$, consider the following:%
	\begin{align*}
		&	\Phi_{f}^{n+1}(\pinfty)(\sigma) \\
		& \eeq  \left(\iverson{\neg\guard}\curlywedge f\right)(\sigma) \ccurlyvee \left(\iverson{\guard}\curlywedge \wlp{C}{\Phi_{f}^n(\pinfty)}\right)(\sigma)  \\
		& \eeq  (\iverson{\neg\guard}\curlywedge f)(\sigma) \ccurlyvee
		\iverson{\guard}(\sigma)\curlywedge
		\inf_{\tau\in  \eval{C} (\sigma)} \Phi_{f}^n(\pinfty)(\tau) \tag{by I.H.\ on $C$} \\
		& \eeq 
		\begin{cases}
			\inf_{\tau\in  \eval{C} (\sigma)} \Phi_{f}^n(\pinfty)(\tau) & \text{ if } \sigma\mmodels \guard\\
			f(\sigma) & \text{ otherwise }
		\end{cases}\\
		& \eeq 
		\begin{cases}
			\inf_{\tau\in  \eval{C} (\sigma)} ~
			\inf_{\tau'\in\eval{\neg\guard} ~ F_{\{\tau\}}^n(\emptyset)}f (\tau') & \text{ if } \sigma\mmodels \guard\\
			f(\sigma) & \text{ otherwise }
		\end{cases}
		\tag{by I.H.\ on $n$}  \\
		& \eeq 
		\begin{cases}
			\inf_{\tau'\in\eval{\neg\guard} ~ F_{\eval{C} (\sigma)}^n(\emptyset)}f (\tau') & \text{ if } \sigma\mmodels \guard\\
			f(\sigma) & \text{ otherwise }
		\end{cases} \\
		& \eeq 
		\begin{cases}
			\inf_{\tau'\in\eval{\neg\guard} ~ F_{(\eval{C} \circ \iguard )(\sigma)}^n(\emptyset)}f (\tau') & \text{ if } \sigma\mmodels \guard\\
			f(\sigma) & \text{ otherwise }
		\end{cases} \\
		& \eeq \inf_{
			\tau'\in\eval{\neg\guard}(\{\sigma\} \cup F_{(\eval{C}\circ\eval{\guard})(\sigma)}^n(\emptyset) ) } f(\tau')\\
		&
		\eeq \inf_{\tau\in\eval{\neg\guard}  F_{\{\sigma\}}^{n+1}(\emptyset)}f (\tau)~.
	\end{align*}
	This concludes the induction on $n$. Now we have:
	\begin{align*}
		\wlp{\WHILEDO{\guard}{C}}{f}(\sigma) 
		\eeq & \bigl(\gfp  X\mydot \iverson{\neg \guard} \curlywedge f \ccurlyvee \iverson{\guard} \curlywedge \wlp{C}{X}\bigr)(\sigma)\\
		\eeq & \inf_{n\in\mathbb{N}}\Phi_{f}^{n}(\pinfty)(\sigma) \tag{by Kleene's fixpoint theorem}\\
		\eeq&	\inf_{n\in\mathbb{N}}~\inf_{\tau\in\eval{\neg\guard}  F_{\{\sigma\}}^n(\emptyset)}f (\tau)\tag{by Equation~\ref{eq:wlp-soundness-induction}}\\
		\eeq&	\inf_{\tau\in	\cup_{n\in\mathbb{N}}(\eval{\neg\guard}  F_{\{\sigma\}}^n(\emptyset))}f (\tau)\\
		\eeq&	\inf_{\tau\in	\eval{\neg\guard} (\cup_{n\in\mathbb{N}}F_{\{\sigma\}}^n(\emptyset))}f (\tau) \tag{by continuity of $\eval{\neg\guard} $}\\
		\eeq&	\inf_{\tau\in\eval{\neg \guard} (\lfp X\mydot \{\sigma\} \cup (\eval{C} \circ \eval{\guard})X )}f (\tau)\tag{by Kleene's fixpoint theorem}\\		
		\eeq&	\inf_{\tau\in\eval{\WHILEDO{\guard}{C}} (\sigma)}f (\tau)~,
	\end{align*}
	and this concludes the proof.
\end{proof}

%
%

%
%


\section{Proofs of Section~\ref{se:sp}}
\subsection{Proof of Soundness for \textnormal{$\spsymbol$}, Thereom~\ref{thm:sp-soundness}}
\label{proof:sp-soundness}

\spsoundness*
\begin{proof}
	We prove Theorem~\ref{thm:sp-soundness} by induction on the structure of $C$.
	For the induction base, we have the atomic statements:%
	\paragraph{The effectless program $\SKIP$:} We have%
	\begin{align*}
		\sp{\SKIP}{f}(\tau)
		\eeq & f (\tau )\\
		\eeq &	\sup_{\sigma\in\States,\tau \in \{\sigma\}}f (\sigma)\\
		\eeq &	\sup_{\sigma\in\States,\tau\in \eval{\SKIP} (\sigma)}f (\sigma)~.
	\end{align*}%
	\paragraph{The assignment $\ASSIGN{x}{e}$:} We have%
	\begin{align*}
		\sp{\ASSIGN{x}{\ee}}{f}(\tau)
		\eeq & (\SupV{\alpha}~ \iverson{x = e\subst{x}{\alpha}}\curlywedge f\subst{x}{\alpha})(\tau) \\
		\eeq & (\sup_{\alpha}~ \iverson{x = e\subst{x}{\alpha}}\curlywedge f\subst{x}{\alpha})(\tau) \\
		\eeq & \sup_{\alpha\colon\tau(x)=\tau(e\subst{x}{\alpha})}~ (f\subst{x}{\alpha})(\tau) \\
		\eeq & \sup_{\alpha\colon \tau(x)=\tau(e\subst{x}{\alpha})}~ f(\tau\subst{x}{\alpha}) \\
		\eeq & \sup_{\alpha\colon \tau\subst{x}{\alpha}\subst{x}{\tau(\ee\subst{x}{\alpha})}=\tau}~ f(\tau\subst{x}{\alpha}) \\
		\eeq & \sup_{\alpha\colon\tau\subst{x}{\alpha}\subst{x}{\tau\subst{x}{\alpha}(\ee)}=\tau}~ f(\tau\subst{x}{\alpha}) \\
		\eeq &	\sup_{\sigma\in\States,\sigma\subst{x}{\sigma(\ee)}=\tau}~ f(\sigma) \tag{By taking $\sigma=\tau\subst{x}{\alpha}$}\\
		\eeq &	\sup_{\sigma\in\States,\tau\in\{\sigma\subst{x}{\sigma(\ee)}\}}f (\sigma)\\
		\eeq &	\sup_{\sigma\in\States,\tau\in\eval{\ASSIGN{x}{\ee}} (\sigma)}f (\sigma)~.
	\end{align*}%

	This concludes the proof for the atomic statements.%
	\paragraph{Induction Hypothesis:} 
	For arbitrary but fixed programs $C$, $C_1$, $C_2$, we proceed with the inductive step on the composite statements.%
	\paragraph{The sequential composition $\COMPOSE{C_1}{C_2}$:} We have%
	\begin{align*}
		\sp{\COMPOSE{C_2}{C_1}}{f}(\tau) 
		\eeq & \sp{C_2}{\sp{C_1}{f}}(\tau) \\
		\eeq & \sup_{\sigma'\in\States,\tau\in\eval{C_2} (\sigma')} \sp{C_1}{f}(\sigma') \tag{by I.H.~on $C_2$}\\
		\eeq & \sup_{\sigma\in\States,\tau\in\eval{C_2} (\sigma')\land \sigma'\in\eval{C_1} (\sigma)} {f}(\sigma) \tag{by I.H.~on $C_2$}\\
		\eeq &	\sup_{\sigma\in\States,\tau\in\eval{C_2} (\eval{C_1} (\sigma))}f (\sigma)\\
		\eeq &	\sup_{\sigma\in\States, \tau\in\eval{\COMPOSE{C_1}{C_2}} (\sigma)}f (\sigma)~.
	\end{align*}%
	\noindent{}%
	\paragraph{The conditional branching $\ITE{\guard}{C_1}{C_2}$:} We have%
	\begin{align*}
		&\sp{\ITE{\guard}{C_1}{C_2}}{f}(\tau) \\
		& \eeq\bigl(\sp{C_1}{\iverson{\guard}\curlywedge f} \ccurlyvee \sp{C_2}{\iverson{\neg\guard}\curlywedge f}\bigr)(\tau)\\
		& \eeq 
		\sup_{\sigma\in\States,\tau\in \eval{C_1} (\sigma)}(\iverson{\guard}\curlywedge f)(\sigma)
		\ccurlyvee
		\sup_{\sigma\in\States,\tau\in \eval{C_2} (\sigma)}(\iverson{\neg\guard}\curlywedge f) (\sigma)\tag{by I.H.~on $C_1, C_2$}\\
		& \eeq 
		\sup_{\sigma\in\States,\tau\in (\eval{C_1} \circ\eval{\guard})(\sigma)}f(\sigma)
		\ccurlyvee
		\sup_{\sigma\in\States,\tau\in (\eval{C_2} \circ\eval{\neg\guard})(\sigma)}f(\sigma)\\
		&\eeq
		\sup_{\sigma\in\States,\tau\in  (\eval{C_1} \circ \eval{\guard}) (\sigma) \cup ( \eval{C_2} \circ \eval{\neg\guard})(\sigma)}f (\sigma)\\
		&\eeq	\sup_{\sigma\in\States,\tau\in\eval{\ITE{\guard}{C_1}{C_2}} (\sigma)}f (\sigma)~.
	\end{align*}%
	\noindent{}%
	\paragraph{The nondeterministic choice $\NDCHOICE{C_1}{C_2}$:} We have%
	\begin{align*}
		\sp{\NDCHOICE{C_1}{C_2}}{f}(\tau)
		\eeq & \bigl(\sp{C_1}{f} \ccurlyvee \sp{C_2}{f}\big)(\tau) \\
		\eeq &	\sup_{\sigma\in\States, \tau\in  \eval{C_1} (\sigma)} f(\sigma) \ccurlyvee \sup_{\sigma\in\States,\tau\in  \eval{C_2} (\sigma)} f(\sigma) \tag{by I.H.~on $C_1,C_2$}\\
		\eeq&	\sup_{\sigma\in\States, \tau\in  \eval{C_1} (\sigma) \cup \eval{C_2} (\sigma)}f (\sigma)\\
		\eeq&	\sup_{\sigma\in\States, \tau\in\eval{\NDCHOICE{C_1}{C_2}} (\sigma)}f (\sigma)~.
	\end{align*}%
	\noindent{}%
	
	\paragraph{The loop $\WHILEDO{\guard}{C}$:} 
	Let%
	\begin{align*}
		\Psi_{f}(X) &\eeq f \ccurlyvee \sp{C}{\iverson{\guard}\curlywedge X}~,
	\end{align*}%
	be the $\spsymbol$-characteristic functions of the loop $\WHILEDO{\guard}{C}$ with respect to preanticipation $f$ and
	\begin{align*}
		F_S(X) &\eeq  S\cup (\eval{C}\circ\eval{\guard})X~,
	\end{align*}%
	be the collecting semantics characteristic functions of the loop $\WHILEDO{\guard}{C}$ with respect to any input $S\in\powerset{\Conf}$.
	We now prove by induction on $n$ that, for all $\tau\in\States$
	\begin{equation}
		\label{eq:sp-soundness-induction}
		\Psi_{f}^{n}(\minfty )(\tau)\eeq\sup_{\sigma\in\States,\tau\in F_{\{\sigma\}}^n(\emptyset)}f (\sigma)~.
	\end{equation}%
	For the induction base $n = 0$, consider the following:%
	\begin{align*}
		\Psi_{f}^{0}(\minfty )(\tau)
		& \eeq  \minfty \\
		& \eeq \sup \emptyset \\
		& \eeq \sup_{\sigma\in\States,\tau\in\emptyset} f(\sigma)\\
		& \eeq \sup_{\sigma\in\States,\tau\in F_{\{\sigma\}}^0(\emptyset)}f (\sigma)~.
	\end{align*}%
	As induction hypothesis, we have for arbitrary but fixed $n$ and all $\tau\in\States$%
	\begin{align*}
		\Psi_{f}^{n}(\minfty )(\tau)\eeq\sup_{\sigma\in\States,\tau\in F_{\{\sigma\}}^n(\emptyset)}f (\sigma)~.
	\end{align*}%
	For the induction step $n \longrightarrow n + 1$, consider the following:%
	\begin{align*}
		&	\Psi_{f}^{n+1}(\minfty )(\tau)  \\
		& \eeq \left(f \ccurlyvee \sp{C}{\iverson{\guard}\curlywedge \Psi_{f}^n(\minfty )} \right)(\tau)\\
		& \eeq  f(\tau) \ccurlyvee
		\sup_{\sigma\in\States, \tau\in\eval{C} (\sigma)}\bigl(\iverson{\guard}\curlywedge  \Psi_{f}^n(\minfty )\bigr) (\sigma)
		\tag{by I.H.\ on $C$}\\
		& \eeq  f(\tau) \ccurlyvee
		\sup_{\sigma\in\States, \tau\in\eval{C} (\sigma)}~
		\sup_{\sigma'\in\States,\sigma\in\eval{\guard}  F_{\{\sigma'\}}^n(\emptyset)}f (\sigma')
		\tag{by I.H.\ on $n$} 
		\\
		& \eeq  f(\tau) \ccurlyvee
		\sup_{\sigma'\in\States,\tau\in(\eval{C}  \circ  \eval{\guard}  )F_{\{\sigma'\}}^n(\emptyset)}f (\sigma')
		\\
		& \eeq 
		\sup_{\sigma'\in\States,\tau\in(\eval{C}  \circ  \eval{\guard}  )F_{\{\sigma'\}}^n(\emptyset)\cup \{\sigma'\}}f (\sigma')
		\\
		&
		\eeq \sup_{\sigma\in\States,\tau\in F_{\{\sigma\}}^{n+1}(\emptyset)}f (\sigma)~.
	\end{align*}
	
	This concludes the induction on $n$. Now we have:

	\begin{align*}
		\sp{\WHILEDO{\guard}{C}}{f}(\tau) 
		\eeq & \left(\inguard\curlywedge\bigl(\lfp  X\mydot f \ccurlyvee \sp{C}{\iverson{\guard}\curlywedge X}\bigr)\right)(\tau)\\
		\eeq & \bigl(\inguard\curlywedge\sup_{n\in\mathbb{N}}~\Psi_{f}^{n}(\minfty )\bigr)(\tau) \tag{by Kleene's fixpoint theorem}\\
		\eeq & \sup_{n\in\mathbb{N}}~\bigl(\inguard\curlywedge\Psi_{f}^{n}(\minfty )\bigr)(\tau) \tag{by continuity of $\lambda X\mydot\inguard\curlywedge X$}\\
		\eeq&	\sup_{n\in\mathbb{N}}~\sup_{\sigma\in\States,\tau\in\eval{\neg\guard}  F_{\{\sigma\}}^n(\emptyset)}f (\sigma)\tag{by Equation~\ref{eq:sp-soundness-induction}}\\
		\eeq&	\sup_{\sigma\in\States,\tau\in	\cup_{n\in\mathbb{N}}(\eval{\neg\guard}  F_{\{\sigma\}}^n(\emptyset))}f (\sigma)\\
		\eeq&	\sup_{\sigma\in\States,\tau\in	\eval{\neg\guard} (\cup_{n\in\mathbb{N}}F_{\{\sigma\}}^n(\emptyset))}f (\sigma) \tag{by continuity of $\eval{\neg\guard} $}\\
		\eeq&	\sup_{\sigma\in\States, \tau\in\eval{\neg \guard} (\lfp X\mydot \{\sigma\} \cup (\eval{C} \circ \eval{\guard})X )}f (\sigma)\tag{by Kleene's fixpoint theorem}\\		
		\eeq&	\sup_{\sigma\in\States,\tau\in\eval{\WHILEDO{\guard}{C}} (\sigma)}f (\sigma)~,
	\end{align*}
	and this concludes the proof.
\end{proof}

\subsection{Proof of Soundness for \textnormal{$\slpsymbol$}, Thereom \ref{thm:slp-soundness}}
\label{proof:slp-soundness}

\slpsoundness*
\begin{proof}
	We prove Theorem~\ref{thm:slp-soundness} by induction on the structure of $C$.
	For the induction base, we have the atomic statements:%
	\paragraph{The effectless program $\SKIP$:} We have%
	\begin{align*}
		\slp{\SKIP}{f}(\tau)
		\eeq & f (\tau )\\
		\eeq &	\inf_{\sigma\in\States,\tau \in \{\sigma\}}f (\sigma)\\
		\eeq &	\inf_{\sigma\in\States,\tau\in \eval{\SKIP} (\sigma)}f (\sigma)~.
	\end{align*}%
	\paragraph{The assignment $\ASSIGN{x}{e}$:} We have%
	\begin{align*}
		\slp{\ASSIGN{x}{\ee}}{f}(\tau)
		\eeq & (\InfV{\alpha}~ \iverson{x \neq e\subst{x}{\alpha}} \curlyvee f\subst{x}{\alpha})(\tau) \\
		\eeq & (\inf_{\alpha}~ \iverson{x \neq e\subst{x}{\alpha}} \curlyvee f\subst{x}{\alpha})(\tau) \\
		\eeq & \inf_{\alpha\colon \tau(x)=\tau(e\subst{x}{\alpha})}~ (f\subst{x}{\alpha})(\tau) \\
		\eeq & \inf_{\alpha\colon\tau(x)=\tau(e\subst{x}{\alpha})}~ f(\tau\subst{x}{\alpha}) \\
		\eeq & \inf_{\alpha\colon\tau\subst{x}{\alpha}\subst{x}{\tau(\ee\subst{x}{\alpha})}=\tau}~ f(\tau\subst{x}{\alpha}) \\
		\eeq & \inf_{\alpha\colon\tau\subst{x}{\alpha}\subst{x}{\tau\subst{x}{\alpha}(\ee)}=\tau}~ f(\tau\subst{x}{\alpha}) \\
		\eeq &	\inf_{\sigma\in\States,\sigma\subst{x}{\sigma(\ee)}=\tau}~ f(\sigma) \tag{By taking $\sigma=\tau\subst{x}{\alpha}$}\\
		\eeq &	\inf_{\sigma\in\States,\tau\in\{\sigma\subst{x}{\sigma(\ee)}\}}f (\sigma)\\
		\eeq &	\inf_{\sigma\in\States,\tau\in\eval{\ASSIGN{x}{\ee}} (\sigma)}f (\sigma)~.
	\end{align*}%

	This concludes the proof for the atomic statements.%
	\paragraph{Induction Hypothesis:} 
	For arbitrary but fixed programs $C$, $C_1$, $C_2$, we proceed with the inductive step on the composite statements.%
	\paragraph{The sequential composition $\COMPOSE{C_1}{C_2}$:} We have%
	\begin{align*}
		\slp{\COMPOSE{C_2}{C_1}}{f}(\tau) 
		\eeq & \slp{C_2}{\slp{C_1}{f}}(\tau) \\
		\eeq & \inf_{\sigma'\in\States,\tau\in\eval{C_2} (\sigma')} \slp{C_1}{f}(\sigma') \tag{by I.H.~on $C_2$}\\
		\eeq & \inf_{\sigma\in\States,\tau\in\eval{C_2} (\sigma')\land \sigma'\in\eval{C_1} (\sigma)} {f}(\sigma) \tag{by I.H.~on $C_2$}\\
		\eeq &	\inf_{\sigma\in\States,\tau\in\eval{C_2} (\eval{C_1} (\sigma))}f (\sigma)\\
		\eeq &	\inf_{\sigma\in\States, \tau\in\eval{\COMPOSE{C_1}{C_2}} (\sigma)}f (\sigma)~.
	\end{align*}%
	\noindent{}%
	\paragraph{The conditional branching $\ITE{\guard}{C_1}{C_2}$:} We have%
	\begin{align*}
		&\slp{\ITE{\guard}{C_1}{C_2}}{f}(\tau) \\
		& \eeq\bigl(\slp{C_1}{\inguard\curlyvee f} \ccurlywedge \slp{C_2}{\iguard \curlyvee f}\bigr)(\tau)\\
		& \eeq 
		\inf_{\sigma\in\States,\tau\in \eval{C_1} (\sigma)}(\inguard\curlyvee f)(\sigma)
		\ccurlywedge
		\inf_{\sigma\in\States,\tau\in \eval{C_2} (\sigma)}(\iguard \curlyvee f) (\sigma)\tag{by I.H.~on $C_1, C_2$}\\
		& \eeq 
		\inf_{\sigma\in\States,\tau\in (\eval{C_1} \circ\eval{\guard})(\sigma)}f(\sigma)
		\ccurlywedge
		\inf_{\sigma\in\States,\tau\in (\eval{C_2} \circ\eval{\neg\guard})(\sigma)}f(\sigma)\\
		&\eeq
		\inf_{\sigma\in\States,\tau\in  (\eval{C_1} \circ \eval{\guard}) (\sigma) \cup ( \eval{C_2} \circ \eval{\neg\guard})(\sigma)}f (\sigma)\\
		&\eeq	\inf_{\sigma\in\States,\tau\in\eval{\ITE{\guard}{C_1}{C_2}} (\sigma)}f (\sigma)~.
	\end{align*}%
	\noindent{}%
	\paragraph{The nondeterministic choice $\NDCHOICE{C_1}{C_2}$:} We have%
	\begin{align*}
		\slp{\NDCHOICE{C_1}{C_2}}{f}(\tau)
		\eeq & \bigl(\slp{C_1}{f} \ccurlywedge \slp{C_2}{f}\big)(\tau) \\
		\eeq &	\inf_{\sigma\in\States, \tau\in  \eval{C_1} (\sigma)} f(\sigma) \ccurlywedge \inf_{\sigma\in\States,\tau\in  \eval{C_2} (\sigma)} f(\sigma) \tag{by I.H.~on $C_1,C_2$}\\
		\eeq&	\inf_{\sigma\in\States, \tau\in  \eval{C_1} (\sigma) \cup \eval{C_2} (\sigma)}f (\sigma)\\
		\eeq&	\inf_{\sigma\in\States, \tau\in\eval{\NDCHOICE{C_1}{C_2}} (\sigma)}f (\sigma)~.
	\end{align*}%
	\noindent{}%
	
	\paragraph{The loop $\WHILEDO{\guard}{C}$:} 
	Let%
	\begin{align*}
		\Psi_{f}(X) &\eeq f \ccurlywedge \slp{C}{\inguard\curlyvee X}~,
	\end{align*}%
	be the $\slpsymbol$-characteristic functions of the loop $\WHILEDO{\guard}{C}$ with respect to preanticipation $f$ and
	\begin{align*}
		F_S(X) &\eeq  S\cup (\eval{C}\circ\eval{\guard})X~,
	\end{align*}%
	be the collecting semantics characteristic functions of the loop $\WHILEDO{\guard}{C}$ with respect to any input $S\in\powerset{\Conf}$.
	We now prove by induction on $n$ that, for all $\tau\in\States$
	\begin{equation}
		\label{eq:slp-soundness-induction}
		\Psi_{f}^{n}(\pinfty )(\tau)\eeq\inf_{\sigma\in\States,\tau\in F_{\{\sigma\}}^n(\emptyset)}f (\sigma)~.
	\end{equation}%
	For the induction base $n = 0$, consider the following:%
	\begin{align*}
		\Psi_{f}^{0}(\pinfty )(\tau)
		& \eeq  \pinfty \\
		& \eeq \inf \emptyset \\
		& \eeq \inf_{\sigma\in\States,\tau\in\emptyset} f(\sigma)\\
		& \eeq \inf_{\sigma\in\States,\tau\in F_{\{\sigma\}}^0(\emptyset)}f (\sigma)~.
	\end{align*}%
	As induction hypothesis, we have for arbitrary but fixed $n$ and all $\tau\in\States$%
	\begin{align*}
		\Psi_{f}^{n}(\pinfty )(\tau)\eeq\inf_{\sigma\in\States,\tau\in F_{\{\sigma\}}^n(\emptyset)}f (\sigma)~.
	\end{align*}%
	For the induction step $n \longrightarrow n + 1$, consider the following:%
	\begin{align*}
		&	\Psi_{f}^{n+1}(\pinfty )(\tau)  \\
		& \eeq \left(f \ccurlywedge \slp{C}{\inguard\curlyvee \Psi_{f}^n(\pinfty )} \right)(\tau)\\
		& \eeq  f(\tau) \ccurlywedge
		\inf_{\sigma\in\States, \tau\in\eval{C} (\sigma)}\bigl(\inguard\curlyvee  \Psi_{f}^n(\pinfty )\bigr) (\sigma)
		\tag{by I.H.\ on $C$}\\
		& \eeq  f(\tau) \ccurlywedge
		\inf_{\sigma\in\States, \tau\in\eval{C} (\sigma)}~
		\inf_{\sigma'\in\States,\sigma\in\eval{\guard}  F_{\{\sigma'\}}^n(\emptyset)}f (\sigma')
		\tag{by I.H.\ on $n$} 
		\\
		& \eeq  f(\tau) \ccurlywedge
		\inf_{\sigma'\in\States,\tau\in(\eval{C}  \circ  \eval{\guard}  )F_{\{\sigma'\}}^n(\emptyset)}f (\sigma')
		\\
		& \eeq 
		\inf_{\sigma'\in\States,\tau\in(\eval{C}  \circ  \eval{\guard}  )F_{\{\sigma'\}}^n(\emptyset)\cup \{\sigma'\}}f (\sigma')
		\\
		&
		\eeq \inf_{\sigma\in\States,\tau\in F_{\{\sigma\}}^{n+1}(\emptyset)}f (\sigma)~.
	\end{align*}
	
	This concludes the induction on $n$. Now we have:

	\begin{align*}
		\slp{\WHILEDO{\guard}{C}}{f}(\tau) 
		\eeq & \left(\iguard \curlyvee\bigl(\gfp  X\mydot f \ccurlywedge \slp{C}{\inguard\curlyvee X}\bigr)\right)(\tau)\\
		\eeq & \bigl(\iguard\curlyvee\inf_{n\in\mathbb{N}}~\Psi_{f}^{n}(\pinfty )\bigr)(\tau) \tag{by Kleene's fixpoint theorem}\\
		\eeq & \inf_{n\in\mathbb{N}}~\bigl(\iguard \curlyvee\Psi_{f}^{n}(\pinfty )\bigr)(\tau) \tag{by co-continuity of $\lambda X\mydot\iguard\curlyvee X$}\\
		\eeq&	\inf_{n\in\mathbb{N}}~\inf_{\sigma\in\States,\tau\in\eval{\neg\guard}  F_{\{\sigma\}}^n(\emptyset)}f (\sigma)\tag{by Equation~\ref{eq:slp-soundness-induction}}\\
		\eeq&	\inf_{\sigma\in\States,\tau\in	\cup_{n\in\mathbb{N}}(\eval{\neg\guard}  F_{\{\sigma\}}^n(\emptyset))}f (\sigma)\\
		\eeq&	\inf_{\sigma\in\States,\tau\in	\eval{\neg\guard}  (\cup_{n\in\mathbb{N}}F_{\{\sigma\}}^n(\emptyset))}f (\sigma) \tag{by continuity of $\eval{\neg\guard} $}\\
		\eeq&	\inf_{\sigma\in\States, \tau\in\eval{\neg \guard}  (\lfp X\mydot \{\sigma\} \cup (\eval{C} \circ \eval{\guard})X )}f (\sigma)\tag{by Kleene's fixpoint theorem}\\		
		\eeq&	\inf_{\sigma\in\States,\tau\in\eval{\WHILEDO{\guard}{C}}  (\sigma)}f (\sigma)~,
	\end{align*}
	and this concludes the proof.
\end{proof}
\section{Proofs of Section~\ref{se:healthiness}}
\subsection{Proof of Healthiness Properties of Quantitative Transformers, Theorem~\ref{thm:wpwlpspslphealthiness}}
Each of the properties is proven individually below.
\begin{itemize}
	\item Quantitative universal conjunctiveness: Theorem~\ref{thm:wpconjunctive},~\ref{thm:spconjunctive};
	\item Quantitative universal disjunctiveness: Theorem~\ref{thm:wlpdisjunctive},~\ref{thm:slpdisjunctive};
	\item Strictness: Corollary~\ref{thm:wpstrict},~\ref{thm:spstrict};
	\item Costrictness: Corollary~\ref{thm:wlpstrict},~\ref{thm:slpstrict};
	\item Monotonicity: Corollary~\ref{thm:mono}
\end{itemize}
\begin{restatable}[Quantitative universal conjunctiveness of \textnormal{$\wpsymbol$}]{theorem}{wpconjunctive}
	\label{thm:wpconjunctive}
	For any set of quantities \mbox{$\subseteq \A$},%
	\begin{align*}
		\wp{C}{\sup S} \qeq \sup~ \wp{C}{S}~.
	\end{align*}%
\end{restatable}%
\begin{proof}
	We prove Theorem~\ref{thm:wpconjunctive} by induction on the structure of $C$.
	For the induction base, we have the atomic statements:%
	\paragraph{The effectless program $\SKIP$:} We have%
	\begin{align*}
		\wp{\SKIP}{\sup S}
		\eeq & \sup S \\
		\eeq & \sup_{g \in S} g \\
		\eeq & \sup_{g \in S}~ \wp{\SKIP}{g} \\
		\eeq & \sup~ \wp{\SKIP}{S}~.
	\end{align*}%
	\paragraph{The assignment $\ASSIGN{x}{e}$:} We have%
	\begin{align*}
		\wp{\ASSIGN{x}{\ee}}{\sup S}
		\eeq & (\sup S)\subst{x}{\ee} \\
		\eeq &\left(\lambda \sigma\mydot \sup_{g \in S} g(\sigma) \right)\subst{x}{\ee} \\
		\eeq & \left(\lambda \sigma\mydot \sup_{g \in S} g\subst{x}{\ee}(\sigma) \right) \\
		\eeq & \sup_{g \in S} g\subst{x}{\ee} \\
		\eeq & \sup_{g \in S} ~\wp{\ASSIGN{x}{e}}{g} \\
		\eeq & \sup~ \wp{\ASSIGN{x}{\ee}}{ S}~.
	\end{align*}%
	This concludes the proof for the atomic statements.%
	\paragraph{Induction Hypothesis:} 
	For arbitrary but fixed programs $C$, $C_1$, $C_2$, Theorem~\ref{thm:wpconjunctive} holds.
	
	We proceed with the inductive step on the composite statements.%
	\paragraph{The sequential composition $\COMPOSE{C_1}{C_2}$:} We have%
	\begin{align*}
		\wp{\COMPOSE{C_1}{C_2}}{\sup S} 
		\eeq & \wp{C_1}{\wp{C_2}{\sup S}}   \\
		\eeq & \wp{C_1}{\sup~ \wp{C_2}{S}} \tag{by I.H.~on $C_1$}\\
		\eeq & \sup~ \wp{C_1}{\wp{C_2}{S}} \tag{by I.H.~on $C_2$}\\
		\eeq & \sup~ \wp{\COMPOSE{C_1}{C_2}}{S} ~.
	\end{align*}%
	\paragraph{The conditional branching $\ITE{\guard}{C_1}{C_2}$:} Here we reason in the reverse direction from the cases before. We have%
	\begin{align*}
		& \wp{\ITE{\guard}{C_1}{C_2}}{\sup S} \\
		& \eeq \iverson{\guard}\curlywedge \wp{C_1}{\sup S} \ccurlyvee \iverson{\neg \guard}\curlywedge \wp{C_2}{\sup S}\\
		& \eeq  \iverson{\guard}\curlywedge \sup~ \wp{C_1}{S} \ccurlyvee\iverson{\neg\guard}\curlywedge \sup~ \wp{C_2}{S} \tag{by I.H.~on $C_1$ and $C_2$} \\
		& \eeq  \sup~ \bigl(\iverson{\guard}\curlywedge \wp{C_1}{S} \bigr) \ccurlyvee\sup~\bigl(\iverson{\neg\guard}\curlywedge  \wp{C_2}{S}\bigr)\\
		& \eeq  \sup~ \bigl(\iverson{\guard}\curlywedge \wp{C_1}{S}  \ccurlyvee \iverson{\neg\guard}\curlywedge  \wp{C_2}{S}\bigr)\\
		& \eeq \sup~  \wp{\ITE{\guard}{C_1}{C_2}}{S}~.
	\end{align*}%

	\noindent{}%
	\paragraph{The loop $\WHILEDO{\guard}{C}$:} 
	Let%
	\begin{align*}
		\Phi_{f}(X) &\eeq \iverson{\neg\guard}\curlywedge f \ccurlyvee \iverson{\guard}\curlywedge \wp{C}{X}~,
	\end{align*}%
	be the $\wpsymbol$-characteristic function of the loop $\WHILEDO{\guard}{C}$ with respect to any postanticipation $f\in\A$ and
	\begin{align*}
		F_S(X) &\eeq  S\cup (\eval{C}\circ\eval{\guard})X~,
	\end{align*}%
	be the collecting semantics characteristic functions of the loop $\WHILEDO{\guard}{C}$ with respect to any input $S\in\powerset{\Conf}$. Observe that $\Phi_f(X)$ is continuous by inductive hypothesis on $C$ and by composition of continuous functions.
	We now prove by induction on $n$ that
	\begin{equation}
		\label{eq:wp-continuity-induction}
		\Phi_{\sup S}^{n}(\minfty)\eeq \sup _{g\in S}\Phi_g^n(\minfty)~.
	\end{equation}%
	For the induction base $n = 0$, consider the following:%
	\begin{align*}
		\Phi_{\sup S}^{0}(\minfty)\eeq
		& \eeq  \minfty\\
		& \eeq  \sup _{g\in S}\minfty\\
		& \eeq  \sup _{g\in S}\Phi_g^0(\minfty)~.
	\end{align*}%
	As induction hypothesis, we have for arbitrary but fixed $n$
	\begin{align*}
		\Phi_{\sup S}^{n}(\minfty)\eeq \sup _{g\in S}\Phi_g^n(\minfty)~.
	\end{align*}%
	For the induction step $n \longrightarrow n + 1$, consider the following:%
	\begin{align*}
		&\Phi_{\sup S}^{n+1}(\minfty)  \\
		& \eeq  \iverson{\neg\guard}\curlywedge \sup S \ccurlyvee \iverson{\guard}\curlywedge \wp{C}{\Phi_{\sup S}^n(\minfty)}\\
		& \eeq  \iverson{\neg\guard}\curlywedge \sup S \ccurlyvee \iverson{\guard}\curlywedge \wp{C}{ \sup _{g\in S}\Phi_g^n(\minfty)}\tag{by I.H.\ on $n$} \\
		& \eeq  \iverson{\neg\guard}\curlywedge \sup S \ccurlyvee \iverson{\guard}\curlywedge  \sup _{g\in S}~\wp{C}{\Phi_g^n(\minfty) }\tag{by I.H.\ on $C$} \\
		& \eeq  \sup _{g\in S}~(\iverson{\neg\guard}\curlywedge g )\ccurlyvee  \sup _{g\in S}~\left(\iverson{\guard}\curlywedge \wp{C}{\Phi_g^n(\minfty)}\right)\\
		& \eeq  \sup_{g\in S} \left(  \iverson{\neg\guard}\curlywedge g \ccurlyvee \iverson{\guard}\curlywedge \wp{C}{\Phi_{g}^n(\minfty)} \right)\\
		& \eeq \sup _{g\in S}\Phi_g^{n+1}(\minfty)~.
	\end{align*}
	
	This concludes the induction on $n$. Now we have:

	\begin{align*}
		\wp{\WHILEDO{\guard}{C}}{\sup S} 
		\eeq & \lfp  X\mydot \iverson{\neg\guard}\curlywedge \sup S \ccurlyvee \iverson{\guard}\curlywedge \wp{C}{X}\\
		\eeq & \sup_{n\in\mathbb{N}}~\Phi_{\sup S}^{n}(\minfty)\tag{by Kleene's fixpoint theorem}\\
		\eeq&	\sup_{n\in\mathbb{N}}~\sup _{g\in S}~\Phi_g^n(\minfty)\tag{by Equation~\ref{eq:wp-continuity-induction}}\\
		\eeq&	\sup _{g\in S}~\sup_{n\in\mathbb{N}}~\Phi_g^n(\minfty)\\
		\eeq&	\sup _{g\in S} \wp{\WHILEDO{\guard}{C}}{g}\tag{by Kleene's fixpoint theorem}\\
		\eeq&	\sup \wp{\WHILEDO{\guard}{C}}{S},~
	\end{align*}
	and this concludes the proof.
\end{proof}

\begin{restatable}[Quantitative universal conjunctiveness of \textnormal{$\spsymbol$}]{theorem}{spconjunctive}
	\label{thm:spconjunctive}
	For any set of quantities \mbox{$\subseteq \A$},%
	\begin{align*}
		\sp{C}{\sup S} \qeq \sup~ \sp{C}{S}~.
	\end{align*}%
\end{restatable}%
\begin{proof}
	We prove Theorem~\ref{thm:spconjunctive} by induction on the structure of $C$.
	For the induction base, we have the atomic statements:%
	\paragraph{The effectless program $\SKIP$:} We have%
	\begin{align*}
		\sp{\SKIP}{\sup S}
		\eeq & \sup S \\
		\eeq & \sup_{g \in S} g \\
		\eeq & \sup_{g \in S}~ \sp{\SKIP}{g} \\
		\eeq & \sup~ \sp{\SKIP}{S}~.
	\end{align*}%
	\paragraph{The assignment $\ASSIGN{x}{e}$:} We have%
	\begin{align*}
		\sp{\ASSIGN{x}{e}}{\sup S}
		\eeq & \SupV{\alpha}~ \iverson{x = e\subst{x}{\alpha}}\curlywedge (\sup S)\subst{x}{\alpha} \\
		\eeq & \SupV{\alpha}~ \iverson{x = e\subst{x}{\alpha}}\curlywedge \left(\lambda \sigma\mydot \sup_{g \in S} g(\sigma) \right)\subst{x}{\alpha} \\
		\eeq & \SupV{\alpha}~ \iverson{x = e\subst{x}{\alpha}}\curlywedge \left(\lambda \sigma\mydot \sup_{g \in S} g\subst{x}{\alpha}(\sigma) \right) \\
		\eeq & \SupV{\alpha}~ \iverson{x = e\subst{x}{\alpha}}\curlywedge \sup_{g \in S} g\subst{x}{\alpha} \\
		\eeq & \SupV{\alpha}~ \sup_{g \in S} \iverson{x = e\subst{x}{\alpha}}\curlywedge g\subst{x}{\alpha} \\
		\eeq & \sup_{g \in S} ~ \SupV{\alpha}~ \iverson{x = e\subst{x}{\alpha}}\curlywedge g\subst{x}{\alpha} \\
		\eeq & \sup_{g \in S} ~\sp{\ASSIGN{x}{e}}{g} \\
		\eeq & \sup~ \sp{\ASSIGN{x}{e}}{S}~.
	\end{align*}%
	This concludes the proof for the atomic statements.%
	\paragraph{Induction Hypothesis:} 
	For arbitrary but fixed programs $C$, $C_1$, $C_2$, Theorem~\ref{thm:spconjunctive} holds.
	
	We proceed with the inductive step on the composite statements.%
	\paragraph{The sequential composition $\COMPOSE{C_1}{C_2}$:} We have%
	\begin{align*}
		\sp{\COMPOSE{C_1}{C_2}}{\sup S} 
		\eeq & \sp{C_2}{\sp{C_1}{\sup S}}   \\
		\eeq & \sp{C_2}{\sup~ \sp{C_1}{S}} \tag{by I.H.~on $C_1$}\\
		\eeq & \sup~ \sp{C_2}{\sp{C_1}{S}} \tag{by I.H.~on $C_2$}\\
		\eeq & \sup~ \sp{\COMPOSE{C_1}{C_2}}{S} ~.
	\end{align*}%
	\paragraph{The conditional branching $\ITE{\guard}{C_1}{C_2}$:} We have%
	\begin{align*}
		& \sp{\ITE{\guard}{C_1}{C_2}}{\sup S} \\
		& \eeq \sp{C_1}{\iverson{\guard}\curlywedge \sup S} \ccurlyvee \sp{C_2}{\inguard \curlywedge\sup S} \\
		& \eeq \sp{C_1}{\sup \iverson{\guard}\curlywedge S} \ccurlyvee \sp{C_2}{\sup \inguard \curlywedge S} \\
		& \eeq \sup~ \sp{C_1}{\iverson{\guard}\curlywedge S} \ccurlyvee \sup~ \sp{C_2}{\iverson{\neg\guard}\curlywedge S} \tag{by I.H.~on $C_1$ and $C_2$} \\
		& \eeq \sup~ \bigl(\sp{C_1}{\iverson{\guard}\curlywedge S} \ccurlyvee \sp{C_2}{\iverson{\neg\guard}\curlywedge S} \bigr) \\
		& \eeq \sup~  \sp{\ITE{\guard}{C_1}{C_2}}{S}~.
	\end{align*}%

	\noindent{}%
	\paragraph{The loop $\WHILEDO{\guard}{C}$:} 
	Let%
	\begin{align*}
		\Psi_{f}(X) &\eeq f \ccurlyvee \sp{C}{\iverson{\guard}\curlywedge X}~,
	\end{align*}%
	be the $\spsymbol$-characteristic function of the loop $\WHILEDO{\guard}{C}$ with respect to any preanticipation $f\in\A$ and
	\begin{align*}
		F_S(X) &\eeq  S\cup (\eval{C}\circ\eval{\guard})X~,
	\end{align*}%
	be the collecting semantics characteristic functions of the loop $\WHILEDO{\guard}{C}$ with respect to any input $S\in\powerset{\Conf}$. Observe that $\Psi_f(X)$ is continuous by inductive hypothesis on $C$ and by composition of continuous functions.
	We now prove by induction on $n$ that
	\begin{equation}
		\label{eq:sp-continuity-induction}
		\Psi_{\sup S}^{n}(\minfty )\eeq \sup _{g\in S}\Psi_g^n(\minfty )~.
	\end{equation}%
	For the induction base $n = 0$, consider the following:%
	\begin{align*}
		\Psi_{\sup S}^{0}(\minfty )\eeq
		& \eeq  \minfty \\
		& \eeq  \sup _{g\in S}\minfty \\
		& \eeq  \sup _{g\in S}\Psi_g^0(\minfty )~.
	\end{align*}%
	As induction hypothesis, we have for arbitrary but fixed $n$
	\begin{align*}
		\Psi_{\sup S}^{n}(\minfty )\eeq \sup _{g\in S}\Psi_g^n(\minfty )~.
	\end{align*}%
	For the induction step $n \longrightarrow n + 1$, consider the following:%
	\begin{align*}
		&		\Psi_{\sup S}^{n+1}(\minfty )  \\
		& \eeq  \sup S \ccurlyvee \sp{C}{\iverson{\guard}\curlywedge \Psi_{\sup S}^n(\minfty )}\\
		& \eeq  \sup S \ccurlyvee \sp{C}{\iverson{\guard}\curlywedge  \sup _{g\in S}\Psi_g^n(\minfty )}\tag{by I.H.\ on $n$} \\
		& \eeq  \sup S \ccurlyvee \sp{C}{\sup _{g\in S}\iverson{ \guard}\curlywedge  \Psi_g^n(\minfty )} \\
		& \eeq  \sup S \ccurlyvee\sup _{g\in S} \sp{C}{\iverson{ \guard}\curlywedge  \Psi_g^n(\minfty )}\tag{by I.H.\ on $C$} \\
		& \eeq  \sup_{g\in S} g \ccurlyvee\sup _{g\in S} \sp{C}{\iverson{ \guard}\curlywedge  \Psi_g^n(\minfty )}\\
		& \eeq  \sup_{g\in S} \left(g \ccurlyvee \sp{C}{\iverson{ \guard}\curlywedge  \Psi_g^n(\minfty )}\right)\\
		& \eeq \sup _{g\in S}\Psi_g^{n+1}(\minfty )~.
	\end{align*}
	
	This concludes the induction on $n$. Now we have:

	\begin{align*}
		\sp{\WHILEDO{\guard}{C}}{\sup S} 
		\eeq & \inguard\curlywedge \bigl(\lfp  X\mydot \sup S \ccurlyvee \sp{C}{\iverson{\guard}\curlywedge X}\bigr)\\
		\eeq & \inguard\curlywedge \sup_{n\in\mathbb{N}}~\Psi_{\sup S}^{n}(\minfty )\tag{by Kleene's fixpoint theorem}\\
		\eeq&	\inguard\curlywedge \sup_{n\in\mathbb{N}}~\sup _{g\in S}\Psi_g^{n}(\minfty )\tag{by Equation~\ref{eq:sp-continuity-induction}}\\
		\eeq&	\inguard\curlywedge \sup _{g\in S}~\sup_{n\in\mathbb{N}}\Psi_g^{n}(\minfty )\\
		\eeq&	\inguard\curlywedge \sup _{g\in S}~\sup_{n\in\mathbb{N}}\Psi_g^{n}(\minfty )\\
		\eeq&	\sup _{g\in S} (\inguard\curlywedge \sup_{n\in\mathbb{N}}\Psi_g^{n}(\minfty ))\\
		\eeq&	\sup _{g\in S} \sp{\WHILEDO{\guard}{C}}{g}\tag{by Kleene's fixpoint theorem}\\
		\eeq&	\sup \sp{\WHILEDO{\guard}{C}}{S},~
	\end{align*}
	and this concludes the proof.
\end{proof}
\begin{restatable}[Quantitative universal disjunctiveness of \textnormal{$\wlpsymbol$}]{theorem}{wlpdisjunctive}
	\label{thm:wlpdisjunctive}
	For any set of quantities \mbox{$\subseteq \A$},%
	\begin{align*}
		\wlp{C}{\inf S} \qeq \inf~ \wlp{C}{S}~.
	\end{align*}%
\end{restatable}%
\begin{proof}
	We prove Theorem~\ref{thm:wlpdisjunctive} by induction on the structure of $C$.
	For the induction base, we have the atomic statements:%
	\paragraph{The effectless program $\SKIP$:} We have%
	\begin{align*}
		\wlp{\SKIP}{\inf S}
		\eeq & \inf S \\
		\eeq & \inf_{g \in S} g \\
		\eeq & \inf_{g \in S}~ \wlp{\SKIP}{g} \\
		\eeq & \inf~ \wlp{\SKIP}{S}~.
	\end{align*}%
	\paragraph{The assignment $\ASSIGN{x}{e}$:} We have%
	\begin{align*}
		\wlp{\ASSIGN{x}{\ee}}{\inf S}
		\eeq & (\inf S)\subst{x}{\ee} \\
		\eeq &\left(\lambda \sigma\mydot \inf_{g \in S} g(\sigma) \right)\subst{x}{\ee} \\
		\eeq & \left(\lambda \sigma\mydot \inf_{g \in S} g\subst{x}{\ee}(\sigma) \right) \\
		\eeq & \inf_{g \in S} g\subst{x}{\ee} \\
		\eeq & \inf_{g \in S} ~\wlp{\ASSIGN{x}{e}}{g} \\
		\eeq & \inf~ \wlp{\ASSIGN{x}{\ee}}{ S}~.
	\end{align*}%
	This concludes the proof for the atomic statements.%
	\paragraph{Induction Hypothesis:} 
	For arbitrary but fixed programs $C$, $C_1$, $C_2$, Theorem~\ref{thm:wlpdisjunctive} holds.
	
	We proceed with the inductive step on the composite statements.%
	\paragraph{The sequential composition $\COMPOSE{C_1}{C_2}$:} We have%
	\begin{align*}
		\wlp{\COMPOSE{C_1}{C_2}}{\inf S} 
		\eeq & \wlp{C_1}{\wlp{C_2}{\inf S}}   \\
		\eeq & \wlp{C_1}{\inf~ \wlp{C_2}{S}} \tag{by I.H.~on $C_1$}\\
		\eeq & \inf~ \wlp{C_1}{\wlp{C_2}{S}} \tag{by I.H.~on $C_2$}\\
		\eeq & \inf~ \wlp{\COMPOSE{C_1}{C_2}}{S} ~.
	\end{align*}%
	\paragraph{The conditional branching $\ITE{\guard}{C_1}{C_2}$:} We have%
	\begin{align*}
		& \wlp{\ITE{\guard}{C_1}{C_2}}{\inf S} \\
		& \eeq \iverson{\guard} \curlywedge \wlp{C_1}{\inf S} \ccurlyvee \iverson{\neg \guard} \curlywedge \wlp{C_2}{\inf S}\\
		& \eeq  \iverson{\guard} \curlywedge \inf~ \wlp{C_1}{S} \ccurlyvee\iverson{\neg\guard} \curlywedge \inf~ \wlp{C_2}{S} \tag{by I.H.~on $C_1$ and $C_2$} \\
		& \eeq  \inf~ \bigl(\iverson{\guard} \curlywedge \wlp{C_1}{S} \bigr) \ccurlyvee\inf~\bigl(\iverson{\neg\guard} \curlywedge  \wlp{C_2}{S}\bigr)\\
		& \eeq \lambda \sigma\mydot \begin{cases}
			\inf~ \bigl( \wlp{C_1}{S} \bigr) & \text{ if } \sigma\mmodels\guard\\
			\inf~\bigl( \wlp{C_2}{S}\bigr) & \text{ otherwise }
		\end{cases}\\
		& \eeq  \inf~ \bigl(\iverson{\guard} \curlywedge \wlp{C_1}{S}  \ccurlyvee \iverson{\neg\guard} \curlywedge  \wlp{C_2}{S}\bigr)\\
		& \eeq \inf~  \wlp{\ITE{\guard}{C_1}{C_2}}{S}~.
	\end{align*}%

	\noindent{}%
	\paragraph{The loop $\WHILEDO{\guard}{C}$:} 
	Let%
	\begin{align*}
		\Phi_{f}(X) &\eeq \iverson{\neg\guard} \curlywedge f \ccurlyvee \iverson{\guard} \curlywedge \wlp{C}{X}~,
	\end{align*}%
	be the $\wlpsymbol$-characteristic function of the loop $\WHILEDO{\guard}{C}$ with respect to any postanticipation $f\in\A$ and
	\begin{align*}
		F_S(X) &\eeq  S\cup (\eval{C}\circ\eval{\guard})X~,
	\end{align*}%
	be the collecting semantics characteristic functions of the loop $\WHILEDO{\guard}{C}$ with respect to any input $S\in\powerset{\Conf}$. Observe that $\Phi_f(X)$ is continuous by inductive hypothesis on $C$ and by composition of continuous functions.
	We now prove by induction on $n$ that
	\begin{equation}
		\label{eq:wlp-continuity-induction}
		\Phi_{\inf S}^{n}(\pinfty)\eeq \inf _{g\in S}\Phi_g^n(\pinfty)~.
	\end{equation}%
	For the induction base $n = 0$, consider the following:%
	\begin{align*}
		\Phi_{\inf S}^{0}(\pinfty)\eeq
		& \eeq  \pinfty\\
		& \eeq  \inf _{g\in S}\pinfty\\
		& \eeq  \inf _{g\in S}\Phi_g^0(\pinfty)~.
	\end{align*}%
	As induction hypothesis, we have for arbitrary but fixed $n$
	\begin{align*}
		\Phi_{\inf S}^{n}(\pinfty)\eeq \inf _{g\in S}\Phi_g^n(\pinfty)~.
	\end{align*}%
	For the induction step $n \longrightarrow n + 1$, consider the following:%
	\begin{align*}
		&\Phi_{\inf S}^{n+1}(\pinfty)  \\
		& \eeq  \iverson{\neg\guard} \curlywedge \inf S \ccurlyvee \iverson{\guard} \curlywedge \wlp{C}{\Phi_{\inf S}^n(\pinfty)}\\
		& \eeq  \iverson{\neg\guard} \curlywedge \inf S \ccurlyvee \iverson{\guard} \curlywedge \wlp{C}{ \inf _{g\in S}\Phi_g^n(\pinfty)}\tag{by I.H.\ on $n$} \\
		& \eeq  \iverson{\neg\guard} \curlywedge \inf S \ccurlyvee \iverson{\guard} \curlywedge  \inf _{g\in S}~\wlp{C}{\Phi_g^n(\pinfty) }\tag{by I.H.\ on $C$} \\
		& \eeq  \inf _{g\in S}~(\iverson{\neg\guard} \curlywedge g )\ccurlyvee  \inf _{g\in S}~\left(\iverson{\guard} \curlywedge \wlp{C}{\Phi_g^n(\pinfty)}\right)\\
		& \eeq 
		\lambda\sigma\mydot 
		\begin{cases}
			\inf _{g\in S}~\left(\wlp{C}{\Phi_g^n(\pinfty)}\right)	&\text{ if } \sigma\mmodels \guard\\
			\inf _{g\in S}~( g )	&\text{ otherwise }
		\end{cases}\\
		& \eeq  \inf_{g\in S} \left(  \iverson{\neg\guard} \curlywedge g \ccurlyvee \iverson{\guard} \curlywedge \wlp{C}{\Phi_{g}^n(\pinfty)} \right)\\
		& \eeq \inf _{g\in S}\Phi_g^{n+1}(\pinfty)~.
	\end{align*}
	
	This concludes the induction on $n$. Now we have:

	\begin{align*}
		\wlp{\WHILEDO{\guard}{C}}{\inf S} 
		\eeq & \gfp  X\mydot \iverson{\neg\guard} \curlywedge \inf S \ccurlyvee \iverson{\guard} \curlywedge \wlp{C}{X}\\
		\eeq & \inf_{n\in\mathbb{N}}~\Phi_{\inf S}^{n}(\pinfty)\tag{by Kleene's fixpoint theorem}\\
		\eeq&	\inf_{n\in\mathbb{N}}~\inf _{g\in S}~\Phi_g^n(\pinfty)\tag{by Equation~\ref{eq:wlp-continuity-induction}}\\
		\eeq&	\inf _{g\in S}~\inf_{n\in\mathbb{N}}~\Phi_g^n(\pinfty)\\
		\eeq&	\inf _{g\in S} \wlp{\WHILEDO{\guard}{C}}{g}\tag{by Kleene's fixpoint theorem}\\
		\eeq&	\inf \wlp{\WHILEDO{\guard}{C}}{S},~
	\end{align*}
	and this concludes the proof.
\end{proof}

\begin{restatable}[Quantitative universal disjunctiveness of \textnormal{$\slpsymbol$}]{theorem}{slpdisjunctive}
	\label{thm:slpdisjunctive}
	For any set of quantities \mbox{$\subseteq \A$},%
	\begin{align*}
		\slp{C}{\inf S} \qeq \inf~ \slp{C}{S}~.
	\end{align*}%
\end{restatable}%
\begin{proof}
	We prove Theorem~\ref{thm:slpdisjunctive} by induction on the structure of $C$.
	For the induction base, we have the atomic statements:%
	\paragraph{The effectless program $\SKIP$:} We have%
	\begin{align*}
		\slp{\SKIP}{\inf S}
		\eeq & \inf S \\
		\eeq & \inf_{g \in S} g \\
		\eeq & \inf_{g \in S}~ \slp{\SKIP}{g} \\
		\eeq & \inf~ \slp{\SKIP}{S}~.
	\end{align*}%
	\paragraph{The assignment $\ASSIGN{x}{e}$:} We have%
	\begin{align*}
		\slp{\ASSIGN{x}{e}}{\inf S}
		\eeq & \InfV{\alpha}~ \iverson{x \neq e\subst{x}{\alpha}} \curlyvee (\inf S)\subst{x}{\alpha} \\
		\eeq & \InfV{\alpha}~ \iverson{x \neq e\subst{x}{\alpha}} \curlyvee \left(\lambda \sigma\mydot \inf_{g \in S} g(\sigma) \right)\subst{x}{\alpha} \\
		\eeq & \InfV{\alpha}~ \iverson{x \neq e\subst{x}{\alpha}} \curlyvee \left(\lambda \sigma\mydot \inf_{g \in S} g\subst{x}{\alpha}(\sigma) \right) \\
		\eeq & \InfV{\alpha}~ \iverson{x \neq e\subst{x}{\alpha}} \curlyvee \inf_{g \in S} g\subst{x}{\alpha} \\
		\eeq & \InfV{\alpha}~ \inf_{g \in S} \iverson{x \neq e\subst{x}{\alpha}} \curlyvee g\subst{x}{\alpha} \\
		\eeq & \inf_{g \in S} ~ \InfV{\alpha}~ \iverson{x \neq e\subst{x}{\alpha}} \curlyvee g\subst{x}{\alpha} \\
		\eeq & \inf_{g \in S} ~\slp{\ASSIGN{x}{e}}{g} \\
		\eeq & \inf~ \slp{\ASSIGN{x}{e}}{S}~.
	\end{align*}%
	This concludes the proof for the atomic statements.%
	\paragraph{Induction Hypothesis:} 
	For arbitrary but fixed programs $C$, $C_1$, $C_2$, Theorem~\ref{thm:slpdisjunctive} holds.
	
	We proceed with the inductive step on the composite statements.%
	\paragraph{The sequential composition $\COMPOSE{C_1}{C_2}$:} We have%
	\begin{align*}
		\slp{\COMPOSE{C_1}{C_2}}{\inf S} 
		\eeq & \slp{C_2}{\slp{C_1}{\inf S}}   \\
		\eeq & \slp{C_2}{\inf~ \slp{C_1}{S}} \tag{by I.H.~on $C_1$}\\
		\eeq & \inf~ \slp{C_2}{\slp{C_1}{S}} \tag{by I.H.~on $C_2$}\\
		\eeq & \inf~ \slp{\COMPOSE{C_1}{C_2}}{S} ~.
	\end{align*}%
	\paragraph{The conditional branching $\ITE{\guard}{C_1}{C_2}$:} We have%
	\begin{align*}
		& \slp{\ITE{\guard}{C_1}{C_2}}{\inf S} \\
		& \eeq \slp{C_1}{\inguard\curlyvee \inf S} \ccurlywedge \slp{C_2}{\iguard\curlyvee \inf S} \\
		& \eeq \slp{C_1}{ \inf\inguard\curlyvee S} \ccurlywedge \slp{C_2}{\inf \iguard\curlyvee  S} \\
		& \eeq \inf~ \slp{C_1}{\inguard\curlyvee S} \ccurlywedge \inf~ \slp{C_2}{\iguard\curlyvee  S} \tag{by I.H.~on $C_1$ and $C_2$} \\
		& \eeq \inf~ \bigl(\slp{C_1}{\inguard\curlyvee S} \ccurlywedge \slp{C_2}{\iverson{\guard} \inguard S} \bigr) \\
		& \eeq \inf~  \slp{\ITE{\guard}{C_1}{C_2}}{S}~.
	\end{align*}%

	\noindent{}%
	\paragraph{The loop $\WHILEDO{\guard}{C}$:} 
	Let%
	\begin{align*}
		\Psi_{f}(X) &\eeq f \ccurlywedge \slp{C}{\inguard\curlyvee X}~,
	\end{align*}%
	be the $\slpsymbol$-characteristic function of the loop $\WHILEDO{\guard}{C}$ with respect to any preanticipation $f\in\A$ and
	\begin{align*}
		F_S(X) &\eeq  S\cup (\eval{C}\circ\eval{\guard})X~,
	\end{align*}%
	be the collecting semantics characteristic functions of the loop $\WHILEDO{\guard}{C}$ with respect to any input $S\in\powerset{\Conf}$. Observe that $\Psi_f(X)$ is continuous by inductive hypothesis on $C$ and by composition of continuous functions.
	We now prove by induction on $n$ that
	\begin{equation}
		\label{eq:slp-continuity-induction}
		\Psi_{\inf S}^{n}(\pinfty )\eeq \inf _{g\in S}\Psi_g^n(\pinfty )~.
	\end{equation}%
	For the induction base $n = 0$, consider the following:%
	\begin{align*}
		\Psi_{\inf S}^{0}(\pinfty )\eeq
		& \eeq  \pinfty \\
		& \eeq  \inf _{g\in S}\pinfty \\
		& \eeq  \inf _{g\in S}\Psi_g^0(\pinfty )~.
	\end{align*}%
	As induction hypothesis, we have for arbitrary but fixed $n$
	\begin{align*}
		\Psi_{\inf S}^{n}(\pinfty )\eeq \inf _{g\in S}\Psi_g^n(\pinfty )~.
	\end{align*}%
	For the induction step $n \longrightarrow n + 1$, consider the following:%
	\begin{align*}
		&		\Psi_{\inf S}^{n+1}(\pinfty )  \\
		& \eeq  \inf S \ccurlywedge \slp{C}{\inguard \curlyvee \Psi_{\inf S}^n(\pinfty )}\\
		& \eeq  \inf S \ccurlywedge \slp{C}{\inguard \curlyvee  \inf _{g\in S}\Psi_g^n(\pinfty )}\tag{by I.H.\ on $n$} \\
		& \eeq  \inf S \ccurlywedge \slp{C}{\inf _{g\in S}\inguard \curlyvee  \Psi_g^n(\pinfty )} \\
		& \eeq  \inf S \ccurlywedge\inf _{g\in S} \slp{C}{\inguard \curlyvee  \Psi_g^n(\pinfty )}\tag{by I.H.\ on $C$} \\
		& \eeq  \inf_{g\in S} g \ccurlywedge\inf _{g\in S} \slp{C}{\inguard \curlyvee  \Psi_g^n(\pinfty )}\\
		& \eeq  \inf_{g\in S} \left(g \ccurlywedge \slp{C}{\inguard \curlyvee  \Psi_g^n(\pinfty )}\right)\\
		& \eeq \inf _{g\in S}\Psi_g^{n+1}(\pinfty )~.
	\end{align*}
	
	This concludes the induction on $n$. Now we have:

	\begin{align*}
		\slp{\WHILEDO{\guard}{C}}{\inf S} 
		\eeq & \iguard \curlyvee\bigl(\gfp  X\mydot \inf S \ccurlywedge \slp{C}{\inguard\curlyvee X}\bigr)\\
		\eeq & \iguard \curlyvee\inf_{n\in\mathbb{N}}~\Psi_{\inf S}^{n}(\pinfty )\tag{by Kleene's fixpoint theorem}\\
		\eeq&	\iguard \curlyvee\inf_{n\in\mathbb{N}}~\inf _{g\in S}\Psi_g^{n}(\pinfty )\tag{by Equation~\ref{eq:slp-continuity-induction}}\\
		\eeq&	\iguard \curlyvee\inf _{g\in S}~\inf_{n\in\mathbb{N}}\Psi_g^{n}(\pinfty )\\
		\eeq&	\iguard \curlyvee\inf _{g\in S}~\inf_{n\in\mathbb{N}}\Psi_g^{n}(\pinfty )\\
		\eeq&	\inf _{g\in S} (\iguard \curlyvee\inf_{n\in\mathbb{N}}\Psi_g^{n}(\pinfty ))\\
		\eeq&	\inf _{g\in S} \slp{\WHILEDO{\guard}{C}}{g}\tag{by Kleene's fixpoint theorem}\\
		\eeq&	\inf \slp{\WHILEDO{\guard}{C}}{S},~
	\end{align*}
	and this concludes the proof.
\end{proof}

\begin{restatable}[Strictness of \textnormal{$\wpsymbol$}]{corollary}{wpstrictness}%
	\label{thm:wpstrict}%
	For all programs $C$, $\wpC{C}$ is strict, i.e.\ 
	\begin{align*}
		\wp{C}{\minfty} \eeq \minfty~.
	\end{align*}%
\end{restatable}%
\begin{proof}
	\begin{align*}
		\wp{C}{\minfty}\eeq& \lambda \sigma\mydot \sup_{
			\tau\in \sem{C}{\sigma}} \minfty (\tau) \tag{by Theorem~\ref{thm:wp-soundness}}\\
		\eeq & \minfty~.
	\end{align*}
\end{proof}
\begin{restatable}[Strictness of \textnormal{$\spsymbol$}]{corollary}{spstrictness}%
	\label{thm:spstrict}
	For all programs $C$, $\spC{C}$ is strict, i.e.\ 
	\begin{align*}
		\sp{C}{\minfty} \eeq \minfty~.
	\end{align*}%
\end{restatable}
\begin{proof}
	\begin{align*}
		\sp{C}{\minfty }\eeq& \lambda \tau\mydot \sup_{\sigma\in\States,\tau\in \eval{C} {\sigma}} \minfty  (\sigma) \tag{by Theorem~\ref{thm:sp-soundness}}\\
		\eeq & \minfty ~.
	\end{align*}
\end{proof}
\begin{restatable}[Co-strictness of \textnormal{$\wlpsymbol$}]{corollary}{wlpstrictness}%
	\label{thm:wlpstrict}%
	For all programs $C$, $\wpC{C}$ is co-strict, i.e.\ 
	\begin{align*}
		\wlp{C}{\pinfty} \eeq \pinfty~.
	\end{align*}%
\end{restatable}%
\begin{proof}
	\begin{align*}
		\wlp{C}{\pinfty}\eeq& \lambda \sigma\mydot \inf_{
			\tau\in \sem{C}{\sigma}} \pinfty (\tau) \tag{by Theorem~\ref{thm:wlp-soundness}}\\
		\eeq & \pinfty~.
	\end{align*}
\end{proof}
\begin{restatable}[Co-strictness of \textnormal{$\slpsymbol$}]{corollary}{slpstrictness}%
	\label{thm:slpstrict}
	For all programs $C$, $\slpC{C}$ is co-strict, i.e.\ 
	\begin{align*}
		\slp{C}{\pinfty} \eeq \pinfty~.
	\end{align*}%
\end{restatable}%
\begin{proof}
	\begin{align*}
		\slp{C}{\pinfty }\eeq& \lambda \tau\mydot \inf_{\sigma\in\States,\tau\in \eval{C} {\sigma}} \pinfty  (\sigma) \tag{by Theorem~\ref{thm:slp-soundness}}\\
		\eeq & \pinfty ~.
	\end{align*}
\end{proof}


\begin{corollary}[Monotonicity of Quantitative Transformers]
	\label{thm:mono}
	For all programs $C$, $f, g \in \A$, we have%
	\begin{align*}
		f \ppreceq g \qqimplies \sfsymbol{ttt}\,\llbracket C \rrbracket\left(f\right) \ppreceq \sfsymbol{ttt}\,\llbracket C \rrbracket\left(g\right)~, \quad \textnormal{for } \sfsymbol{ttt} \in \{\wpsymbol,\, \wlpsymbol,\, \spsymbol,\, \slpsymbol\}
	\end{align*}
\end{corollary}%
\begin{proof}
	Direct consequence of universal conjunctiveness and universal disjunctiveness.
\end{proof}

\subsection{Proof of Linearity, Theorem~\ref{thm:wpwlpspslplinear}}
\wpwlpspslplinear*
\begin{proof}
	For $\wpsymbol$ we have:
	\begin{align*}
	& \wp{C}{r\cdot f +  g} \\
	&\eeq \lambda \sigma\mydot \sup_{\tau\in \eval{C} {\sigma}}
	(r\cdot f +  g)(\tau) \tag{by Theorem~\ref{thm:wp-soundness}}
	\\
	&\eeq
	\lambda \sigma\mydot \sup_{\tau\in \eval{C} {\sigma}}
	\bigl( (r\cdot f)(\tau) +  g(\tau)\bigr)\\
	&\ppreceq
	\lambda \sigma\mydot \sup_{\tau\in \eval{C} {\sigma}}
	(r\cdot f)(\tau) +  
	\sup_{\tau\in \eval{C} {\sigma}} g(\tau)\\
	&\eeq
	\lambda \sigma\mydot r\cdot \sup_{\tau\in \eval{C} {\sigma}}
	f(\tau) +  
	\sup_{\tau\in \eval{C} {\sigma}} g(\tau)
	\tag{$\sup (r\cdot A) = r\cdot \sup A$ for $A\subseteq \Reals,r\in\PosReals$}\\
	&\eeq
	r\cdot \lambda \sigma\mydot \sup_{\tau\in \eval{C} {\sigma}}
	f(\tau) +  
	\lambda \sigma\mydot
	\sup_{\tau\in \eval{C} {\sigma}} g(\tau)\\
	&\eeq  r\cdot\wp{C}{f} + \wp{C}{g} \tag{by Theorem~\ref{thm:wp-soundness}}~.
	\end{align*}
	For $\wpsymbol$ we have:
	\begin{align*}
	& \sp{C}{r\cdot f +  g} \\
	&\eeq \lambda \tau\mydot \sup_{\sigma\in\States,\tau\in \eval{C} {\sigma}}
	(r\cdot f +  g)(\sigma) \tag{by Theorem~\ref{thm:sp-soundness}}
	\\
	&\eeq
	\lambda \tau\mydot \sup_{\sigma\in\States,\tau\in \eval{C} {\sigma}}
	\bigl( (r\cdot f)(\sigma) +  g(\sigma)\bigr)\\
	&\ppreceq
	\lambda \tau\mydot \sup_{\sigma\in\States,\tau\in \eval{C} {\sigma}}
	(r\cdot f)(\sigma) +  
	\sup_{\sigma\in\States,\tau\in \eval{C} {\sigma}} g(\sigma)\\
	&\eeq
	\lambda \tau\mydot r\cdot \sup_{\sigma\in\States,\tau\in \eval{C} {\sigma}}
	f(\sigma) +  
	\sup_{\sigma\in\States,\tau\in \eval{C} {\sigma}} g(\sigma)
	\tag{$\sup (r\cdot A) = r\cdot \sup A$ for $A\subseteq \Reals,r\in\PosReals$}\\
	&\eeq
	r\cdot \lambda \tau\mydot \sup_{\sigma\in\States,\tau\in \eval{C} {\sigma}}
	f(\sigma) +  
	\lambda \tau\mydot
	\sup_{\sigma\in\States,\tau\in \eval{C} {\sigma}} g(\sigma)\\
	&\eeq  r\cdot\sp{C}{f} + \sp{C}{g} \tag{by Theorem~\ref{thm:sp-soundness}}~.
	\end{align*}
	For $\wlpsymbol$ we have:
	\begin{align*}
	& r\cdot\wlp{C}{f} + \wlp{C}{g} \\ 
	&\eeq
	r\cdot \lambda \sigma\mydot \inf_{\tau\in \eval{C} {\sigma}}
	f(\tau) +  
	\lambda \sigma\mydot
	\inf_{\tau\in \eval{C} {\sigma}} g(\tau)\tag{by Theorem~\ref{thm:wlp-soundness}}\\
	&\eeq 
	\lambda \sigma\mydot r\cdot \inf_{\tau\in \eval{C} {\sigma}}
	f(\tau) +  
	\inf_{\tau\in \eval{C} {\sigma}} g(\tau)\\
	&\eeq
	\lambda \sigma\mydot \inf_{\tau\in \eval{C} {\sigma}}
	(r\cdot f)(\tau) +  
	\inf_{\tau\in \eval{C} {\sigma}} g(\tau)
	\tag{$\inf (r\cdot A) = r\cdot \inf A$ for $A\subseteq \Reals,r\in\PosReals$}\\
	&\ppreceq 
	\lambda \sigma\mydot \inf_{\tau\in \eval{C} {\sigma}}
	\bigl( (r\cdot f)(\tau) +  g(\tau)\bigr)\\
	&\eeq
	\lambda \sigma\mydot \inf_{\tau\in \eval{C} {\sigma}}
	(r\cdot f +  g)(\tau) 
	\\
	&\eeq
	\wlp{C}{r\cdot f +  g}\tag{by Theorem~\ref{thm:wlp-soundness}}~;
	\end{align*}
	For $\slpsymbol$ we have:
	\begin{align*}
		& r\cdot\slp{C}{f} + \slp{C}{g} \\ 
		&\eeq
		r\cdot \lambda \tau\mydot \inf_{\sigma\in\States,\tau\in \eval{C} {\sigma}}
		f(\sigma) +  
		\lambda \tau\mydot
		\inf_{\sigma\in\States,\tau\in \eval{C} {\sigma}} g(\sigma)
		\tag{by Theorem~\ref{thm:slp-soundness}}\\
		&\eeq 
		\lambda \tau\mydot r\cdot \inf_{\sigma\in\States,\tau\in \eval{C} {\sigma}}
		f(\sigma) +  
		\inf_{\sigma\in\States,\tau\in \eval{C} {\sigma}} g(\sigma)\\
		&\eeq
		\lambda \tau\mydot \inf_{\sigma\in\States,\tau\in \eval{C} {\sigma}}
		(r\cdot f)(\sigma) +  
		\inf_{\sigma\in\States,\tau\in \eval{C} {\sigma}} g(\sigma)
		\tag{$\inf (r\cdot A) = r\cdot \inf A$ for $A\subseteq \Reals,r\in\PosReals$}\\
		&\ppreceq 
		\lambda \tau\mydot \inf_{\sigma\in\States,\tau\in \eval{C} {\sigma}}
		\bigl( (r\cdot f)(\sigma) +  g(\sigma)\bigr)\\
		&\eeq
		\lambda \tau\mydot \inf_{\sigma\in\States,\tau\in \eval{C} {\sigma}}
		(r\cdot f +  g)(\sigma) 
		\\
		&\eeq
		\slp{C}{r\cdot f +  g}\tag{by Theorem~\ref{thm:slp-soundness}}~.
	\end{align*}
\end{proof}

\subsection{Proof of Embedding Classical into Quantitative Transformers, Theorem~\ref{thm:wpwlpspslpembeddingpredicates}}
\wpwlpspslpembeddingpredicates*
\begin{proof}
	For $\wpsymbol$ we have:
	\begin{align*}
		\wp{C}{\indicator{F}}
		\eeq &
		\lambda\sigma\mydot
		\begin{cases}
			\indicator{F}(\tau)& \text{ if } \sem{C}{\sigma}=\{\tau\}\\
			\minfty & \text{ otherwise }
		\end{cases} \tag{by Corollary~\ref{cor:wp-wlp-deterministic}}\\
		\eeq &
		\lambda\sigma\mydot
		\begin{cases}
			\pinfty & \text{ if } \sem{C}{\sigma}=\{\tau\} \land \tau\mmodels F\\
			\minfty & \text{ otherwise }
		\end{cases} \\
		\eeq &
		\indicator{\wpD{C}{F}}~.
	\end{align*}
	For $\wlpsymbol$ we have:
	\begin{align*}
		\wlp{C}{\indicator{F}} \eeq&
		\lambda\sigma\mydot
		\begin{cases}
			\indicator{F}(\tau)& \text{ if } \sem{C}{\sigma}=\{\tau\}\\
			\pinfty & \text{ otherwise }
		\end{cases}\tag{by Corollary~\ref{cor:wp-wlp-deterministic}}\\
		\eeq &
		\lambda\sigma\mydot
		\begin{cases}
			\minfty & \text{ if } \sem{C}{\sigma}=\{\tau\} \land \tau\nnotmodels F\\
			\pinfty & \text{ otherwise }
		\end{cases} \\
		\eeq &
		\indicator{\wlpD{C}{F}}~.
	\end{align*}
	For $\spsymbol$ we have:
	\begin{align*}
		\sp{C}{\indicator{G}}
		\eeq &
		\lambda\tau\mydot
		\sup_{\sigma\in\States,\tau\in \eval{C}{\sigma}}
		\indicator{G}(\sigma) \tag{by Theorem~\ref{thm:sp-soundness}}\\
		\eeq &
		\lambda\tau\mydot
		\begin{cases}
			\pinfty & \text{ if } \exists\sigma\in\States,\tau\in\sem{C}{\sigma}\land \sigma\mmodels G\\
			\minfty & \text{ otherwise }
		\end{cases} \\
		\eeq &
		\indicator{\sp{C}{G}}~,
	\end{align*}%
	For $\slpsymbol$ we have:
	\begin{align*}
		\slp{C}{\indicator{G}}
		\eeq &
		\lambda\tau\mydot
		\inf_{\sigma\in\States,\tau\in \eval{C}{\sigma}}
		\indicator{G}(\sigma)\tag{by Theorem~\ref{thm:slp-soundness}}\\
		\eeq &
		\lambda\tau\mydot
		\begin{cases}
			\minfty & \text{ if } \exists\sigma\in\States,\tau\in\sem{C}{\sigma}\land \sigma\nnotmodels G\\
			\pinfty & \text{ otherwise }
		\end{cases} \\
		\eeq &
		\lambda\tau\mydot
		\begin{cases}
			\pinfty & \text{ if } \forall\sigma\in\States,\tau\notin\sem{C}{\sigma}\lor \sigma\mmodels G\\
			\minfty & \text{ otherwise }
		\end{cases} \\
		\eeq  &
		\lambda\tau\mydot
		\begin{cases}
			\pinfty & \text{ if } \forall\sigma\in\States,\tau\in\sem{C}{\sigma}\implies\sigma\mmodels G\\
			\minfty & \text{ otherwise }
		\end{cases} \\
		\eeq & \indicator{\slp{C}{\psi}}~.
	\end{align*}%
\end{proof}

\subsection{Proof of Liberal-Non-liberal Duality, Theorem~\ref{thm:wpwlpspslpduality}}

\wpwlpspslpduality*
\begin{proof}
	For $\wpsymbol$ and $\wlpsymbol$ we have:
	\begin{align*}
		\wp{C}{f}\eeq &
		\lambda \sigma\mydot \sup_{\tau\in \eval{C} {\sigma}}
		f(\tau) \tag{by Theorem~\ref{thm:wp-soundness}}\\
		\eeq &
		\lambda \sigma\mydot - \inf_{\tau\in \eval{C} {\sigma}}
		-f(\tau) \tag{$\sup A = -\inf (-A)$}\\
		\eeq &	-\wlp{C}{-f}~.
	\end{align*}
	For $\spsymbol$ and $\slpsymbol$ we have:
	\begin{align*}
		\sp{C}{g}\eeq &
		\lambda \tau\mydot \sup_{\sigma\in\States,\tau\in \eval{C} {\sigma}}
		g(\sigma)\tag{by Theorem~\ref{thm:sp-soundness}}\\
		\eeq &
		\lambda \tau\mydot - \inf_{\sigma\in\States,\tau\in \eval{C} {\sigma}}
		-g(\sigma) \tag{$\sup A = -\inf (-A)$}\\
		\eeq &	-\slp{C}{-g}~.
	\end{align*}
\end{proof}

\section{Proofs of Section~\ref{se:relationship}}
\subsection{Proof of Galois Connection between $\wlpsymbol$ and $\spsymbol$, Theorem~\ref{thm:wlp-sp-relationship}}

\wlpsprelationship*%
\begin{proof}
	\begin{align*}
		g\ppreceq \wlp{C}{f}
		\iff & \forall \sigma\in\States \mydot g(\sigma)\leq \wlp{C}{f}(\sigma) \\
		\iff & \forall \sigma \in\States \mydot g(\sigma) \leq	 \inf_{\tau \in\sem{C}{\sigma}}f(\tau)   \tag{by Theorem~\ref{thm:wlp-soundness}}  \\
		\iff & \forall \sigma,\tau\in\States\colon\tau \in \sem{C}{\sigma}\mydot
		g(\sigma) \leq f(\tau) \\
		\iff & \forall \tau\in\States \mydot \sup_{\sigma\in\States,\tau \in \sem{C}{\sigma}}
		g(\sigma) \leq f(\tau) \\
		\iff & \forall \tau\in\States \mydot \sp{C}{g}(\tau)\leq f(\tau) \tag{by Theorem~\ref{thm:sp-soundness}}\\\\
		\iff & \sp{C}{g}\ppreceq f~.
	\end{align*}%
\end{proof}
\subsection{Proof of Galois Connection between \textnormal{$\wpsymbol$} and \textnormal{$\slpsymbol$}, Theorem \ref{thm:wp-slp-relationship}}

\wpslprelationship*%
\begin{proof}
	\begin{align*}
		\wp{C}{f} \ppreceq g
		\iff & \forall \sigma\in\States \mydot \wp{C}{f}(\sigma)\leq g(\sigma) \\
		\iff & \forall \sigma \in\States \mydot 	 \sup_{\tau \in \sem{C}{\sigma}}f(\tau) \leq g(\sigma)   \tag{by Theorem~\ref{thm:wp-soundness}}  \\
		\iff & \forall \sigma,\tau\in\States\colon\tau \in \sem{C}{\sigma}\mydot  f(\tau)\leq
		g(\sigma)  \\
		\iff & \forall \tau\in\States \mydot  f(\tau)\leq \inf_{\sigma\in\States,\tau\in \sem{C}{\sigma}}
		g(\sigma) \\
		\iff & \forall \tau\in\States \mydot f(\tau)\leq \slp{C}{g}(\tau) \tag{by Theorem~\ref{thm:sp-soundness}}\\
		\iff & f \ppreceq \slp{C}{g}~.
	\end{align*}%
\end{proof}
%
\section{Proofs of Section~\ref{se:loops}}
\subsection{Proof of Induction Rules for Loops, Theorem~\ref{thm:quantitativeinductionrules}}
\quantitativeinductionrules*
\begin{proof}
	We prove each rule individually.\\
	For $\mathrm{while-}\wlpsymbol$ we have:
	\begin{align*}
		& i\ppreceq \iverson{\neg \guard} \curlywedge f \ccurlyvee \iverson{\guard} \curlywedge \wlp{C}{i} \tag{Premise of the rule}\\
		& \implies i \ppreceq \gfp X\mydot \iverson{\neg \guard} \curlywedge f \ccurlyvee \iverson{\guard} \curlywedge \wlp{C}{X}  \tag{by Park's Induction~\cite{Park69}}\\ 
		& \implies i \ppreceq \wlp{\WHILEDO{\guard}{C}}{f} \tag{by~\Cref{def:wlp}}\\
		& \implies g\ppreceq \wlp{\WHILEDO{\guard}{C}}{f} \tag{$g\ppreceq i$ and transitivity of $\ppreceq$}
	\end{align*}
	For $\mathrm{while-}\spsymbol$ we have:
	\begin{align*}
		& g \curlyvee \sp{C}{\iverson{\guard} \curlywedge i}
        \ppreceq i \tag{Premise of the rule}\\
		& \implies\lfp X\mydot g \curlyvee \sp{C}{\iverson{\guard} \curlywedge X} \ppreceq i \tag{by Park's Induction~\cite{Park69}}\\
		& \implies \inguard \curlywedge\lfp X\mydot g \curlyvee \sp{C}{\iverson{\guard} \curlywedge X}\ppreceq \inguard \curlywedge i \tag{by monotonicity of $\lambda X. \inguard  \curlywedge X$}\\
		& \implies \sp{\WHILEDO{\guard}{C}}{g}\ppreceq \iverson{\neg\guard}\curlywedge i \tag{by~\Cref{def:sp}}\\
		& \implies \sp{\WHILEDO{\guard}{C}}{g}\ppreceq f \tag{$\inguard\curlywedge i \ppreceq f$ and transitivity of $\ppreceq$}
	\end{align*}
	For $\mathrm{while-}\wpsymbol$ we have:
	\begin{align*}
		& \iverson{\neg \guard} \curlywedge f \ccurlyvee \iverson{\guard} \curlywedge \wp{C}{i} \ppreceq i \tag{Premise of the rule}\\
		& \implies\lfp X\mydot \iverson{\neg \guard} \curlywedge f \ccurlyvee \iverson{\guard} \curlywedge \wp{C}{X} \ppreceq i  \tag{by Park's Induction~\cite{Park69}}\\ 
		& \implies 
		\wp{\WHILEDO{\guard}{C}}{f}\ppreceq i \tag{by~\Cref{def:wp}}\\
		& \implies \wp{\WHILEDO{\guard}{C}}{f}\ppreceq g \tag{$i\ppreceq g$ and transitivity of $\ppreceq$}
	\end{align*}
	For $\mathrm{while-}\slpsymbol$ we have:
	\begin{align*}
		& i \ppreceq g \ccurlywedge \slp{C}{\inguard \curlyvee i} \tag{Premise of the rule}\\
		& \implies i \ppreceq \gfp X\mydot g \curlywedge \slp{C}{\inguard \curlyvee X} \tag{by Park's Induction~\cite{Park69}}\\
		& \implies \iguard \curlyvee i \ppreceq \iguard \curlyvee\gfp X\mydot g \curlywedge \slp{C}{\inguard \curlyvee X}
		\tag{by monotonicity of $\lambda X. \iguard \curlyvee X$}\\
		& \implies \iguard \curlyvee i\ppreceq \slp{\WHILEDO{\guard}{C}}{g}\tag{by~\Cref{def:slp}}\\
		& \implies f \ppreceq \slp{\WHILEDO{\guard}{C}}{g} \tag{$f\ppreceq \iguard \curlyvee i $ and transitivity of $\ppreceq$}
	\end{align*}
\end{proof}

\subsection{Proof of \Cref{prop:spslpconvergence}}
\spslpconvergence*
\begin{proof} We prove each statement individually. Let $^\spsymbol \Psi_f$ and $^\slpsymbol \Psi_f$ be, respectively, the $\spsymbol$--charac{\-}teristic and $\slpsymbol$--charac{\-}teristic functions of $\WHILEDO{\guard}{C}$.\\
	For $\spsymbol$ we have:
	\begin{align*}
		& \iguard \curlywedge f\ppreceq f \\
		& \sp{C}{\iguard \curlywedge f}\ppreceq \sp{C}{f} \tag{by Monotonicity of $\spsymbol$}\\
		& \sp{C}{\iguard \curlywedge f}\ppreceq f
		\tag{by hypothesis and transitivity of $\ppreceq$}\\
		& f \curlyvee\sp{C}{\iguard \curlywedge f}\ppreceq f\tag{by Monotonicity of $\lambda X\mydot f\curlyvee X$}\\
		& ^\spsymbol \Psi_f^2(\minfty)\ppreceq ^\spsymbol \Psi_f(\minfty)\tag{by \Cref{def:sp}}
	\end{align*}
	Hence, the Kleene's iterates have converged immediately and the least fixpoint is exactly:
	\begin{align*}
		\lfp X\mydot ^\spsymbol \Psi_f(X)\eeq ^\spsymbol \Psi_f(\minfty)\eeq f~,
	\end{align*}
and thus we conclude:
\begin{align*}
\sp{\WHILEDO{\guard}{C}}{f} \eeq & 
\inguard \curlywedge	\lfp X\mydot ^\spsymbol \Psi_f(X) \tag{by \Cref{def:sp}}\\
 \eeq & \inguard \curlywedge f\tag{$\lfp X\mydot ^\spsymbol \Psi_f(X)\eeq f$}~.
\end{align*}
	For $\slpsymbol$ we have:
	\begin{align*}
		& f \ppreceq \inguard\curlyvee f \tag{$\dagger$}\\
		& \slp{C}{f}\ppreceq\slp{C}{\inguard \curlyvee f} \tag{by Monotonicity of $\slpsymbol$}\\
		& f\ppreceq \slp{C}{\inguard \curlyvee f}
		\tag{by hypothesis and transitivity of $\ppreceq$}\\
		& f\ppreceq f \curlywedge\slp{C}{\inguard \curlyvee f}\tag{by Monotonicity of $\lambda X\mydot f\curlywedge X$}\\
		& ^\slpsymbol \Psi_f(\pinfty)\ppreceq ^\slpsymbol \Psi_f^2(\pinfty)\tag{by \Cref{def:slp}}
	\end{align*}
	Hence, the Kleene's iterates have converged immediately and the greatest fixpoint is exactly:
	\begin{align*}
		\gfp X\mydot ^\slpsymbol \Psi_f(X)= ^\slpsymbol \Psi_f(\pinfty)=f~,
	\end{align*}
and thus we conclude:
\begin{align*}
	\sp{\WHILEDO{\guard}{C}}{f} \eeq & 
	\iguard \curlyvee	\gfp X\mydot ^\slpsymbol \Psi_f(X) \tag{by \Cref{def:slp}}\\
	\eeq & \iguard \curlyvee f\tag{$\gfp X\mydot ^\slpsymbol \Psi_f(X)\eeq f$}~.
\end{align*}
\end{proof}
\section{Full calculations of Section~\ref{se:examples}}
\label{app:examples}
\subsection{Full calculations of Example~\ref{ex:1}}
\begin{example}
	The strongest post of $C=\ITE{hi>7}{\ASSIGN{lo}{99}}{\ASSIGN{lo}{80}}$ for the preanticipation $hi=\lambda \sigma\mydot\sigma(hi)$ are:
\begin{align*}
	& \sp{\ITE{hi>7}{\ASSIGN{lo}{99}}{\ASSIGN{lo}{80}}}{hi}\\
	& \eeq \sp{\ASSIGN{lo}{99}}{\iverson{hi>7}\curlywedge hi}\curlyvee\sp{\ASSIGN{lo}{80}}{\iverson{hi\leq7}\curlywedge hi}\\
	& \eeq \SupV{\alpha}~\iverson{lo=99}\curlywedge (\iverson{hi>7}\curlywedge hi)\subst{lo}{\alpha} \ccurlyvee
	\SupV{\alpha}~\iverson{lo=80}\curlywedge (\iverson{hi\leq7}\curlywedge hi)\subst{lo}{\alpha}\\
	& \eeq \iverson{lo=99}\curlywedge \iverson{hi>7}\curlywedge hi\ccurlyvee
	\iverson{lo=80}\curlywedge \iverson{hi\leq7}\curlywedge hi ~
\end{align*}
and
\begin{align*}
	& \slp{\ITE{hi>7}{\ASSIGN{lo}{99}}{\ASSIGN{lo}{80}}}{hi}\\
	& \eeq \slp{\ASSIGN{lo}{99}}{\iverson{hi\leq7}\curlyvee hi}\ccurlywedge\slp{\ASSIGN{lo}{80}}{\iverson{hi>7}\curlyvee hi}\\
	& \eeq \bigl(\InfV{\alpha}~\iverson{lo \neq 99}\curlyvee (\iverson{hi\leq7}\curlyvee hi)\subst{lo}{\alpha}\bigr) \ccurlywedge
	\bigl(\InfV{\alpha}~\iverson{lo \neq 80}\curlyvee (\iverson{hi>7}
	\curlyvee  hi)\subst{lo}{\alpha}\bigr)\\
	& \eeq \bigl(\iverson{lo \neq 99}\curlyvee \iverson{hi\leq7}\curlyvee  hi\bigr)\ccurlywedge \bigl(
	\iverson{lo \neq 80}\curlyvee \iverson{hi>7}\curlyvee  hi\bigr)~.
\end{align*}
\end{example}
\subsection{Full calculations of Example~\ref{ex:2}}
\begin{example}
	The strongest post of $C=\COMPOSE{\ASSIGN{hi}{hi+5}}{\WHILEDO{lo<hi}{\ASSIGN{lo}{lo+1}}}$ for the preanticipation $hi=\lambda \sigma\mydot\sigma(hi)$ are:
	\begin{align*}
		\sp{C}{hi} & \eeq \iverson{lo\geq hi}\curlywedge(hi-5) \\
		\slp{C}{hi} & \eeq \iverson{lo< hi}\curlyvee (hi-5)\\
	\end{align*}
In fact, we have:
\begin{align*}
	& \sp{
		\COMPOSE{
			\ASSIGN{hi}{hi+5}
		}
		{\WHILEDO{lo<hi}
			{\ASSIGN{lo}{lo+1}}
		}
	}
	{hi}\\
	& \eeq \sp{
		\WHILEDO{lo<hi}
		{\ASSIGN{lo}{lo+1}}
	}
	{\sp{\ASSIGN{hi}{hi+5}}{hi}}\\
	& \eeq \sp{
		\WHILEDO{lo<hi}
		{\ASSIGN{lo}{lo+1}}
	}
	{\SupV{\alpha}~\iverson{hi=\alpha+5}\curlywedge \alpha}\\
	& \eeq \sp{
		\WHILEDO{lo<hi}
		{\ASSIGN{lo}{lo+1}}
	}
	{hi-5}\tag{$\alpha=hi-5$ is selected}\\
	& \eeq 
	\iverson{lo\geq hi}\curlywedge\lfp X \mydot\Psi_{hi-5}(X)\\
	& \eeq 
	\iverson{lo\geq hi}\curlywedge\Psi_{hi-5}^\omega(\minfty) \tag{by Kleene's fixpoint theorem}
\end{align*}
Let us compute some Kleene's iterates:
\begin{align*}
	\Psi_{hi-5}(\minfty)&\eeq (hi-5) \ccurlyvee \sp{\ASSIGN{lo}{lo+1}}{ \iverson{lo<hi}\curlywedge  \minfty}\\
	&\eeq (hi-5) \ccurlyvee \sp{\ASSIGN{lo}{lo+1}}{ \minfty}\\
	&\eeq (hi-5) \ccurlyvee  (\minfty) \tag{by \Cref{thm:wpwlpspslphealthiness}~(\ref{thm:wpwlpspslphealthiness:strictness})}\\
	&\eeq (hi-5)\\
	\Psi_{hi-5}^2(\minfty)&\eeq (hi-5) \ccurlyvee \sp{\ASSIGN{lo}{lo+1}}{\iverson{lo<hi}\curlywedge (hi-5)}\\
	&\eeq (hi-5) \ccurlyvee (\SupV{\alpha}~\iverson{lo=\alpha+1}\curlywedge \iverson{\alpha<hi} \curlywedge(hi-5))\\
	&\eeq (hi-5) \ccurlyvee (\iverson{lo<hi+1} \curlywedge(hi-5))\tag{$\alpha=lo-1$ is selected}\\
	&\eeq (hi-5)
\end{align*}
The iteration sequence has converged (in just 2 iterations), so we obtain:
\begin{align*}
	&\sp{
		\COMPOSE{
			\ASSIGN{hi}{hi+5}
		}
		{\WHILEDO{lo<hi}
			{\ASSIGN{lo}{lo+1}}
		}
	}
	{hi}\\
	&\eeq\iverson{lo\geq hi}\curlywedge\Psi_{hi-5}^\omega(\minfty)\\
	&\eeq \iverson{lo\geq hi}\curlywedge(hi-5)
\end{align*}
Similarly, for $\slpsymbol$ we have:
\begin{align*}
	& \slp{
		\COMPOSE{
			\ASSIGN{hi}{hi+5}
		}
		{\WHILEDO{lo<hi}
			{\ASSIGN{lo}{lo+1}}
		}
	}
	{hi}\\
	& \eeq \slp{
		\WHILEDO{lo<hi}
		{\ASSIGN{lo}{lo+1}}
	}
	{\slp{\ASSIGN{hi}{hi+5}}{hi}}\\
	& \eeq \slp{
		\WHILEDO{lo<hi}
		{\ASSIGN{lo}{lo+1}}
	}
	{\InfV{\alpha}~\iverson{hi \neq \alpha+5}\curlyvee \alpha}\\
	& \eeq \slp{
		\WHILEDO{lo<hi}
		{\ASSIGN{lo}{lo+1}}
	}
	{hi-5}\tag{$\alpha=hi-5$ is selected}\\
	& \eeq 
	\iverson{lo< hi}\curlyvee\gfp X \mydot\Psi_{hi-5}(X)\\
	& \eeq 
	\iverson{lo< hi}\curlyvee\Psi_{hi-5}^\omega(\pinfty) \tag{by Kleene's fixpoint theorem}
\end{align*}
Let us compute some Kleene's iterates:
\begin{align*}
	\Psi_{hi-5}(\pinfty)&\eeq (hi-5) \ccurlywedge \slp{\ASSIGN{lo}{lo+1}}{ \iverson{lo \geq hi} \curlyvee \pinfty}\\
	&\eeq (hi-5) \ccurlywedge \slp{\ASSIGN{lo}{lo+1}}{\pinfty}\\
	&\eeq (hi-5) \ccurlywedge \pinfty \tag{by \Cref{thm:wpwlpspslphealthiness}~(\ref{thm:wpwlpspslphealthiness:costrictness})}\\
	&\eeq (hi-5)\\
	\Psi_{hi-5}^2(\pinfty)&\eeq (hi-5) \ccurlywedge \slp{\ASSIGN{lo}{lo+1}}{\iverson{lo\geq hi}\curlyvee (hi-5)}\\
	&\eeq (hi-5) \ccurlywedge (\InfV{\alpha}~\iverson{lo\neq\alpha+1}\curlyvee \iverson{\alpha\geq hi} \curlyvee(hi-5))\\
	&\eeq (hi-5) \ccurlywedge (\iverson{lo\geq hi+1} \curlyvee (hi-5))\tag{$\alpha=lo-1$ is selected}\\
	&\eeq (hi-5)
\end{align*}
Again, the iteration sequence has converged in 2 iterations, so we conclude:
\begin{align*}
	&\slp{
		\COMPOSE{
			\ASSIGN{hi}{hi+5}
		}
		{\WHILEDO{lo<hi}
			{\ASSIGN{lo}{lo+1}}
		}
	}
	{hi}\\
	&\eeq\iverson{lo< hi}\curlyvee\Psi_{hi-5}^\omega(\pinfty)\\
	&\eeq \iverson{lo< hi}\curlyvee (hi-5)
\end{align*}
\end{example}

\section{Extended comparison with~\cite{Aguirre2020WeakestPI}}
\label{app:related-work}
In this section, we show how our $\textsf{w(l)p}$, \emph{restricted to the fragment of loop-free programs}, can be derived by instantiating~\cite[Corollary 4.6]{Aguirre2020WeakestPI}. Consider:
\begin{itemize}
\item the powerset monad $\mathcal{P}$;
\item the lattice of extended reals $\Rinf$;
\item the Eilenberg-Moore algebra $\sup\colon\powerset \Rinf\to\Rinf$.
\end{itemize}
As a consequence of~\cite[Corollary 4.6]{Aguirre2020WeakestPI}, we obtain an abstract operation $\awpsymbol\colon (A\to\powerset{B})\to(B\to\Rinf)\to(A\to\Rinf)$ such that:
\begin{align*}
    \awpsymbol (C)(f)(a)=\sup_{b\in C(a)} f(b)
\end{align*}
Note that $\awpsymbol$ preserves all joins in the position of $f$. By taking as monad the collecting semantics starting from a single state $\eval{C}\colon\States\to\powerset{\States}$ which maps states into set of states, for all loop-free programs $C$, $f\in\A,\sigma\in\States$ we have:
\begin{align*}
    \awpsymbol (\eval{C})(f)(\sigma) = \sup_{\tau\in \sem{C}{\sigma}} f(\tau) = \wp{C}{f}~.
\end{align*}
Similarly, if we consider the Eilenberg-Moore algebra $\inf$, we obtain an abstract operator $\awlpsymbol$ such that:
\begin{align*}
    \awlpsymbol (\eval{C})(f)(\sigma) = \inf_{\tau\in \sem{C}{\sigma}} f(\tau) = \wlp{C}{f}~.
\end{align*}

\end{document}